\documentclass[10pt]{article}
\pdfoutput=1

\usepackage{jheppub}
\usepackage{amssymb}
\usepackage{amsmath}
\usepackage{amsthm}
\usepackage{mathtools}
\usepackage[usenames,dvipsnames]{xcolor}
\usepackage{epsfig}
\usepackage{dcolumn}
\usepackage{upgreek}
\usepackage{setspace}
\usepackage{enumitem}
\usepackage{array,multirow,bigdelim,arydshln}
\usepackage{appendix}
\usepackage[export]{adjustbox}
\usepackage{xparse}
\usepackage[utf8]{inputenc}
\usepackage{microtype}
\usepackage{blkarray}
\usepackage{bigstrut}
\usepackage{nccmath}
\usepackage{stmaryrd}
\usepackage{shuffle}

\hypersetup{
	colorlinks=true,
	urlcolor=Maroon,
	linkcolor=Maroon,
	citecolor=Maroon,
	pdftitle={Perturbiner Methods for Effective Field Theories and the Double Copy},
	pdfauthor={Sebastian Mizera and Barbara Skrzypek},
	pdfdisplaydoctitle=true,
	pdfstartview=FitH
}

\def \be {\begin{equation}}
\def \ee {\end{equation}}
\def \nn {\nonumber}

\def \A {\mathbb{A}}
\def \F {\mathbb{F}}
\def \X {\mathbb{X}}
\def \Y {\mathbb{Y}}
\def \Z {\mathbb{Z}}
\def \XX {\mathbf{X}}
\def \YY {\mathbf{Y}}
\def \ZZ {\mathbf{Z}}
\def \XXX {\mathcal{X}}
\def \YYY {\mathcal{Y}}
\def \ZZZ {\mathcal{Z}}

\DeclareMathOperator{\Tr}{Tr}
\newcommand{\overbar}[1]{\mkern 1.5mu\overline{\mkern-1.5mu#1\mkern-1.5mu}\mkern 1.5mu}

\newcommand\topstrut[1][1.2ex]{\setlength\bigstrutjot{#1}{\bigstrut[t]}}
\newcommand\botstrut[1][0.9ex]{\setlength\bigstrutjot{#1}{\bigstrut[b]}}

\allowdisplaybreaks

\title{Perturbiner Methods\\ for Effective Field Theories and the Double Copy}

\author[a,b]{Sebastian Mizera}\emailAdd{smizera@pitp.ca}
\author[a,b,c]{and Barbara Skrzypek}\emailAdd{bskrzypek@g.harvard.edu}

\affiliation[a]{Perimeter Institute for Theoretical Physics, Waterloo, ON N2L 2Y5, Canada}
\affiliation[b]{Department of Physics \& Astronomy, University of Waterloo, Waterloo, ON N2L 3G1, Canada}
\affiliation[c]{Department of Physics, Harvard University, Cambridge, MA 02138, USA}

\abstract{Perturbiner expansion provides a generating function for all Berends--Giele currents in a given quantum field theory. We apply this method to various effective field theories with and without color degrees of freedom. In the colored case, we study the $\text{U}(N)$ non-linear sigma model of Goldstone bosons (NLSM) in a recent parametrization due to Cheung and Shen, as well as its extension involving a coupling to the bi-adjoint scalar. We propose a Lagrangian and a Cachazo--He--Yuan formula for the latter valid in multi-trace sectors and systematically calculate its amplitudes. Furthermore, we make a similar proposal for a higher-derivative correction to NLSM that agrees with the subleading order of the abelian Z-theory. In the colorless cases, we formulate perturbiner expansions for the special Galileon and Born--Infeld theories. Finally, we study Kawai--Lewellen--Tye-like double-copy relations for Berends--Giele currents between the above colored and colorless theories. We find that they hold up to pure gauge terms, but without the need for further field redefinitions.}

\begin{document}

\maketitle
\addtocontents{toc}{\protect\setcounter{tocdepth}{1}}
\numberwithin{equation}{section}
\setcounter{page}{2}

\section{Introduction}

It is often said that tree-level scattering amplitudes encode all solutions of the classical field equations in massless quantum field theories. This statement was made concrete by Rosly and Selivanov \cite{Rosly:1996vr,Rosly:1997ap,Selivanov:1997aq,Selivanov:1997ts,Rosly:1998vm,Selivanov:1998hn}, who introduced an ansatz for such solutions as an infinite expansion in terms of plane-wave states, which, at the same time, can be thought of as a generating function for all tree-level scattering amplitudes in a given theory. This ansatz is called the \emph{perturbiner} expansion.

Perturbiner methods have been used to analyze various aspects of scattering amplitudes. In particular, they provide a method for deriving Berends--Giele recursion relations \cite{Berends:1987me}, and they simplify the extraction of kinematic numerators in super Yang--Mills theory \cite{Mafra:2015gia,Lee:2015upy,Mafra:2015vca}. Furthermore, they have been used to obtain a systematic procedure for the low-energy expansion of string-theory integrals \cite{Mafra:2016mcc}, compute Kawai--Lewellen--Tye (KLT) matrices \cite{Mafra:2016ltu}, and manifest the MHV vertex expansion of Yang--Mills theory on the level of the Lagrangian \cite{Gorsky:2005sf}. In a companion paper \cite{GQS}, Garozzo, Queimada, and Schlotterer use perturbiners to explore the connection between low-energy expansion of bosonic string amplitudes and the color-kinematics duality \cite{Bern:2008qj,Bern:2010ue}.

In this work we apply perturbiner methods to compute currents in effective field theories: the $\text{U}(N)$ non-linear sigma model (NLSM), special Galileon (sGal), and Born--Infeld (BI) theories. In particular, we use a recent formulation due to Cheung and Shen \cite{Cheung:2016prv}, who constructed the Lagrangians for these theories featuring only a finite number of interaction vertices, which at the same time manifests their double-copy properties \cite{Cachazo:2014xea}. It was shown that such a prescription can be understood as an embedding of Yang--Mills and Einstein gravity theories in higher space-time dimensions \cite{Cheung:2016prv,Cheung:2017yef}. We use an analogous construction that starts with Yang--Mills-scalar theory \cite{Chiodaroli:2014xia} in higher dimension and naturally induces couplings between NLSM and bi-adjoint scalars, known from \cite{Cachazo:2016njl}. This provides a realization of the Lagrangian computing mixed-species amplitudes studied in \cite{Cachazo:2016njl}, including all multi-trace contributions.

We further explore the idea of dimensional reduction to construct Lagrangians of effective field theories. We consider the abelian sector of Z-theory, which is a string-theoretic model that in the low-energy limit reduces to the NLSM \cite{Carrasco:2016ldy}. We focus on the first subleading $\alpha'$ correction to the NLSM at order $\alpha'^2$. Following \cite{Mizera:2017sen} we start with the $F^3$ gauge theory in higher dimension and show evidence that, after dimensional reduction, it reduces to the subleading $\alpha'$ order of the abelian sector of Z-theory. This gives an explicit realization of the correction to the effective Lagrangian of abelian Z-theory. Its perturbiner expansion gives an efficient way of computing abelian Z-theory amplitudes up to the first subleading $\alpha'$ order.

For theories without color degrees of freedom one needs to modify the perturbiner ansatz, which normally relies on the Lie algebra generators as a book-keeping device. Following the original Berends--Giele construction \cite{Berends:1987me} we consider color-dressed perturbiners which can be applied to any quantum field theory, yielding, for example, a color-dressed version of perturbiners in the Yang--Mills theory. We apply this perturbiner ansatz to the special Galileon and Born--Infeld theories in the formulation of \cite{Cheung:2016prv,Cheung:2017yef}, which resembles that of the NLSM and in particular has only a finite number of interaction vertices. As a consequence, it gives Berends--Giele recursion relations with a finite number of terms.

It is well-known that tree-level amplitudes in the effective field theories of our interest satisfy double-copy relations \cite{Cachazo:2014xea}. We explore the possibility that these relations extend to the off-shell objects, namely the Berends--Giele currents computed using the perturbiner expansion. We find that these currents coming from the specific representation of NLSM, sGal, and BI from \cite{Cheung:2016prv,Cheung:2017yef} satisfy a version of KLT relations \cite{Kawai:1985xq} up to pure gauge terms, but without further field redefinitions. To be precise, we find the KLT formula of the schematic form:
\be\label{eq:intro}
\big(J_{12\cdots m}^{\Lambda \bar{\Lambda}}\big)^{\text{theory}_1 \otimes \text{theory}_2} = \sum_{ \rho, \tau \in S_{m-1} } \big(J_{1\rho}^{\Lambda}\big)^{\text{theory}_1} S[\rho | \tau]_1 \, \big(J_{1\tau}^{\bar{\Lambda}}\big)^{\text{theory}_2} \;+\; \big(\Delta^{\Lambda \bar{\Lambda}}_{12\cdots m}\big)^{\text{theory}_1 \otimes \text{theory}_2}.
\ee
The left-hand side features the current $J_{12\cdots m}^{\Lambda \bar{\Lambda}}$ in a colorless theory depending on the particle labels $12\cdots m$ and possibly Lorentz indices $\Lambda\bar\Lambda$. On the right-hand side we have currents $J_{1\rho}^{\Lambda}$ and $J_{1\tau}^{\bar{\Lambda}}$ of two colored theories. The sum goes over $(m{-}1)!$ permutations $\rho$ and $\tau$ entering the definition of the currents. Here $S[\rho|\tau]_1$ denotes an inverse matrix of bi-adjoint scalar \emph{currents}. In addition, we allow terms $\Delta^{\Lambda \bar{\Lambda}}_{12\cdots m}$ that are pure gauge (removable by gauge transformations). Notice that this is a non-trivial statement, since currents are off-shell quantities and are defined only up to field redefinitions. We find that no field redefinitions are necessary to make \eqref{eq:intro} work. When the currents are used to compute amplitudes with $n=m{+}1$ external states, \eqref{eq:intro} become the $(n{-}2)!$ KLT relations known from \cite{BjerrumBohr:2010ta,BjerrumBohr:2010zb}.

This paper is structured as follows. In Section~\ref{sec:perturbiner-methods} we review perturbiner methods for theories with and without color degrees of freedom. We illustrate how to use them to derive off-shell recursion relations on the examples of Yang--Mills and bi-adjoint scalar theories. In Section~\ref{sec:EFTs-with-colors} we apply perturbiner methods to effective field theories with colors: NLSM and its extension including interactions with the bi-adjoint scalar. We propose a CHY formula for the latter in the multi-trace sector. We also consider a subleading correction to the abelian Z-theory effective Lagrangian. In Section~\ref{sec:EFTs-without-colors} we use perturbiner methods to formulate recursion relations for the special Galileons and Born--Infeld theories. We put these two types of perturbiner expansions together in Section~\ref{sec:double-copy}, where we study the KLT double-copy for currents. We conclude with a discussion of the results and future directions in Section~\ref{sec:discussion}. More lengthy examples of amplitudes are listed in Appendix~\ref{app:example-amplitudes}.

\section{\label{sec:perturbiner-methods}Perturbiner Methods}

In this section we introduce two different types of perturbiner expansions: color-stripped and color-dressed, which can be used to construct recursion relations for various theories with and without color degrees of freedom.

\subsection{Color-Stripped Perturbiners}

Color-stripped perturbiners can be used to construct Berends--Giele currents and partial amplitudes for theories with colors. Here we do so for gauge theory and the bi-adjoint scalar, whose amplitudes will be used later in the text. Method of deriving currents for other theories, such as the NLSM, will follow exactly the same steps.

\subsubsection{\label{sec:color-stripped-YM}Yang--Mills Theory}

We begin by reviewing the case of $\text{U}(N)$ Yang--Mills theory in general space-time dimension, whose perturbiner expansion was first studied from a superspace perspective in \cite{Mafra:2015gia,Lee:2015upy,Mafra:2015vca}. The gauge-theory Lagrangian $\mathcal{L}^{\text{YM}} = -\frac{1}{4} \Tr \F_{\mu\nu} \F^{\mu\nu}$ leads to the following equations of motion for the Lie algebra-valued gauge field $\A^\mu = A^\mu_a T^a$:
\be\label{eq:A-eom}
\Box \A^\mu = [ \A_\nu , \partial^\nu \A^\mu + \F^{\nu\mu}],
\ee
in the Lorenz gauge $\partial_\mu \A^\mu = 0$, where $\F^{\mu\nu} = F^{\mu\nu}_a T^a = \partial^\mu \A^\nu - \partial^\nu \A^\mu - [\A^\mu, \A^\nu]$ is the field strength. In terms of the covariant derivative, $\nabla_\mu = \partial_\mu - [\A_{\mu}, \cdot\,]$, it satisfies $[\nabla_\mu, \F^{\mu\nu}] = 0$. We use the normalization in which the $\text{U}(N)$ generators satisfy $\Tr (T^{a} T^{b}) = \delta^{ab}$ and $[T^a, T^b] = {f}_{abc} T^c$, where ${f}_{abc}$ are the structure constants. We will often move the Lorentz and color indices up and down for the sake of clarity.

In order to linearize the above equations, we introduce an ansatz for the gauge field in terms of plane-wave states:
\be\label{eq:perturiner-gauge-field}
\A^{\mu}(x) := \sum_{P} A^\mu_P\,  T^P e^{k_P \cdot x} = \sum_i A_i^\mu\, T^{a_i} e^{k_i \cdot x} + \sum_{i,j} A_{ij}^\mu\, T^{a_i} T^{a_j} e^{k_{ij} \cdot x} + \ldots.
\ee
This is the perturbiner expansion. Here the sum goes over non-empty words $P = 12\cdots m$, such that
\be\label{eq:T-k}
T^{P} := T^{a_1} T^{a_2} \cdots T^{a_m}, \qquad k_{P}^\mu := k_{1}^\mu + k_2^\mu + \cdots + k_m^\mu.
\ee
The infinite number of colors $a_i$ and momenta $k_i^\mu$ should, for now, be understood as auxiliary labels of the expansion and do not satisfy any constraints. We take the momenta $k_i^\mu$ to be imaginary for later convenience (this does not cause problems in tree-level computations). Notice that \eqref{eq:perturiner-gauge-field} is valued not in $\text{U}(N)$, but in its universal enveloping algebra. It will of course become Lie algebra-valued after solving for constraints given by the equations of motion. Similarly, we can write down the perturbiner expansion for the field strength:
\be\label{eq:perturiner-field-strength}
\F^{\mu\nu}(x) := \sum_{P} F_P^{\mu\nu}\, T^P e^{k_P \cdot x}.
\ee
The reason we refer to the above perturbiner expansions as ``color-stripped" is because the coefficients $A_P^{\mu}$ and $F_P^{\mu\nu}$ appearing in the expansion do not have any color degrees of freedom. These coefficients are Berends--Giele currents \cite{Berends:1987me,Mafra:2015gia}. To obtain recursion relations, we substitute them into the equations of motion \eqref{eq:A-eom} and the definition of the field strength and collecting terms of the same order, i.e., with the same number of Lie algebra generators $T^{a_i}$ on both sides. At the linear order, we have:
\be
\sum_{i} k_i^2 A_i^\mu\, T^{a_i} e^{k_i \cdot x} = 0, \qquad \sum_{i} F^{\mu\nu}_i\, T^{a_i} e^{k_i \cdot x} = \sum_{i} \left( k_i^\mu A_i^\nu - k_i^\nu A_i^\mu \right) T^{a_i} e^{k_i \cdot x},
\ee
which is equivalent to imposing the momenta to be null, $k_i^2 = 0$ and that $F_i^{\mu\nu} = k_i^\mu A_i^\nu - k_i^\nu A_i^\mu$. At the quadratic order we find the following constraint:
\begin{align}
\sum_{i,j} k_{ij}^2 \, A_{ij}^\mu \, T^{a_i} T^{a_j} e^{k_{ij} \cdot x} &= \sum_{i,j} \left[ A_i^\nu\, T^{a_i} e^{k_i \cdot x} , \left( k_j^\nu A_j^\mu + F^{\nu\mu}_j \right) T^{a_j} e^{k_j \cdot x} \right]\nn\\
&=\sum_{i,j} \Big( (A_i \cdot k_j) A^\mu_j + A_i^\nu F^{\nu\mu}_j \Big) [T^{a_i}, T^{a_j}]\, e^{k_{ij} \cdot x}.
\end{align}
Here we find that $A_{ij}^\mu = -A_{ji}^\mu$, so that the generators on the left-hand side organize into commutators. In order to see this, we decompose the sums on both sides into those over $i<j$ and $j<i$ (for $i=j$ we have $A_{ii}^\mu = 0$ straightforwardly), which after relabeling $i \leftrightarrow j$ in the second sum gives:
\be
\sum_{i<j} k_{ij}^2 \, A_{ij}^\mu \, [T^{a_i}, T^{a_j}] \, e^{k_{ij} \cdot x} = \sum_{i<j} \Big( (A_i \cdot k_j) A^\mu_j + A_i^\nu F^{\nu\mu}_j - (i \leftrightarrow j ) \Big) [T^{a_i}, T^{a_j}]\, e^{k_{ij} \cdot x}.
\ee
Comparing the coefficients of each $[T^{a_i}, T^{a_j}]$ we find that $A_{ij}^\mu$ satisfies the recursion relation:
\be
A_{ij}^\mu = \frac{1}{k_{ij}^2} \Big( (A_i \cdot k_j) A^\mu_j + A_i^\nu F^{\nu\mu}_j - (i \leftrightarrow j )\Big).
\ee
Similarly, for the field strength we have at the quadratic order:
\begin{align}
\sum_{i,j} F_{ij}^{\mu\nu} \, T^{a_i} T^{a_j} e^{k_{ij} \cdot x} &= \sum_{i,j} \left( k_{ij}^\mu A_{ij}^\nu - k_{ij}^\nu A_{ij}^\mu \right) T^{a_i} T^{a_j} e^{k_{ij}\cdot x} - \sum_{i,j} \left[ A_i\, T^{a_i} e^{k_i\cdot x}, A_j\, T^{a_j} e^{k_j\cdot x} \right]\nn\\
&= \sum_{i,j} \Big( \frac{1}{2} k_{ij}^\mu A_{ij}^\nu - \frac{1}{2} k_{ij}^\nu A_{ij}^\mu - A_i^\mu A_j^\nu \Big) [T^{a_i}, T^{a_j}] \, e^{k_{ij}\cdot x},
\end{align}
where in the second line we used antisymmetry of $A_{ij}^\mu$ in $ij$. After using the same property of $F_{ij}^{\mu\nu}$ and comparing coefficients of each commutator we obtain:
\be
F_{ij}^{\mu\nu} = k_{ij}^\mu A_{ij}^\nu - k_{ij}^\nu A_{ij}^\mu - \Big( A_i^\mu A_j^\nu - (i\leftrightarrow j)\Big).
\ee
Following the same procedure at the cubic order, using symmetries $A_{ijk}^\mu + A_{jik}^\mu + A_{jki}^\mu = A_{ijk}^\mu + A_{ikj}^\mu + A_{jik}^\mu = 0$ and similar ones for $F_{ijk}^{\mu\nu}$, we find:
\begin{gather}
A_{ijk}^\mu = \frac{1}{k_{ijk}^2} \Big[ \Big( (A_{i} \cdot k_{jk}) A_{jk}^\mu +  A_{i}^\nu F^{\nu\mu}_{jk} - (i \leftrightarrow jk) \Big) + \Big( (A_{ij} \cdot k_{k}) A_{k}^\mu + A_{ij}^\nu F^{\nu\mu}_{k} + (ij \leftrightarrow k) \Big) \Big],\\
F_{ijk}^{\mu\nu} = k_{ijk}^\mu A_{ijk}^\nu - k_{ijk}^\nu A_{ijk}^\mu - \Big[ \Big( A_{i}^\mu A_{jk}^\nu - (i \leftrightarrow jk) \Big) + \Big( A_{ij}^\mu A_{k}^\nu - (ij \leftrightarrow k) \Big) \Big].
\end{gather}
The pattern continues to recursions for higher-rank perturbiners. We have the shuffle symmetries \cite{Lee:2015upy,Berends:1988zn}:
\be
A_{P \shuffle Q}^\mu = F_{P \shuffle Q}^{\mu\nu} = 0 \qquad \text{for all}\qquad P,Q \neq \varnothing
\ee
Recall that a shuffle product $ P \shuffle Q$ of two words $P$ and $Q$ is defined as a sum over all permutations of $P \cup Q$ that preserve orderings of both words $P$ and $Q$, for example $A^{\mu}_{1 \shuffle 2} = A^{\mu}_{12} + A^{\mu}_{21}$, $A^{\mu}_{1 \shuffle 23} = A^{\mu}_{123} + A^{\mu}_{213} + A^{\mu}_{231}$, and $A^{\mu}_{12 \shuffle 3} = A^{\mu}_{123} + A^{\mu}_{132} + A^{\mu}_{312}$. With these symmetries, one finds the following recursion relations for perturbiner coefficients
\cite{Mafra:2015gia}:
\begin{gather}\label{eq:A-recursion}
A^\mu_P = \frac{1}{2s_P} \sum_{P=QR} \Big( (A_Q \cdot k_R) A_R^\mu + A_Q^{\nu} F_R^{\nu\mu} - (Q \leftrightarrow R) \Big),\\
F^{\mu\nu}_P = k_P^\mu A_P^\nu - k_P^\nu A_P^\mu \,-\! \sum_{P=QR}  \Big( A_Q^\mu A_R^\nu - (Q \leftrightarrow R) \Big).
\end{gather}
Here $s_P := \frac{1}{2} k_P^2$ are the Mandelstam invariants. The sum goes over all deconcatenations of the word $P=p_1 p_2 \cdots p_m$ into non-empty ordered words, i.e., $Q=p_1 p_2 \cdots p_j$ and $R = p_{j{+}1} p_{j{+}2} \cdots p_m$ for $j=1,2,\ldots, m{-}1$. The antisymmetrization $(Q \leftrightarrow R)$ is a consequence of the commutators in the equations of motion and the definition of the field strength. (Note that the separation of the perturbiners into those for $\A^\mu$ and $\F^{\mu\nu}$ is arbitrary and one could equally well construct a recursion for $\A^\mu$ alone, at a cost of introducing a triple deconcatenation in \eqref{eq:A-recursion}, encoding the quartic vertex \cite{Berends:1987me}.)

The above recursion computes Berends--Giele currents in gauge theory. The boundary conditions come from imposing that the linear order in \eqref{eq:perturiner-gauge-field} are solutions of the free-field equations, i.e., the one-particle states $A_i^{\mu} := \varepsilon_i^\mu$ are the polarization vectors, which satisfy the transversality condition $k_i \cdot \varepsilon_i = 0$. In fact, one can check that $k_P^\mu A_P^\mu = 0$ holds for any $P$. Color-ordered amplitudes, say for the canonical ordering $\mathbb{I}_n = 12\cdots n$, can then be computed as
\be\label{eq:A-amplitude}
\mathcal{A}^{\text{YM}}_n(\mathbb{I}_n) = \lim\limits_{k_n^2 \to 0} s_{12\cdots n{-}1} A^{\mu}_{12\cdots n{-}1} A^{\mu}_n.
\ee
After contracting the two perturbiners, we take the off-shell leg with momentum $k_n := {-}(k_1{+}k_2{+}\cdots+k_{n{-}1})$ to be on-shell. The factor $s_{12\cdots n{-}1}$ is inserted to cancel the otherwise-divergent propagator inside $A^{\mu}_{12\cdots n{-}1}$. For instance, we can compute the 3-pt amplitude as follows. First, we compute the rank-$2$ current using \eqref{eq:A-recursion}:
\begin{align}
A^{\mu}_{12} &= \frac{1}{2s_{12}} \Big( \varepsilon_1 \cdot k_2\, \varepsilon_2^\mu + \varepsilon_1^\nu \left( k_2^\nu \varepsilon_2^\mu  - k_2^\mu \varepsilon_2^\nu \right) - (1 \leftrightarrow 2) \Big) \nn\\
&= \frac{1}{s_{12}} \Big( \varepsilon_1 \cdot k_2\, \varepsilon_2^\mu - \varepsilon_2 \cdot k_1\, \varepsilon_1^\mu + \frac{1}{2}\varepsilon_1 \cdot \varepsilon_2 (k_1^\mu - k_2^\mu)\Big),\label{eq:A12-example}
\end{align}
which features only one deconcatenation $Q=1, R=2$ of $P=12$. We then use \eqref{eq:A-amplitude} with $n=3$:
\begin{align}
\mathcal{A}^{\text{YM}}_3(\mathbb{I}_3) &= \lim\limits_{k_3^2 \to 0} \Big( \varepsilon_1 \cdot k_2\, \varepsilon_2^\mu - \varepsilon_2 \cdot k_1\, \varepsilon_1^\mu + \frac{1}{2}\varepsilon_1 \cdot \varepsilon_2 (k_1^\mu - k_2^\mu)\Big) \varepsilon_3^\mu \nn\\
&= \varepsilon_1 \cdot k_2 \, \varepsilon_2 \cdot \varepsilon_3 + \varepsilon_2 \cdot k_3 \, \varepsilon_3 \cdot \varepsilon_1 + \varepsilon_3 \cdot k_1 \, \varepsilon_1 \cdot \varepsilon_2.\label{eq:YM-3-pt}
\end{align}
In the last line it was necessary to use momentum conservation and the transversality condition $\varepsilon_i \cdot k_i = 0$.

\subsubsection{\label{sec:bi-adjoint-perturbiners}Bi-Adjoint Scalar Theory}

Another theory of our interest is the so-called bi-adjoint scalar theory (BA) \cite{BjerrumBohr:2012mg,Cachazo:2013iea}. It will appear later in this work in the context of double-copy relations, as well as coupling to the NLSM. The theory consists of a single massless scalar field $\Phi = \phi_{a\tilde{a}} T^{a} \widetilde{T}^{\tilde{a}}$ valued in the Lie algebra of $\text{U}(N) \times \text{U}(\widetilde{N})$, with the Lagrangian
\be
{\cal L}^{\text{BA}} = \frac{1}{2} \phi_{a\tilde{a}} \Box \phi^{a\tilde{a}} - \frac{1}{3} f^{abc} \tilde{f}^{\tilde{a}\tilde{b}\tilde{c}}\, \phi_{a\tilde{a}}\, \phi_{b\tilde{b}}\, \phi_{c\tilde{c}}.
\ee
Here $f^{abc}$ and $\tilde{f}^{\tilde{a}\tilde{b}\tilde{c}}$ the structure constants of the two groups. This leads to the equation of motion:
\be\label{eq:Phi-eom}
\Box \Phi = \llbracket \Phi, \Phi \rrbracket,
\ee
where we used a double commutator, $\llbracket \Phi, \Phi \rrbracket := 2(\phi_{a\tilde{a}} \,\phi_{b\tilde{b}} - \phi_{a\tilde{b}} \,\phi_{b\tilde{a}}) T^{a} T^{b} \widetilde{T}^{\tilde{a}} \widetilde{T}^{\tilde{b}}$. Because of the presence of two color groups, it has an interesting color-stripped perturbiner expansion, which was introduced in \cite{Mafra:2016ltu}:
\be\label{eq:Phi-perturbiner}
\Phi(x) := \sum_{P, Q} \phi_{P|Q}\, T^{P} \widetilde{T}^{Q} e^{k_{P}\cdot x} =  \sum_{i,j} \phi_{i|j}\, T^{a_i} \widetilde{T}^{\tilde{a}_j} e^{k_i \cdot x} + \sum_{i,j,k,l} \phi_{ij|kl}\, T^{a_i} T^{a_j} \widetilde{T}^{\tilde{a}_k} \widetilde{T}^{\tilde{a}_l}\, e^{k_{ij}\cdot x} +  \ldots.
\ee
Here the sum goes over words $P,Q$, and $\widetilde{T}^Q$ is defined analogously to \eqref{eq:T-k}. In order to have a well-defined multiparticle interpretation, one imposes that perturbiner coefficients, or the color-stripped Berends--Giele double currents, are non-vanishing only if $P$ and $Q$ are permutations of each other, i.e.,
\be\label{eq:Phi-vanishing}
\phi_{P|Q} = 0 \quad\text{if}\quad P \setminus Q \neq \varnothing,
\ee
which also implies that $k_P = k_Q$.

Recursion relations for Berends--Giele currents can be derived in a way analogous to the one given in the previous section. They are given as sums over deconcatenations of the words $P$ and $Q$ \cite{Mafra:2016ltu}:
\be
\phi_{P|Q} = \frac{1}{s_{P}} \sum_{P = RS} \sum_{Q = TU} \Big( \phi_{R|T}\, \phi_{S|U} - (R \leftrightarrow S) \Big),
\ee
Boundary conditions $\phi_{i|j} = \delta_{ij}$ give normalized solutions of the linearized field equations and follow from \eqref{eq:Phi-vanishing}. We also have shuffle symmetries in both sets of indices $\phi_{P\shuffle Q | R} = \phi_{P | Q \shuffle R} = 0$ for all $P,Q,R$. Using the above recursions, we have the following rank-$2$ and rank-$3$ examples:
\begin{gather}
\phi_{12|12} = \frac{1}{s_{12}} \left( \phi_{1|1} \phi_{2|2} - \phi_{2|1} \phi_{1|2} \right) = \frac{1}{s_{12}}, \qquad \phi_{12|21} = \frac{1}{s_{12}} \left( \phi_{1|2} \phi_{2|1} - \phi_{2|2} \phi_{1|1} \right) = -\frac{1}{s_{12}},\nn\\
\phi_{123|123} = \frac{1}{s_{123}} \left( \phi_{12|12} \phi_{3|3} + \phi_{1|1} \phi_{23|23} \right) = \frac{1}{s_{123}} \left( \frac{1}{s_{12}} + \frac{1}{s_{23}} \right),\label{eq:phi-current-examples}\\
\phi_{123|132} = \frac{1}{s_{123}} \left( \phi_{12|13} \phi_{3|2} + \phi_{1|1} \phi_{23|32} \right) = - \frac{1}{s_{123} s_{23}}\nn.
\end{gather}
Doubly-partial amplitudes in bi-adjoint scalar theory are computed using
\be\label{eq:bi-adjoint-amplitude}
m(P n | Q n) = \lim_{k_n^2 \to 0} s_{P}\, \phi_{P|Q}\, \phi_{n|n}.
\ee
Here we shifted the label $n$ into the last slot using cyclic invariance in both permutations. For example, we have:
\be
m(123|123) = -m(123|132) = 1, \qquad m(1234|1234) = \frac{1}{s_{12}} + \frac{1}{s_{23}}, \quad m(1234|1324) = - \frac{1}{s_{23}},\label{eq:bi-adjoint-34}
\ee
which are straightforwardly obtained from the currents given in \eqref{eq:phi-current-examples}.

\subsection{Color-Dressed Perturbiners}

In order to study theories without color ordering, such as special Galileon or Born--Infeld theory, we need to introduce a notion of perturbiner expansion in such a setting. Recall that in the perturbiners used in the previous sections, e.g., \eqref{eq:perturiner-gauge-field} and \eqref{eq:Phi-perturbiner}, the matrix products $T^P$ were used to organize the expansion. In the absence of Lie algebra generators, we will instead use only the plane waves $e^{k_P \cdot x}$ in order to separate the terms in the expansion. 

In order to illustrate the idea, let us consider the simplest example of a cubic scalar theory with equation of motion $\Box \varphi = \varphi^2$. We use the following perturbiner expansion:\footnote{An alternative formulation for color-dressed perturbiners using auxiliary parameters living in a universal enveloping algebra of $\text{U}(N)$ was given in Appendix~B of \cite{Mafra:2016ltu}.}
\be\label{eq:colorless-perturbiner}
\varphi(x) := \sum_{ \mathcal{P} } \varphi_{{\cal P}}\, e^{k_{{\cal P}}\cdot x} =  \sum_i \varphi_i\, e^{k_i \cdot x} + \sum_{i<j} \varphi_{ij}\, e^{k_{ij}\cdot x} + \sum_{i<j<l} \varphi_{ijl}\, e^{k_{ijl}\cdot x} + \ldots.
\ee
Here the sum goes over non-empty ordered words ${\cal P}=p_1 p_2 \ldots p_m$ with $p_1<p_2<\cdots <p_m$ to avoid double counting and combinatorial factors. Then, plugging \eqref{eq:colorless-perturbiner} into the field equation, we obtain:
\begin{align}\label{eq:phi-expansion}
\Box \varphi &= \left( \sum_i \varphi_i\, e^{k_i \cdot x} + \sum_{i<j} \varphi_{ij}\, e^{k_{ij}\cdot x} + \ldots \right) \left( \sum_p \varphi_p\, e^{k_p \cdot x} + \sum_{p<q} \varphi_{pq}\, e^{k_{pq}\cdot x} + \ldots \right)\nn\\
&= \sum_{i} \sum_{p} \varphi_i\, \varphi_p\, e^{k_{ip}\cdot x} 
+ \sum_{i<j} \sum_p \varphi_{ij}\, \varphi_p\, e^{k_{ijp}\cdot x}
+ \sum_{i} \sum_{p<q} \varphi_{i}\, \varphi_{pq}\, e^{k_{ipq}\cdot x} + \ldots.
\end{align}
The sums on the right-hand side need to be reorganized into ordered sums, e.g.,
\be
\sum_{i} \sum_{p} = \sum_{i<p} + \sum_{p<i}\;, \qquad
\sum_{i<j} \sum_p = \sum_{i<j<p} +\sum_{i<p<j} +\sum_{p<i<j} \; ,
\ee
before matching with the expansion of $\Box \varphi$ on the left-hand side. Here we need to take perturbiners $\varphi_i$ to be nilpotent, i.e., $\varphi_i^2 = 0$, so that no diagonal terms contribute. Along with given plane waves $e^{k_r\cdot x}, e^{k_{rs}\cdot x}$ and $e^{k_{rst}\cdot x}$, this kind of bookkeeping allows us to write the first few recursions from \eqref{eq:phi-expansion}:
\begin{gather}
k_r^2\, \varphi_r = 0, \qquad k_{rs}^2\, \varphi_{rs} = \varphi_r \varphi_s + \varphi_s \varphi_r,\\
k_{rst}^2\, \varphi_{rst} = \varphi_{rs} \varphi_t + \varphi_{rt} \varphi_s + \varphi_{st} \varphi_r +
\varphi_{r} \varphi_{st} + \varphi_{s} \varphi_{rt} + \varphi_{t} \varphi_{rs}.
\end{gather}
This generalizes straightforwardly to arbitrary perturbiner as follows:
\be\label{eq:phi-perturbiner}
\varphi_{\cal P} = \frac{1}{2 s_{{\cal P}} } \sum_{{\cal P} = {\cal Q} \cup {\cal R}} \varphi_{{\cal Q}}\, \varphi_{{\cal R}} .
\ee
The sum is over ${\cal P} = {\cal Q} \cup {\cal R}$ instructs to distribute the letters of the ordered words ${\cal P}$ into
non-empty ordered words ${\cal Q}$ and $ {\cal R}$, e.g.,
\begin{align}
{\cal P} &= 12 \;\;\quad \Rightarrow \quad ({\cal Q},{\cal R}) = (1,2) ,\, (2,1),\\
{\cal P} &= 123 \quad \Rightarrow \quad ({\cal Q},{\cal R}) = (12,3) ,\, (13,2),\, (23,1),\, (1,23),\, (2,13),\, (3,12),\label{eq:123-decomp}
\end{align}
where a doubling of terms related by ${\cal Q}\leftrightarrow {\cal R}$ occurs.
With the initial conditions $\varphi_i=1$, the above recursion \eqref{eq:phi-perturbiner} yields the following currents up to rank-$4$:
\begin{gather}
\varphi_{12} = \frac{1}{s_{12}}, \qquad \varphi_{123} = \frac{1}{s_{123}} \left( \frac{1}{s_{12}}+\frac{1}{s_{13}}+\frac{1}{s_{23}} \right)\\
\varphi_{1234} = \frac{1}{s_{1234}} \left\{ \frac{1}{s_{12} s_{34}}+\frac{1}{s_{13} s_{24}}+\frac{1}{s_{14}s_{23}}  + \left[ \frac{1}{s_{234}} \left(   \frac{1}{s_{23}}+\frac{1}{s_{24}}+\frac{1}{s_{34}} \right) +(1\leftrightarrow 2,3,4) \right]\right\}.
\end{gather}
By construction, the expressions for $\varphi_{12\ldots n}$ following from \eqref{eq:phi-perturbiner} are permutation invariant in $1,2,\ldots,n$. The amplitudes are computed as before with:
\be\label{eq:phi-amplitude}
\mathcal{A}^{\varphi^3}_n = \lim\limits_{k_n^2 \to 0} s_{12\cdots n{-}1}\, \varphi_{12\cdots n{-}1}\, \varphi_{n}.
\ee
Hence we see that the above expressions for the currents give rise to correct amplitudes in the cubic scalar theory.

\subsubsection{Yang--Mills Theory}

As another application of color-dressed perturbiners of the form \eqref{eq:colorless-perturbiner}, we consider an alternative perturbiner formulation of Yang--Mills theory. A color-dressed perturbiner ansatz in this case reads:
\be
A^{\mu, a}(x) := \sum_{\cal P} A_{\cal P}^{\mu, a}\, e^{k_{\cal P}\cdot x}, \qquad
F^{\mu \nu, a}(x) := \sum_{\cal P} F_{\cal P}^{\mu \nu, a}\, e^{k_{\cal P}\cdot x},
\ee
with an additional adjoint color index $a$ along with the coefficients of the expansion. This allows us to derive the recursions: 
\begin{gather}
A_{{\cal P}}^{\mu,a} = \frac{1}{2 s_{\cal P}} {f}_{abc}\! \sum_{{\cal P} = {\cal Q} \cup {\cal R}} \Big( A_{\cal Q}^b \!\cdot\! k_{\cal R}\,  A_{{\cal R}}^{\mu,c} + A^{\nu,b}_{\cal Q}\, F^{\nu \mu, c}_{\cal R} \Big),\\
F_{{\cal P}}^{\mu \nu,a} = k^\mu_{\cal P}A_{{\cal P}}^{\nu,a} - k^\nu_{\cal P} A_{{\cal P}}^{\mu,a} - {f}_{abc}\!
\sum_{{\cal P} = {\cal Q} \cup {\cal R}}
A_{{\cal Q}}^{\mu,b}
A_{{\cal R}}^{\nu,c}
\end{gather}
from the equation of motion \eqref{eq:A-eom} and the definition of the field strength. The initial conditions $A_{i}^{\mu,a} = \varepsilon^\mu_i \delta^{a ,a_i}$ require a Kronecker delta dependence on the adjoint index $a_i$ of the $i$-th particle. For example, this gives the following color-dressed rank-$2$ current:
\begin{align}
A_{12}^{\mu,a}  &= \frac{ {f}_{abc} }{2 s_{12}} \Big( \varepsilon_1 \cdot k_2\, \delta^{b a_1} \varepsilon_2^\mu\, \delta^{c a_2} + \varepsilon_{1}^{\nu} \,\delta^{b a_1}  (k_2^\nu \varepsilon_2^\mu - k_2^\mu \varepsilon_2^\nu ) \delta^{c a_2} + (1\leftrightarrow 2) \Big)\nn\\
&= \frac{ f_{aa_1a_2} }{s_{12}} \Big( \varepsilon_1 \cdot k_2\, \varepsilon_2^\mu - \varepsilon_2 \cdot k_1\, \varepsilon_1^\mu + \frac{1}{2}\varepsilon_1 \cdot \varepsilon_2 (k_1^\mu - k_2^\mu)
\Big).\label{eq:rank-3-YM}
\end{align}
The coefficient of $f_{aa_1a_2}$ may be recognized as the color-ordered current $A_{12}^\mu$ given in \eqref{eq:A12-example}. Indeed, higher-rank currents are related to those from Section~\ref{sec:color-stripped-YM} by $A_{123}^{\mu,a}  = f_{a_1 a_2 b} f_{b a_3 a} A_{123}^\mu + (2\leftrightarrow 3)$. More generally, let us introduce half-ladder contractions,
\be\label{eq:half-ladder}
{\cal F}_{12\cdots m}^a  = f_{a_1 a_2 b} {f}_{b a_3 c} \cdots  {f}_{y a_{m{-}1} z} {f}_{z a_m a}
\ee
and other permutation given by relabeling. We then have
\be\label{eq:A-color-dressed}
A_{12 \cdots m}^{\mu,a} = {\cal F}_{12\cdots m}^a A_{12\cdots m}^\mu + {\rm sym}(2,3,\ldots ,m).
\ee
Permutation invariance of the color-dressed currents \eqref{eq:A-color-dressed} can be verified by means of Jacobi identities $ f_{a_1 a_2 b} {f}_{b a_3 a} + {\rm cyc}(a_1, a_2, a_3)=0$ and shuffle symmetry $A_{P \shuffle Q}^\mu=0$ of the color-ordered currents. Full amplitudes are computed with
\begin{align}
\mathcal{A}^{\text{YM}}_{n} &= \lim\limits_{k_n^2 \to 0} s_{12\cdots n{-}1}\, A_{12\cdots n{-}1}^{\mu,a}\, A_{n}^{\mu,a}\nn\\
&= \sum_{\rho \in S_{n-2}} {\cal F}^{a_n}_{1\rho(23\cdots n-1)}\, {\cal A}_n^{\text{YM}} (1 \rho(23\cdots n-1) n).
\end{align}
They come out naturally organized into the Del Duca--Dixon--Maltoni half-ladder basis \cite{DelDuca:1999rs} of partial amplitudes. In the second line we expressed it in terms of the color-ordered amplitudes computed with \eqref{eq:A-color-dressed}. For example, using \eqref{eq:rank-3-YM} we have straightforwardly:
\be
    \mathcal{A}_3^{\text{YM}} = \lim\limits_{k_3^2 \to 0}s_{12}A_{12}^{\mu,a}A_3^{\mu,a} = f_{a_1 a_2 a_3} A^{\text{YM}}_3(\mathbb{I}_3),
\ee
where $A^{\text{YM}}_3(\mathbb{I}_3)$ is the 3-pt amplitude from \eqref{eq:YM-3-pt}. In practical computations one should use the color-stripped perturbiners \eqref{eq:A-recursion} directly.

\subsubsection{Bi-Adjoint Scalar Theory}

A color-stripped perturbiner formulation for bi-adjoint scalar theory was given before in Section~\ref{sec:bi-adjoint-perturbiners}. For completeness, here we briefly discuss the color-dressed formulation with both adjoint indices of $\text{U}(N) {\times} \text{U}(\widetilde{N})$ present. (In Section~\ref{sec:extended-NLSM} we will implicitly describe an intermediate formulation in which only one color is stripped away.)

Recall that the equations of motion for this theory \eqref{eq:Phi-eom} can be written as $\Box \Phi^{a\tilde a} = {f}_{abc} {\tilde{f}}_{\tilde a\tilde b \tilde c} \Phi^{b\tilde b} \Phi^{c \tilde c}$, where $\Phi = \Phi_{a \tilde a} T^a \widetilde{T}^{\tilde a}$ denotes the bi-adjoint field. Inserting the color-dressed perturbiner ansatz,
\be
\Phi^{a\tilde a}(x) := 
\sum_{\cal P} \phi_{\cal P}^{a \tilde a} e^{k_{\cal P}\cdot x} = \sum_{i} \phi_i^{a\tilde a}\, e^{k_i\cdot x} + \sum_{i<j} \phi_{ij}^{a \tilde a}\, e^{k_{ij}\cdot x} + \sum_{i<j<k} \phi_{ijk}^{a \tilde a}\, e^{k_{ijk}\cdot x} + \ldots,
\ee
leads to the recursion:
\be
\phi_{\cal P}^{a \tilde a} = \frac{1}{2 s_{\cal P}} {f}_{abc}\, {\tilde f}_{\tilde a \tilde b \tilde c}\! \sum_{{\cal P} = {\cal Q} \cup {\cal R}}  \phi_{{\cal Q}}^{b \tilde b}\, \phi_{{\cal R}}^{c \tilde c}.
\ee
The initial conditions $\phi_i^{a  \tilde a} = \delta^{a a_i}\delta^{\tilde a \tilde a_i}$ make sure the adjoint indices are matched with the pair of color degrees of freedom $a_i, \tilde a_i$ of the $i$-th leg. Explicit examples of rank-$2$ and $3$ currents are:
\be\label{eq:phi-rank-2-3}
\phi_{12}^{a \tilde a} = \frac{ f_{a_1 a_2 a} \tilde f_{\tilde a_1 \tilde a_2 \tilde a} }{s_{12}} , \qquad
\phi_{123}^{a\tilde a} = \frac{1}{s_{123}} \left( \frac{ f_{a_1 a_2 b} {f}_{b a_3 a} \tilde f_{\tilde a_1 \tilde a_2 \tilde b} {\tilde{f}}_{\tilde b \tilde a_3 \tilde a }}{s_{12}}  + \text{cyc}(1,2,3) \right).
\ee
We can express them in terms of fully color-stripped currents $\phi_{P|Q}$ from \eqref{eq:phi-current-examples}. The relation between them and \eqref{eq:phi-rank-2-3} is as follows:
\begin{align}
\phi_{12}^{a\tilde a} &= f_{a_1 a_2 a} \tilde f_{\tilde a_1 \tilde a_2 \tilde a}\, \phi_{12|12},\\
\phi_{123}^{a\tilde a} &= {\cal F}_{123}^a \widetilde{\cal F}_{123}^{\tilde a} \, \phi_{123|123}
+ {\cal F}_{123}^a \widetilde{\cal F}_{132}^{\tilde a}\, \phi_{123|132}
+ {\cal F}_{132}^a \widetilde{\cal F}_{123}^{\tilde a}\, \phi_{132|123}
+ {\cal F}_{132}^a \widetilde{\cal F}_{132}^{\tilde a}\, \phi_{132|132}
\end{align}
with ${\cal F}_{123}^a = f_{a_1 a_2 b} {f}_{b a_3 a}$ and $\widetilde {\cal F}_{123}^{\tilde a} = \tilde f_{\tilde a_1 \tilde a_2 \tilde b}  {\tilde f}_{\tilde b \tilde a_3 \tilde a }$. More generally, writing the half-ladder structure-constant contractions from \eqref{eq:half-ladder} and their equivalents for the other color group, we have the following rank-$n$ extension:
\be
\phi_{12\cdots m}^{a \tilde a} = \sum_{\rho,\sigma \in S_{m{-}1}} {\cal F}_{1\rho(23\cdots m)}^a \widetilde {\cal F}_{1\sigma(23\cdots m)}^{\tilde a}\, \phi_{1 \rho(23\cdots m) | 1 \sigma(23\cdots m)}.
\ee
The amplitudes are computed using a straightforward generalization of the colorless case \eqref{eq:phi-amplitude},
\begin{align}
\mathcal{A}^{\text{BA}}_n &= \lim\limits_{k_n^2 \to 0} s_{12\cdots n{-}1}\, \phi_{12\cdots n{-}1}^{a\tilde a}\, \phi_{n}^{a\tilde a}\nn\\
&= \sum_{\rho,\tau \in S_{n-2}} {\cal F}^{a_n}_{1\rho(23\cdots n-1)}\, \widetilde{\cal F}^{\tilde{a}_n}_{1\tau(23\cdots n-1)}\, m (1 \rho(23\cdots n{-}1) n \,|\, 1 \tau(23\cdots n{-}1) n).
\end{align}
In the second line we related it to the doubly-partial amplitudes that can be computed using the fully color-stripped perturbiner via \eqref{eq:bi-adjoint-amplitude}.

\section{\label{sec:EFTs-with-colors}Effective Field Theories With Colors}

In this section we apply perturbiner methods to theories with color degrees of freedom. Our main example is the $\text{U}(N)$ NLSM. We also discuss its coupling to bi-adjoint scalar, as well an extension with higher-dimensional operators.

\subsection{\label{sec:NLSM}Non-Linear Sigma Model}

In its conventional representation, the NLSM is defined through a Lagrangian with an infinite number of interaction vertices \cite{Cronin:1967jq,Weinberg:1966fm,Weinberg:1968de}. With the goal of manifesting the color-kinematics duality, recently Cheung and Shen introduced an alternative formulation featuring only cubic interactions \cite{Cheung:2016prv}. In order to do so, one needs to consider a triplet of fields: $(\X^\mu, \Y, \Z^\mu)$ transforming in the adjoint of the $\text{U}(N)$. The Lagrangian reads
\be\label{eq:NLSM-Lagrangian}
\mathcal{L}^{\text{NLSM}} = \Tr \Big( \X_\mu \Box \Z^\mu + \frac{1}{2} \Y \Box \Y +  \frac{1}{\sqrt{2}} (\partial_\mu \X_\nu \!-\! \partial_\nu \X_\mu) [\Z^\mu, \Z^\nu] + \frac{1}{\sqrt{2}} \Z^\mu [\Y, \partial_\mu \Y] \Big).
\ee
This leads to the following equations of motion:
\begin{align}
\Box \X^\mu &= -\sqrt{2} [\Z_\nu, \partial^\mu \X^\nu {-} \partial^\nu \X^\mu] - \frac{1}{\sqrt{2}}[\Y, \partial^\mu \Y],\\
\Box \Y &= -\sqrt{2} [\partial_\mu \Y, \Z^\mu] - \frac{1}{\sqrt{2}} [\Y, \partial_\mu \Z^\mu],\\
\Box \Z^\mu &= -\sqrt{2} \partial_\nu [\Z^\mu, \Z^\nu].
\end{align}
The transversality condition $\partial_\mu \Z^\mu = 0$ holds whenever $\Z^\mu$ is an off-shell source \cite{Cheung:2016prv}. Note that the field $\X^\mu$ enters the Lagrangian only through the field strength $\partial_\mu \X_\nu \!-\! \partial_\nu \X_\mu$, which means it is a gauge field with a redundancy $\X^\mu \to \X^\mu + \partial^\mu \lambda$.  Bearing this in mind, we use the perturbiner expansion analogous to \eqref{eq:perturiner-gauge-field} for the triplet:
\be\label{eq:XYZ-perturbiners}
\begin{pmatrix} \X^\mu, & \Y, & \Z^\mu \end{pmatrix} = \sum_{P} \begin{pmatrix} X^\mu_P, & Y_P, & Z^\mu_P \end{pmatrix} T^P e^{k_P \cdot x}.
\ee
Plugging it into the equations of motion and discarding all the terms proportional to $k_Q \cdot Z_Q$, we find:
\begin{align}
X^\mu_P &= -\frac{1}{\sqrt{2}s_P}\sum_{P=QR} \bigg[ k_R^\mu \left( \frac{1}{2} Y_Q Y_R + Z_Q {\cdot} X_R\right) + k_Q {\cdot} Z_R \, X_Q^\mu - (Q \leftrightarrow R) \bigg],\\
Y_P &= -\frac{1}{\sqrt{2}s_P} \sum_{P=QR} \Big( Y_Q\, k_Q {\cdot} Z_R - (Q \leftrightarrow R) \Big),\label{eq:NLSM-recursion}\\
Z^\mu_P &= -\frac{1}{\sqrt{2}s_P} 
\sum_{P=QR} \Big( Z_Q^\mu\, k_Q {\cdot} Z_R - (Q \leftrightarrow R) \Big).
\end{align}
Recall that $s_P := \frac{1}{2}k_P^2$. Color-ordered amplitudes in the NLSM can be computed using
\be\label{eq:NLSM-amplitude}
\mathcal{A}^{\text{NLSM}}_n(12\cdots n) = \lim\limits_{k_n^2 \to 0} s_{12\cdots n{-}1} \Big( X^{\mu}_{12\cdots n{-}1} Z^{\mu}_n \,+\, Z^{\mu}_{12\cdots n{-}1} X^{\mu}_n \,+\, Y_{12\cdots n{-}1} Y_n \Big).
\ee
This gives off-shell recursion relations for NLSM amplitudes. (Other recursion relations were constructed previously in \cite{Kampf:2012fn} and later also in \cite{Kampf:2013vha,Cheung:2015ota,Cachazo:2016njl}.) Here we have summed over all possible states running in the $n$-th leg. Note that the first two terms involve mixed fields, which reflects the off-diagonal propagator in the Lagrangian \eqref{eq:NLSM-Lagrangian}. One can show that there are only two choices for external states that compute non-vanishing amplitudes: $(a)$ with exactly $n{-}2$ external $\Z$'s and two $\Y$'s, and $(b)$ with $n{-}1$ external $\Z$'s and one $\X$. The corresponding boundary conditions for perturbiners are:
\begin{align}
(a):& \qquad \big( X^\mu_i,\; Y_i,\; Z^\mu_i \big) = \begin{cases}
	\big( 0,\; 1,\;\, 0\,\; \big)  \qquad &\text{if } i \text{ is a } \Y\text{-state} \\
	\big( 0,\; 0,\; k^\mu_i \big)  \qquad &\text{if } i \text{ is a } \Z\text{-state},
\end{cases}\label{eq:case-a}\\
(b):& \qquad \big( X^\mu_i,\; Y_i,\; Z^\mu_i \big) = \begin{cases}
\big( \overline{k}_i^\mu,\; 0,\; \,0\; \big)  \qquad &\text{if } i \text{ is an } \X\text{-state},\\
\big( \;0\;,\; 0,\; k^\mu_i \big)  \qquad &\text{if } i \text{ is a } \Z\text{-state}.
\end{cases}
\end{align}
The resulting amplitude is independent of these choices. We used the notation $\overline{k}_i^{\mu}$ to denote the conjugate momentum such that $k_i \cdot \overline{k}_i = -1$. Notice that because of these boundary conditions, only one out of the three terms in \eqref{eq:NLSM-amplitude} contributes to the amplitude. The case ($a$) was already introduced in \cite{Cheung:2016prv} based on previous considerations \cite{Cheung:2015aba}.

In order to make the assignment of external states more transparent, let us introduce the following notation. Whenever $i$-th particle is a $\X$ or $\Y$-state, we use the overline, $\overline{i}$, and underline, $\underline{i}$, respectively. Labels without additional decoration correspond to $\Z$-states. For instance, $Y_{1\underline{2}3}$ denotes the Y-current where particle $2$ is a $\Y$-state, and $1$ and $3$ are $\Z$-states. 

\subsubsection{Four-Point Examples}

Let us illustrate how to use the recursion relations \eqref{eq:NLSM-recursion} to compute $4$-pt amplitudes. This computation can be made in a couple of different ways as follows.

\begin{itemize}[leftmargin=*]
	\item Case $(a)$ with $(Z_1^\mu, Y_{\underline{2}}, Z_3^\mu, Y_{\underline{4}}) = (k_1^\mu, 1, k_3^\mu, 1)$ and all other $(X_i^\mu, Y_i, Z_i^\mu)$ vanishing.
	The relevant rank-$2$ currents following from the recursion \eqref{eq:NLSM-recursion} read
	\be
	Y_{1\underline{2}} = -Y_{\underline{2}3} =  \frac{1}{\sqrt{2}}, \qquad Z_{1\underline{2}}^\mu = Z_{\underline{2}3}^\mu = 0,
	\ee
	from which we find the rank-$3$ scalar current 
	\be
	Y_{1\underline{2}3} = -\frac{1}{2s_{123}} \Big( s_{12} + s_{23} + 2 s_{13} \Big).
	\ee
	Therefore, the amplitude can be computed as
	\be
	\mathcal{A}^{\text{NLSM}}_4(\mathbb{I}_4) = \lim_{k_4^2 \to 0} s_{123} Y_{1\underline{2}3} Y_{\underline{4}} = - \frac{1}{2} s_{13}.
	\ee
	
	\item Case $(a)$ with $(Y_{\underline{1}}, Z_2^\mu, Y_{\underline{3}}, Z_4^\mu) = (1,k_2^\mu, 1, k_4^\mu)$ and all other $(X_i^\mu, Y_i, Z_i^\mu)$ vanishing.
	We will need the following rank-$2$ currents
	\be
	Y_{\underline{1}2} = -Y_{2\underline{3}} = -\frac{1}{\sqrt{2}}, \qquad Z_{\underline{1}2}^\mu = Z_{2\underline{3}}^\mu = X_{\underline{1}2}^\mu = X_{2\underline{3}}^\mu = 0,
	\ee
	such that
	\be
	X_{\underline{1}2\underline{3}}^\mu = - \frac{k_2^\mu}{2s_{123}},
	\ee
	giving us the amplitude
	\be
	\mathcal{A}^{\text{NLSM}}_4(\mathbb{I}_4) = \lim_{k_4^2 \to 0} s_{123} X^\mu_{\underline{1}2\underline{3}} Z^\mu_4 = - \frac{1}{2}k_2 {\cdot} k_4 = - \frac{1}{2} s_{13}.
	\ee
	
	\item Case $(b)$ with $(Z_1^\mu, Z_2^\mu, Z_3^\mu, X_{\overline{4}}^\mu) = (k_1^\mu, k_2^\mu, k_3^\mu, \overline{k}_4^\mu)$ and all other $(X_i^\mu, Y_i, Z_i^\mu)$ vanishing.
	The relevant rank-$2$ currents following from the recursion \eqref{eq:NLSM-recursion} read
	\be\label{4pt-example1}
	Z_{12}^\mu = -\frac{1}{\sqrt{2}}(k_1^\mu - k_2^\mu), \qquad Z_{23}^\mu = -\frac{1}{\sqrt{2}}(k_2^\mu - k_3^\mu),
	\ee
	which enter the computation of the rank-$3$ current
	\be\label{4pt-example2}
	Z_{123}^\mu = \frac{1}{2s_{123}} \Big( (s_{12}+s_{23})(k_1^\mu - k_2^\mu + k_3^\mu) - 2s_{13}\, k_2^\mu \Big).
	\ee
	This gives the amplitude
	\be
	\mathcal{A}^{\text{NLSM}}_4(\mathbb{I}_4) = \lim_{k_4^2 \to 0} s_{123} Z^\mu_{123} X^\mu_{\overline{4}} = \frac{1}{2}s_{13}\, k_4 {\cdot} \overline{k}_4 = -\frac{1}{2}s_{13}.
	\ee
	
\end{itemize}
As expected, all methods lead to the same answer for the $4$-pt amplitude.

\subsubsection{Five-Point Examples}

Odd-point amplitudes in the NLSM vanish. Let us see how this occurs in our setup.

\begin{itemize}[leftmargin=*]

\item Case ($a$) with $(Y_{\underline{1}},Z_2^\mu,Z_3^\mu,Z_4^\mu,Y_{\underline{5}}) = (1,k_2^\mu,k_3^\mu,k_4^\mu,1)$ and all other $(X_i^\mu, Y_i, Z_i^\mu)$ vanishing.
We will need the rank-$2$ currents:
\be
Y_{\underline{1}2} = -\frac{1}{\sqrt{2}}, \quad  Z_{23}^\mu = - \frac{1}{\sqrt{2}}(k_2^\mu - k_3^\mu), \quad Z_{34}^\mu = -\frac{1}{\sqrt{2}}(k_3^\mu - k_4^\mu),\quad Y_{23} = Y_{34} =  Z_{\underline{1}2}^\mu = 0.
\ee
With these we can compute the relevant rank-$3$ currents:
\be
Y_{\underline{1}23} = \frac{s_{12}+s_{23}}{2s_{123}}, \quad Z_{234}^\mu = \frac{1}{2s_{234}}\left( (s_{23}{+}s_{34})(k_2^\mu - k_3^\mu + k_4^\mu) - 2s_{24}\, k_3^\mu\right),\quad Y_{234}=  Z_{\underline{1}23}^\mu = 0.
\ee
Note that the above $Z$-currents can be simply obtained by relabeling \eqref{4pt-example1} and \eqref{4pt-example2}. Putting everything together we find:
\be
Y_{\underline{1}234} = \frac{1}{2\sqrt{2}} \left( \frac{s_{13}}{s_{123}} + \frac{s_{24}}{s_{234}}-1\right).
\ee
What matters is that the pole in $s_{1234}$ has cancelled. As a result, the amplitude vanishes simply as a consequence of the factor $s_{1234} = k_5^2 \to 0$ in the numerator:
\be
\mathcal{A}^{\text{NLSM}}_5(\mathbb{I}_5) = \lim_{k_5^2 \to 0} s_{1234} Y_{\underline{1}234} Y_{\underline{5}} = 0.
\ee

\item Case ($a$) with $(Y_{\underline{1}},Z_2^\mu,Z_3^\mu,Y_{\underline{4}},Z_5^\mu) = (1,k_2^\mu,k_3^\mu,1,k_5^\mu)$ and all other $(X_i^\mu, Y_i, Z_i^\mu)$ vanishing.
We have the following rank-$2$ currents:
\be
Y_{\underline{1}2} = -Y_{3\underline{4}} = -\frac{1}{\sqrt{2}}, \quad Z_{23}^\mu = -\frac{1}{\sqrt{2}}(k_2^\mu-k_3^\mu),\quad Y_{23} = Z_{\underline{1}2}^\mu = Z_{3\underline{4}}^\mu = X_{\underline{1}2}^\mu = X_{23}^\mu = X_{3\underline{4}}^\mu  = 0.
\ee
These give rise to the rank-$3$ currents:
\be
Y_{\underline{1}23} = \frac{s_{12}+s_{23}}{2s_{123}}, \quad Y_{23\underline{4}} = \frac{s_{23}+s_{34}}{2s_{234}}, \quad Z_{\underline{1}23}^\mu = Z_{23\underline{4}}^\mu = X_{\underline{1}23}^\mu = X_{23\underline{4}}^\mu = 0.
\ee
Hence we have:
\be
X^\mu_{\underline{1}23\underline{4}} = \frac{1}{4\sqrt{2}s_{1234}} \left( - \frac{s_{13}}{s_{123}}(k_{123}^\mu-k_4^\mu) + \frac{s_{24}}{s_{234}}(-k_1^\mu + k_{234}^\mu) + k_1^\mu {-} k_2^\mu {+} k_3^\mu {-} k_4^\mu\right).
\ee
Notice that the denominator $s_{1234}$ is present, therefore it is not manifest that the corresponding amplitude will vanish. However, a computation reveals
\be
\mathcal{A}^{\text{NLSM}}_5(\mathbb{I}_5) = \lim_{k_5^2 \to 0} s_{1234} X_{\underline{1}23\underline{4}}^\mu Z_{5}^\mu = 0,
\ee
where we used momentum conservation.

\end{itemize}

\subsubsection{Six-Point Example}

Let us consider an example computation of the $6$-pt amplitude using the following assignment of boundary conditions.

\begin{itemize}[leftmargin=*]

\item Case ($a$) with $(Y_{\underline{1}},Z_2^\mu,Z_3^\mu,Z_4^\mu,Y_{\underline{5}},Z_6^\mu) = (1,k_2^\mu,k_3^\mu,k_4^\mu,1,k_6^\mu)$ and all other $(X_i^\mu, Y_i, Z_i^\mu)$ vanishing. The relevant rank-$2$ and $3$ currents were computed before. Explicitly, at rank-$3$ we have:
\begin{gather}
Y_{\underline{1}23} = \frac{s_{12}+s_{23}}{2s_{123}}, \qquad Y_{34\underline{5}} = \frac{s_{34}+s_{45}}{2s_{345}}, \qquad Z_{234}^\mu = \frac{1}{2s_{234}}\left( (s_{23}+s_{34})(k_2^\mu - k_3^\mu + k_4^\mu) - 2s_{24}\, k_3^\mu\right),\nn\\
Y_{234} = Z_{\underline{1}23}^\mu = Z_{34\underline{5}}^\mu = X_{\underline{1}23}^\mu = X_{234}^\mu = X_{34\underline{5}}^\mu = 0.
\end{gather}
This gives the following rank-$4$ currents:
\begin{gather}
Y_{\underline{1}234} = \frac{1}{2\sqrt{2}}\left( \frac{s_{13}}{s_{123}} + \frac{s_{24}}{s_{234}} - 1\right),\qquad Y_{234\underline{5}} = -\frac{1}{2\sqrt{2}}\left( \frac{s_{24}}{s_{234}} + \frac{s_{35}}{s_{345}} - 1\right),\\
Z_{\underline{1}234}^\mu = Z_{234\underline{5}}^\mu = X_{\underline{1}234}^\mu = X_{234\underline{5}}^\mu = 0.
\end{gather}
Hence we can compute the rank-$5$ $X$-current needed for the amplitude:
\be
X_{\underline{1}234\underline{5}}^\mu = \frac{1}{4s_{12345}s_{234}} \left( \frac{s_{24}s_{35}-(s_{23}{+}s_{34})(s_{34}{+}s_{45})}{s_{345}}k_2^\mu + \frac{s_{13}s_{24}-(s_{12}{+}s_{23})(s_{23}{+}s_{34})}{s_{123}}k_4^\mu + s_{24}\, k_3^\mu\right).\nn
\ee
Finally, plugging this result into \eqref{eq:NLSM-amplitude} we find:
\begin{align}
\mathcal{A}^{\text{NLSM}}_6(\mathbb{I}_6) &= \lim_{k_6^2 \to 0} s_{12345} X_{\underline{1}234\underline{5}}^\mu Z_{6}^\mu\nn\\
&= \frac{1}{4}\bigg( \frac{(s_{12}{+}s_{23})(s_{45}{+}s_{56})}{s_{123}} + \frac{(s_{23}{+}s_{34})(s_{56}{+}s_{61})}{s_{234}} + \frac{(s_{34}{+}s_{45})(s_{61}{+}s_{12})}{s_{345}}\nn\\
&\qquad\qquad\qquad\qquad\qquad\qquad\qquad\qquad\qquad-(s_{12}{+}s_{23}{+}s_{34}{+}s_{45}{+}s_{56}{+}s_{61}) \bigg).
\end{align}

\end{itemize}
The same computation can be repeated with other types of boundary conditions. We checked explicitly up to $9$-pt that the recursion relations \eqref{eq:NLSM-recursion} give rise to correct NLSM amplitudes.

\subsection{\label{sec:extended-NLSM}Extended Non-Linear Sigma Model}

Let us now consider an extension of the standard NLSM that includes couplings to the bi-adjoint scalar, which we denote with $\text{NLSM}\oplus\text{BA}$. This theory was first found in the soft limit of the NLSM pions and used to construct new recursion relations for their amplitudes \cite{Cachazo:2016njl}. The same theory was shown to appear in the low-energy limit of Z-theory when both abelian and non-abelian states are combined \cite{Carrasco:2016ygv}. An explicit form of the interaction vertices between pions and bi-adjoint scalars was given in \cite{Low:2017mlh,Low:2018acv}. Furthermore, it was checked in \cite{Cheung:2016prv} that an addition of a single cubic vertex $- \frac{1}{3} \tilde{f}^{\tilde{a}\tilde{b}\tilde{c}}\, \Y_{\tilde{a}} [\Y_{\tilde{b}}, \Y_{\tilde{c}}]$ to the Lagrangian \eqref{eq:NLSM-Lagrangian} reproduces the soft theorem of \cite{Cachazo:2016njl}, which involves certain single-trace $\text{NLSM}\oplus\text{BA}$ amplitudes with three external bi-adjoint scalars. However, more complicated amplitudes in the $\text{NLSM}\oplus\text{BA}$ theory have never been computed using a Lagrangian with a finite number of vertices.

In this section we propose a Lagrangian for the full extended NLSM theory in all the multi-trace sectors for the group $\text{U}(\widetilde{N})$. We follow a strategy similar to that of \cite{Cheung:2016prv,Cheung:2017yef}, where it was shown that the $(\X,\Y,\Z)$ formulation of the NLSM in $D$ space-time dimensions can be thought of as a specific embedding of Yang--Mills theory in $2D{+}1$ dimensions. In order to extend this analysis to a theory coupled to bi-adjoint scalars, we instead start with the Lagrangian for the YM$\oplus$BA theory \cite{Chiodaroli:2014xia} in $2D$ space-time dimensions:
\be\label{eq:YMS-Lagrangian}
\mathcal{L}^{\text{YM}\oplus\text{BA}} =  \Tr \left( -\frac{1}{4} \F_{MN} \F^{MN} - \frac{1}{2} \nabla_{M} \Y_{\tilde{a}} \nabla^{M} \Y^{\tilde{a}} -\frac{1}{4} [\Y_{\tilde{a}}, \Y_{\tilde{b}}] [\Y^{\tilde{a}}, \Y^{\tilde{b}}] - \frac{1}{3} \tilde{f}^{\tilde{a}\tilde{b}\tilde{c}}\, \Y_{\tilde{a}} [\Y_{\tilde{b}}, \Y_{\tilde{c}}]  \right),
\ee
where $\Y_{\tilde{a}} = Y_{a\tilde{a}} T^{a}$ represents a field transforming in the adjoint of $\text{U}(N) \times \text{U}(\widetilde{N})$, and couples minimally to the gauge field through the covariant derivative. We used Lorentz indices $M,N = 0,1,\ldots,2D{-}1$.

The next step is to consider a reparametrization of the gauge field $\A^M$ in terms of fields $\X^\mu$ and $\Z^{\mu}$, both living in $D$ space-time dimensions \cite{Cheung:2017yef},
\be\label{eq:A-into-XY}
\A^{\!M} = \frac{1}{\sqrt{2}} \begin{pmatrix}
	\Z^{\mu} \!+\! \X^{\mu} \\
	i (\Z^{\mu} \!-\! \X^{\mu})
\end{pmatrix}.
\ee
We further impose that the space-time metric is block-diagonal and that the derivatives act only in the first $D$ dimensions:
\be\label{eq:metric-sub}
\eta^{MN} = \begin{pmatrix}
	\eta^{\mu\nu} & 0 \\
	0 & \eta^{\mu\nu}
\end{pmatrix} \qquad\text{and}\qquad \partial^{M} = \begin{pmatrix}
	\partial^\mu \\
	0
\end{pmatrix}.
\ee
Plugging the substitutions into the YM$\oplus$BA Lagrangian \eqref{eq:YMS-Lagrangian} we find:
\begin{align}\label{eq:extended-NLSM-Lagrangian}
\mathcal{L}^{\text{YM}\oplus\text{BA}} \bigg|_{\substack{\scriptscriptstyle\eqref{eq:A-into-XY}\\ \scriptscriptstyle\eqref{eq:metric-sub}}}
&=  \Tr \bigg( \X_\mu \Box \Z^\mu + \frac{1}{2} \Y_{\tilde{a}} \Box \Y^{\tilde{a}} + \frac{1}{\sqrt{2}} (\X_\mu {+} \Z_\mu) \Big( [\Z_\nu, \partial^\mu \X^\nu] + [\X_\nu, \partial^\mu \Z^\nu] + [\Y_{\tilde a}, \partial^\mu \Y^{\tilde a}] \Big) \nn\\
&\qquad\qquad - \frac{1}{2} [\X_{\mu},\X_{\nu}][\Z^{\mu},\Z^{\nu}] - \frac{1}{2} [\X_{\mu},\Z_{\nu}][\Z^{\mu},\X^{\nu}] - [\X_{\mu},\Y_{\tilde a}][\Z^{\mu},\Y^{\tilde a}]\nn \\
&\qquad\qquad - \frac{1}{4} [\Y_{\tilde{a}}, \Y_{\tilde{b}}] [\Y^{\tilde{a}}, \Y^{\tilde{b}}] - \frac{1}{3} \tilde{f}^{\tilde{a}\tilde{b}\tilde{c}}\, \Y_{\tilde{a}} [\Y_{\tilde{b}}, \Y_{\tilde{c}}] \bigg) =: \mathcal{L}^{\text{NLSM}\oplus\text{BA}}.
\end{align}
First of all, let us comment on how this Lagrangian relates to \eqref{eq:NLSM-Lagrangian}. It was shown in \cite{Cheung:2017yef} that in order to determine which terms contribute to pure NLSM amplitudes, one associates weights $w(\X) = 2$, $w(\Y) = 1$, $w(\Z)=0$. Then all interaction vertices with weights adding up to $>2$ do not contribute and can be dropped. Hence only three interaction terms from the first line survive and it is straightforward to manipulate them into \eqref{eq:NLSM-Lagrangian} after integration by parts. The elimination of terms from weight-counting is responsible for the fact that in NLSM we have $\Z$-transversality, i.e. $\partial_{\mu}\Z^{\mu} = 0$. In the extended NLSM, on the other hand, we retain these terms and thus break the $\Z$-transversality that would otherwise hold if weight-counting were relevant. Lastly, note that compared to \eqref{eq:NLSM-Lagrangian}, we have a change $\Y \to \Y^{\tilde{a}}$. Since in the pure NLSM amplitudes we use at most two external $\Y$-fields, this means that $\mathcal{L}^{\text{NLSM}\oplus\text{BA}}$ introduces at most a global factor of $\Tr(T^{a_i} T^{a_j}) = \delta^{a_i a_j}$ compared to $\mathcal{L}^{\text{NLSM}}$ but does not affect the form of the amplitude.

We find that in order to compute amplitudes in the single-trace sector of $\text{NLSM}\oplus\text{BA}$, it is enough to consider the terms appearing in $\mathcal{L}^{\text{NLSM}}$ in addition to the cubic term in $\Y$, i.e., selecting all the terms from \eqref{eq:extended-NLSM-Lagrangian} with weights $\leq 3$. However, for multi-trace contributions in the color group $\text{U}(\widetilde{N})$ one needs all the terms in \eqref{eq:extended-NLSM-Lagrangian}.

A derivation of the perturbiner expansion proceeds analogously to the one presented in the previous section. We plug in the ansatz \eqref{eq:XYZ-perturbiners} with $Y_P \to Y_P^{\tilde{a}}$ into the equations of motion computed from \eqref{eq:extended-NLSM-Lagrangian} and find:
\begin{align}\label{Extended-NLSM-perturbiner}
X^\mu_P &= -\frac{1}{2 s_P} \bigg\{ \frac{1}{\sqrt{2}} \sum_{P=QR} \left[ k_R^{\mu} \left( Y_Q^{\tilde a} Y_R^{\tilde a} + X_Q \!\cdot\! Z_R + Z_Q \!\cdot\! X_R \right) + X_Q^{\mu} (2k_Q {+} k_R) \!\cdot\! (X_R {+} Z_R)  - (Q \leftrightarrow R)\right]\nn\\
& \qquad \qquad +  \sum_{P=QRS} \left[ X_Q^{\mu} \left( Y_R^{\tilde a} Y_S^{\tilde a} + X_R \!\cdot\! Z_S + Z_R \!\cdot\! X_S \right) + (QRS \to SQR) - 2 (QRS \to RSQ) \right] \bigg\},\nn\\
Y_P^{\tilde{a}} &= -\frac{1}{2 s_P} \bigg\{ \frac{1}{\sqrt{2}} \sum_{P=QR} \left[ Y_Q^{\tilde a} (2k_Q {+} k_R)\!\cdot\!(X_R {+} Z_R) - \sqrt{2} \tilde{f}^{\tilde a \tilde b \tilde c} Y_Q^{\tilde b} Y_R^{\tilde c} - (Q \leftrightarrow R)\right]\\
& \qquad \qquad +  \sum_{P=QRS} \left[ Y_Q^{\tilde a} (X_R \!\cdot\! Z_S + Z_R \!\cdot\! X_S ) + Y_Q^{\tilde a} Y_R^{\tilde b} Y_S^{\tilde b} + (QRS \to SQR) - 2 (QRS \to RSQ) \right] \bigg\},\nn\\
Z^\mu_P &= -\frac{1}{2 s_P} \bigg\{ \frac{1}{\sqrt{2}} \sum_{P=QR} \left[ k_R^\mu \left( Y_Q^{\tilde a} Y_R^{\tilde a} + X_Q \!\cdot\! Z_R + Z_Q \!\cdot\! X_R  \right) + Z_Q^\mu (2k_Q {+} k_R) \!\cdot\! (X_R {+} Z_R)  - (Q \leftrightarrow R)\right]\nn\\
& \qquad \qquad + \sum_{P=QRS} \left[ Z_Q^\mu \left( Y_R^{\tilde a} Y_S^{\tilde a} + X_R \!\cdot\! Z_S + Z_R \!\cdot\! X_S \right) + (QRS \to SQR) - 2 (QRS \to RSQ) \right] \bigg\}.\nn
\end{align}
In order to set boundary conditions, we generalize the case $(a)$ from \eqref{eq:case-a}, by allowing an arbitrary number of $\Z$ and $\Y$ fields on the external states. We have
\be
\big( X^\mu_i,\; Y_i^{\tilde{a}},\; Z^\mu_i \big) = \begin{cases}
	\big( 0,\; \delta^{\tilde{a} \tilde{a}_i},\;\, 0\,\; \big)  \qquad &\text{if } i \text{ is a } \Y\text{-state}, \\
	\big( 0,\;\;\; 0,\;\;\; k^\mu_i\, \big)  \qquad &\text{if } i \text{ is a } \Z\text{-state}.
\end{cases}
\ee
Here $\tilde{a}_i$ denotes the $\text{U}(\widetilde{N})$ colour of the $i$-th particle. In the case with exactly two external $\Y$'s, the amplitude can be treated as either that of pure NLSM, or alternatively NLSM with two bi-adjoint scalars. We do not consider the boundary conditions of type $(b)$.

\subsubsection{\label{sec:Extended-NLSM-examples}Low-Point Examples}

Let us illustrate how to do computations with recursion relations \eqref{Extended-NLSM-perturbiner} on a $4$-pt example first. We consider the case with all external bi-adjoint scalars, which is the first example containing double-trace amplitudes with respect to $\text{U}(\widetilde{N})$. (The amplitudes are stripped with respect to the other group $\text{U}(N)$.)

\begin{itemize}[leftmargin=*]
    \item 
    We take the boundary conditions: $(Y_{\underline{1}}^{\tilde{a}},Y_{\underline{2}}^{\tilde{b}},Y_{\underline{3}}^{\tilde{c}},Y_{\underline{4}}^{\tilde{d}}) = (\delta^{\tilde{a}\tilde{a}_1},\delta^{\tilde{b}\tilde{a}_2},\delta^{\tilde{c}\tilde{a}_3},\delta^{\tilde{d}\tilde{a}_4})$ with all other $(X_i^{\mu},Y_i^{\tilde{a}},Z_i^{\mu})$ vanishing. At rank-$2$ we have
    \begin{gather}
    Y_{\underline{12}}^{\tilde{a}} = \frac{\tilde{f}^{\tilde{a}\tilde{b}\tilde{c}}}{2s_{12}}\Big(-\delta_{\tilde{c}\tilde{a}_1}\delta_{\tilde{b}\tilde{a}_2}+\delta_{\tilde{b}\tilde{a}_1}\delta_{\tilde{c}\tilde{a}_2}\Big) = \frac{f^{\tilde{a}\tilde{a}_1\tilde{a}_2}}{s_{12}}, \qquad Y_{\underline{23}}^{\tilde{a}} = \frac{\tilde{f}^{\tilde{a}\tilde{b}\tilde{c}}}{2s_{23}}\Big(-\delta_{\tilde{b}\tilde{a}_3}\delta_{\tilde{c}\tilde{a}_2}+\delta_{\tilde{b}\tilde{a}_2}\delta_{\tilde{c}\tilde{a}_3}\Big) = \frac{f^{\tilde{a}\tilde{a}_2\tilde{a}_3}}{s_{23}},\nn\\
    Z_{\underline{12}}^{\mu} = X_{\underline{12}}^{\mu} = \frac{\delta^{\tilde{a}_1\tilde{a}_2}}{2\sqrt{2}s_{12}} (k_1^\mu - k_2^\mu), \qquad Z_{\underline{23}}^{\mu} = X_{\underline{23}}^{\mu} = \frac{\delta^{\tilde{a}_2\tilde{a}_3}}{2\sqrt{2}s_{23}} (k_2^\mu - k_3^\mu).
    \end{gather}
    From these, we find the rank-$3$ scalar current:
    \begin{align}
    Y_{\underline{123}}^{\tilde{a}} = &\,-\frac{1}{2s_{123}}\Big(\delta_{\tilde{a}\tilde{a}_1}\delta_{\tilde{a}_2\tilde{a}_3}+\delta_{\tilde{a}\tilde{a}_3}\delta_{\tilde{a}_1\tilde{a}_2}-2\delta_{\tilde{a}\tilde{a}_2}\delta_{\tilde{a}_2\tilde{a}_3}\Big) + \frac{1}{2s_{12}s_{123}}\Big(2\tilde{f}^{\tilde{a}\tilde{a}_3\tilde{c}}\tilde{f}^{\tilde{c}\tilde{a}_2\tilde{a}_1}+(s_{13}{-}s_{23})\delta_{\tilde{a}\tilde{a}_3}\delta_{\tilde{a}_1\tilde{a}_2}\Big) \nn\\
    &+ \frac{1}{2s_{23}s_{123}}\Big(2\tilde{f}^{\tilde{a}\tilde{a}_1\tilde{c}}\tilde{f}^{\tilde{c}\tilde{a}_2\tilde{a}_3}+(s_{13}{-}s_{12})\delta_{\tilde{a}\tilde{a}_1}\delta_{\tilde{a}_2\tilde{a}_3}\Big).
    \end{align}
    We compute the full amplitude for external bi-adjoint scalars as
    \begin{align}\label{eq:A-extended-1234}
        \mathcal{A}_4^{\text{NLSM}\oplus\text{BA}}(\Y_1 \Y_2 \Y_3 \Y_4) &= \lim_{k_4^2\to 0}s_{123}Y_{\underline{123}}^{\tilde{a}}Y_{\underline{4}}^{\tilde{a}}\nn\\
        &= \bigg[ \left(\frac{1}{s_{12}}{+}\frac{1}{s_{23}}\right)\text{Tr}(\tilde{T}^{\tilde{a}_1}\tilde{T}^{\tilde{a}_2}\tilde{T}^{\tilde{a}_3}\tilde{T}^{\tilde{a}_4}) -\frac{1}{s_{12}}\text{Tr}(\tilde{T}^{\tilde{a}_1}\tilde{T}^{\tilde{a}_2}\tilde{T}^{\tilde{a}_4}\tilde{T}^{\tilde{a}_3})\nn\\
        &\qquad-\frac{1}{s_{23}}\text{Tr}(\tilde{T}^{\tilde{a}_1}\tilde{T}^{\tilde{a}_3}\tilde{T}^{\tilde{a}_2}\tilde{T}^{\tilde{a}_4}) + (1234 \to 4321)\bigg]\\
        &\qquad+\text{Tr}(\tilde{T}^{\tilde{a}_1}\tilde{T}^{\tilde{a}_3})\text{Tr}(\tilde{T}^{\tilde{a}_2}\tilde{T}^{\tilde{a}_4}) -\left(1+\frac{s_{23}}{s_{12}}\right)\text{Tr}(\tilde{T}^{\tilde{a}_1}\tilde{T}^{\tilde{a}_2})\text{Tr}(\tilde{T}^{\tilde{a}_3}\tilde{T}^{\tilde{a}_4})\nn\\ &\qquad-\left(1+\frac{s_{12}}{s_{23}}\right)\text{Tr}(\tilde{T}^{\tilde{a}_1}\tilde{T}^{\tilde{a}_4})\text{Tr}(\tilde{T}^{\tilde{a}_2}\tilde{T}^{\tilde{a}_3}).\nn
    \end{align}
    Both single and double-trace amplitudes with respect to the group $\text{U}(\widetilde{N})$ contribute. (Partial order with respect to $\text{U}(N)$ is the identity $\mathbb{I}_4$.) The coefficients of single-traces are the familiar bi-adjoint amplitudes, as in \eqref{eq:bi-adjoint-34}. Double-trace amplitudes involve contact terms as well as exchanges of the $X/Z$-state, as expected from the Lagrangian \eqref{eq:extended-NLSM-Lagrangian}.

 \item
  In contrast with the pure NLSM, the extended NLSM theory has non-vanishing odd-pt amplitudes. Let us consider an example computation for an amplitude with two external pions and three bi-adjoint scalars. (The one with three external pions and two bi-adjoint scalars is proportional to the $5$-pt pure NLSM amplitude).
  We use boundary conditions $(Z_1^{\mu},Z_2^{\mu},Y_{\underline{3}}^{\tilde{a}},Y_{\underline{4}}^{\tilde{b}},Y_{\underline{5}}^{\tilde{c}}) = (k_1^{\mu},k_2^{\mu},\delta^{\tilde{a}\tilde{a}_3},\delta^{\tilde{b}\tilde{a}_4},\delta^{\tilde{c}\tilde{a}_5})$ with the remaining $(X_i^{\mu},Y_i^{\tilde{a}},Z_i^{\mu})$ vanishing. The non-zero rank-2 currents are
  \be
    Y_{2\underline{3}}^{\tilde{a}} = \frac{1}{\sqrt{2}}\delta^{\tilde{a}\tilde{a}_3}, \qquad Y_{\underline{34}}^{\tilde a} = \frac{\tilde{f}^{\tilde a \tilde{a}_3 \tilde{a}_4}}{s_{34}}, \qquad Z_{12}^{\mu} = \frac{1}{\sqrt{2}}(k_2^{\mu}-k_1^{\mu}),\qquad Z_{\underline{34}}^{\mu} = X_{\underline{34}}^{\mu} = \frac{\delta^{\tilde{a}_3 \tilde{a}_4}}{2\sqrt{2} s_{34}}(k_3^{\mu} - k_4^\mu).\nn 
  \ee
  This gives us the following rank-3 currents:
  \begin{gather}
      Y_{12\underline{3}}^{\tilde{a}} = \frac{\delta^{\tilde{a}\tilde{a}_3}}{2s_{123}}(s_{12}+s_{23}), \qquad Y_{2\underline{34}}^{\tilde{a}} = \frac{\tilde{f}^{\tilde{a}\tilde{a}_3 \tilde{a}_4}}{\sqrt{2} s_{34}},\\
      Z_{2\underline{34}}^{\mu} = \frac{\delta^{\tilde{a}_3\tilde{a}_4}}{8s_{234}}\left(\frac{s_{23}}{s_{34}}(-3k_2^{\mu}{+}k_3^{\mu}{-}3k_4^{\mu}) + \frac{s_{24}}{s_{34}}(3k_2^{\mu}{+}3k_3^{\mu}{-}k_4^{\mu})-2(k_2^{\mu}{-}k_3^{\mu}{+}k_4^{\mu})\right),\\
      X_{2\underline{34}}^{\mu} = \frac{\delta^{\tilde{a}_3\tilde{a}_4}}{8s_{234}}\left(\frac{s_{23}}{s_{34}}(k_2^{\mu}+k_3^{\mu}-3k_4^{\mu})-\frac{s_{24}}{s_{34}}(k_2^{\mu}-3k_3^{\mu}+k_4^{\mu}) + 2(k_2^{\mu}+k_3^{\mu}-k_4^{\mu})\right).
  \end{gather}
  All of these contribute to the rank-$4$ current:
  \begin{align}
      Y_{12\underline{34}}^{\tilde{a}} &= \frac{\tilde{f}^{\tilde{a}\tilde{a}_3\tilde{a}_4}}{2s_{1234}}\left( \frac{s_{12}+s_{23}+s_{24}}{s_{34}} + \frac{s_{12} + s_{23}}{s_{123}}\right).
  \end{align}
  This current allows us to arrive at the following amplitude:
  \begin{align}\label{eq:A-extended-345}
      \mathcal{A}_5^{\text{NLSM}\oplus\text{BA}}(\Z_1 \Z_2 \Y_3 \Y_4 \Y_5) &= \lim_{k_5^2\to 0}s_{1234}Y_{12\underline{34}}^{\tilde{a}}Y_{\underline{5}}^{\tilde{a}}\\
      & = \frac{1}{2}\left(\frac{s_{12}+s_{23}}{s_{45}}+\frac{s_{51}+s_{12}}{s_{34}} - 1\right) \left(\text{Tr}(\tilde{T}^{\tilde{a}_3}\tilde{T}^{\tilde{a}_4}\tilde{T}^{\tilde{a}_5})-\text{Tr}(\tilde{T}^{\tilde{a}_3}\tilde{T}^{\tilde{a}_5}\tilde{T}^{\tilde{a}_4})\right).\nn
  \end{align}
  As a cross-check let us calculate a cyclically-shifted amplitude, which involves computations of other currents. Choosing boundary conditions to be $(Z_1^{\mu},Y_{\underline{2}}^{\tilde{a}},Y_{\underline{3}}^{\tilde{b}},Y_{\underline{4}}^{\tilde{c}},Z_5^{\mu}) = (k_1^{\mu},\delta^{\tilde{a}\tilde{a}_2},\delta^{\tilde{b}\tilde{a}_3},\delta^{\tilde{c}\tilde{a}_3},k_5^{\mu})$ and all other $(X_i^{\mu},Y_i^{\tilde{a}},Z_i^{\mu})$ vanishing, the amplitude is now computed using the $X$-current, which we find to be
  \begin{align}
  X_{1\underline{234}}^{\mu} =& \frac{\tilde{f}^{\tilde{a}_2 \tilde{a}_3 \tilde{a}_4}}{8 s_{23} s_{34} s_{234} s_{1234}} \bigg[ \Big( s_{13} \left(s_{34}{-}s_{23}\right)+s_{12} \left(s_{23}{+}s_{34}\right)-\left(s_{23}{+}s_{34}\right) \left(s_{14}{-}2
   \left(s_{23}{+}s_{24}{+}s_{34}\right)\Big) \right) k_1^\mu\nn\\
   &\qquad\qquad\qquad +\Big( s_{12} \left(s_{23}{+}s_{34}\right)+\left(3 s_{14}{+}2 \left(s_{23}{+}s_{24}{+}s_{34}\right)\right) \left(s_{23}{+}s_{34}\right)+s_{13}
   \left(3 s_{23}{+}s_{34}\right) \Big) k_2^\mu \nn\\
   &\qquad\qquad\qquad+\Big( 2 s_{34}^2+\left(s_{12}{+}s_{13}{+}3 s_{14}{+}2 s_{24}\right) s_{34}-s_{23} \left(3 s_{12}{+}s_{13}{+}s_{14}{+}2
   \left(s_{23}{+}s_{24}\right)\right) \Big) k_3^\mu \nn\\
   &\qquad\qquad\qquad - \Big(3 s_{12} \left(s_{23}{+}s_{34}\right)+\left(s_{14}{+}2 \left(s_{23}{+}s_{24}{+}s_{34}\right)\right)
   \left(s_{23}{+}s_{34}\right)+s_{13} \left(s_{23}{+}3 s_{34}\right) \Big) k_4^\mu \bigg].\nn
  \end{align}
 There are no overlapping poles, even though this fact is not manifest above. It gives the following amplitude:
  \begin{align}
      \mathcal{A}_5^{\text{NLSM}\oplus\text{BA}}(\Z_1 \Y_2 \Y_3 \Y_4 \Z_5) &= \lim_{k_5^2\to 0}s_{1234}X_{1\underline{234}}^{\mu}Z_5^{\mu}\\
      & = \frac{1}{2}\left(\frac{s_{51}+s_{12}}{s_{34}}+\frac{s_{45}+s_{51}}{s_{23}}-1\right) \left(\text{Tr}(\tilde{T}^{\tilde{a}_2}\tilde{T}^{\tilde{a}_3}\tilde{T}^{\tilde{a}_4})-\text{Tr}(\tilde{T}^{\tilde{a}_2}\tilde{T}^{\tilde{a}_4}\tilde{T}^{\tilde{a}_3})\right).\nn
  \end{align}
  After relabelling $i\to i+1$, it is equal to \eqref{eq:A-extended-345}, as expected, and agrees with the results of \cite{Cachazo:2016njl}.
  \end{itemize}

Computations of higher-multiplicity amplitudes involve more complicated currents, and hence will not be displayed here explicitly. We list all independent amplitudes computed with the perturbiners \eqref{Extended-NLSM-perturbiner} up to $6$-pt in Appendix~\ref{app:example-amplitudes}. We use the following notation for doubly-partial $n$-pt amplitudes with ordering $\alpha$ with respect to $\text{U}(N)$, and $m$ traces with orderings $\beta_1, \beta_2, \ldots, \beta_m$ with respect to $\text{U}(\widetilde{N})$: 
\be
{\cal A}^{\text{NLSM}\oplus\text{BA}}_n(\alpha \| \beta_1 | \beta_2 | \cdots |\beta_m).
\ee
A given external state is a bi-adjoint scalar if it appears in any $\beta_i$, and a NLSM pion otherwise. For example, from \eqref{eq:A-extended-1234} and \eqref{eq:A-extended-345} we can read off:
\be
{\cal A}^{\text{NLSM}\oplus\text{BA}}_4(\mathbb{I}_4 || 12 | 34) = -\left(1 + \frac{s_{23}}{s_{12}} \right), \qquad {\cal A}^{\text{NLSM}\oplus\text{BA}}_5(\mathbb{I}_5 || 345) = \frac{1}{2}\left( \frac{s_{12}+s_{23}}{s_{45}} + \frac{s_{51}+s_{12}}{s_{34}} - 1 \right).\nn
\ee

\subsubsection{Cachazo--He--Yuan Formulation for Multi-Trace Amplitudes}

Amplitudes in the extended theory $\text{NLSM}\oplus\text{BA}$ were originally found using their CHY representation, where they appear in the soft limit of pure NLSM amplitudes \cite{Cachazo:2016njl}. This leads to the following CHY formula for single-trace amplitudes:
\be\label{eq:CHY-single-trace}
{\cal A}^{\text{NLSM}\oplus\text{BA}}_n(\alpha \| \beta) = \frac{1}{2^{|\bar{\beta}|/2}} \!\int d\mu^{\text{CHY}}_n\; \text{PT}(\alpha) \left( \text{PT}(\beta) \left( \text{Pf} A_{\bar{\beta}} \right)^2 \right).
\ee
Here we used the standard notation for the CHY measure and Parke--Taylor factors, see, e.g., \cite{Cachazo:2018hqa}. In the left-integrand $\text{PT}(\alpha)$ we have a single Parke--Taylor factor encoding the color ordering $\alpha$ of the first color group $\text{U}(N)$. In the right-integrand $\text{PT}(\beta) \left( \text{Pf} A_{\bar{\beta}} \right)^2$ there is another Parke--Taylor factor responsible for the color ordering $\beta$ of the group $\text{U}(\widetilde{N})$, as well a permutation-invariant Pfaffian of the antisymmetric matrix $A_{\bar{\beta}}$. By $A_{\bar{\beta}}$ we denote a matrix with off-diagonal entries $[A_{\bar{\beta}}]_{ij} = s_{ij}/\sigma_{ij}$, whose rows and columns are associated to the complementary set of variables $\overline{\beta} = \mathbb{I}_n \!\setminus\! \beta$. Bi-adjoint external states are those in $\beta$, while NLSM pions are in $\overline{\beta}$. Since a Pfaffian of an odd-by-odd matrix vanishes, the number of pions $|\overline{\beta}|$ needs to be even to obtain a non-zero amplitude. A decomposition of the right-integrand in terms of Parke--Taylor factors was given in \cite{Carrasco:2016ygv}.

The above CHY integrand can be obtained from a dimensional embedding of YM$\oplus$BA described at the beginning of this section. Its on-shell equivalent boils down to using the replacement rules:\footnote{Alternative replacement scheme with $\varepsilon_i \cdot k_j \to 0,\; \varepsilon_i \cdot \varepsilon_j \to s_{ij}$ employed in \cite{Cachazo:2014xea} is related by a gauge transformation.}
\be\label{eq:replacement-rules}
\varepsilon_i \cdot k_j \to \frac{s_{ij}}{\sqrt{2}}, \qquad \varepsilon_i \cdot \varepsilon_j \to 0\qquad \text{for all}\quad i,j.
\ee
Note that this replacement changes the mass dimension of an amplitude. It is straightforward to show that the CHY formula \eqref{eq:CHY-single-trace} can be obtained from the formula for YM$\oplus$BA amplitudes given in \cite{Cachazo:2014nsa}. Let us consider a generalization of this statement to multi-trace sector of the second color group $\text{U}(\widetilde{N})$. Starting with the YM$\oplus$BA half-integrand \cite{Cachazo:2014xea}, we find that $m$-trace contribution can be calculated up to a sign with the CHY formula:
\be\label{eq:CHY-multi-trace}
{\cal A}^{\text{NLSM}\oplus\text{BA}}_n(\alpha \| \beta_1 | \beta_2 | \cdots |\beta_m) = \int d\mu^{\text{CHY}}_n\; \text{PT}(\alpha) \bigg( \text{PT}(\beta_1) \cdots \text{PT}(\beta_m) \, \text{Pf}^\prime \Pi_{\beta_1, \ldots, \beta_m} \bigg).
\ee
It features an antisymmetric matrix $\Pi_{\beta_1, \ldots, \beta_m}$, which decomposes into $16$ blocks as follows:
\be
\Pi_{\beta_1, \ldots, \beta_m} = 
\begin{blockarray}{ccccc}
{ \small{j \in \overline{\beta}} } & { b \in \{\beta_1, \ldots, \beta_m\} } & { \small{j \in \overline{\beta}} }  & { b' \in \{ \beta_1, \ldots, \beta_m \}^\prime } & \\
	\begin{block}{[cccc]c}
		{ \displaystyle\delta_{ij}\frac{s_{ij}}{\sigma_{ij}} } & { \bullet } & {\bullet}  & {\bullet} & {\small{i \in \overline{\beta}}} \topstrut \vspace{.5em}\\ 
		{ \displaystyle\sum_{k\in \beta_a} \frac{s_{kj}}{\sigma_{kj}}} & { \displaystyle\delta_{ab}\!\!\!\!\!\sum_{\substack{k \in \beta_a, l \in \beta_b\\ k\neq l}} \frac{s_{kl}}{\sigma_{kl}}} & {\bullet}  & {\bullet} & {a \in \{ \beta_1, \ldots, \beta_m \}} \\
		{ \displaystyle\delta_{ij}\frac{s_{ij} }{\sqrt{2} \sigma_{ij}} } & { \displaystyle\sum_{l \in \beta_b} \frac{s_{il}}{\sqrt{2} \sigma_{il}} } & {\displaystyle 0}  & {\bullet} & {i \in \overline{\beta}} \\
		{ \displaystyle\sum_{k' \in \beta_{a'}} \!\!\!\frac{\sigma_{k'} s_{k'j}}{\sigma_{k'j}} } & { \displaystyle\sum_{\substack{k' \in \beta_{a'}, l \in \beta_b\\ k'\neq l }} \!\!\!\!\!\!\!\frac{\sigma_{k'} s_{k' l}}{\sigma_{k'l}} } & {\displaystyle \sum_{k' \in \beta_{a'}} \!\!\!\frac{\sigma_{k'} s_{k'j}}{\sqrt{2}\sigma_{k'j}}}  & {\displaystyle \delta_{a'b'}\!\!\!\!\!\!\!\!\!\!\sum_{\substack{k' \in \beta_{a'}, l' \in \beta_{b'}\\ k' \neq l'}} \!\!\!\!\!\!\!\!\frac{\sigma_{k'} s_{k'l'} \sigma_{l'}}{\sigma_{k'l'}}}  & {a' \in \{ \beta_1, \ldots, \beta_m \}^\prime} \botstrut\\
	\end{block}
\end{blockarray}\nn
\ee
The columns and rows are labelled by NLSM pions $i,j \in \overline{\beta} = \overline{\beta_1 \cup \cdots \cup \beta_m}$, as well as $m$ traces $a,a',b,b' \in \{ \beta_1, \ldots, \beta_m\}$. The factors of $\delta_{ij}$ are included in order to set diagonal entries to zero. The prime in \eqref{eq:CHY-multi-trace} instructs us to remove two columns and two rows associated to traces, one from the set $\{\beta_1, \ldots, \beta_m \}$ and one from $\{\beta_1, \ldots, \beta_m \}^\prime$. The resulting reduced Pfaffian $\text{Pf}^\prime \Pi_{\beta_1, \ldots, \beta_m}$ is independent of this choice \cite{Cachazo:2014xea}. Notice that as one of the momenta of a pion becomes soft, the matrix $\Pi_{\beta_1, \ldots, \beta_m}$ develops a zero column and row, and hence the amplitude vanishes. This is the so-called Adler zero \cite{Adler:1964um}.

Let us consider a couple of examples. For single-trace amplitudes we have $m=1$ and hence $a,a',b,b' \in \varnothing$ after removing columns and rows. Recognizing that $\delta_{ij} s_{ij}/\sigma_{ij}$ are nothing but the entries of $A_{\bar{\beta}}$, we find in case:
\be
\text{Pf}^\prime \Pi_{\beta} = \text{Pf}\, \raisebox{-0.35em}{$\begin{blockarray}{cc}
	\begin{block}{[cc]}
		{ \displaystyle A_{\bar{\beta}} } & { \displaystyle - A_{\bar{\beta}}^{\intercal} / \sqrt{2} }  \topstrut \\
		{ \displaystyle A_{\bar{\beta}} / \sqrt{2} } & {0} \botstrut\\
	\end{block}
\end{blockarray}$} = \frac{1}{(-2)^{|\bar{\beta}|/2}}\left( \text{Pf} A_{\bar\beta} \right)^2,
\ee
which is exactly the same factor, as the one appearing in the right-integrand of \eqref{eq:CHY-single-trace}, up to a sign. Another special case is when there are no external pions, i.e., $\overline{\beta} = \varnothing$. In this situation we are left with an $(2m{-}2) \times (2m{-}2)$ matrix (after removing two columns and rows):
\be
\Pi_{\substack{\beta_1, \ldots, \beta_m\\ \bar{\beta} = \varnothing}} = 
\begin{blockarray}{ccc}
{ b \in \{\beta_1, \ldots, \beta_m\} } &  { b' \in \{ \beta_1, \ldots, \beta_m \}^\prime } & \\
	\begin{block}{[cc]c}
	{ \displaystyle \delta_{ab}\!\!\!\!\!\sum_{\substack{k \in \beta_a, l \in \beta_b\\ k\neq l}} \frac{s_{kl}}{\sigma_{kl}}} & {\bullet} & {a \in \{ \beta_1, \ldots, \beta_m \}} \topstrut\\
	{ \displaystyle\sum_{\substack{k' \in \beta_{a'}\!,\, l \in \beta_b\\ k'\neq l }} \!\!\!\!\!\!\!\frac{\sigma_{k'} s_{k' l}}{\sigma_{k'l}} }  & {\displaystyle \delta_{a'b'}\!\!\!\!\!\!\!\!\!\sum_{\substack{k' \in \beta_{a'}\!,\, l' \in \beta_{b'}\\ k' \neq l'}} \!\!\!\!\!\!\!\!\frac{\sigma_{k'} s_{k'l'} \sigma_{l'}}{\sigma_{k'l'}}}  & {a' \in \{ \beta_1, \ldots, \beta_m \}^\prime} \botstrut\\
	\end{block}
\end{blockarray}
\ee
For instance, for a double-trace, $m=2$, we can remove the columns and rows associated to $\beta_2$, and find the reduced Pfaffian:
\be
\text{Pf}^\prime \Pi_{\substack{\beta_1, \beta_2\\ \bar{\beta}=\varnothing}} = \raisebox{-0.35em}{$\begin{blockarray}{cc}
	\begin{block}{[cc]}
	{ \displaystyle 0} & { \displaystyle -\!\!\sum_{\substack{k,l'\! \in \beta_{1}\\ k\neq l' }} \frac{\sigma_{l'} s_{l' k}}{\sigma_{l'k}} } \topstrut\\
	{ \displaystyle\sum_{\substack{k'\!,l \in \beta_{1}\\ k'\!\neq l }} \frac{\sigma_{k'} s_{k' l}}{\sigma_{k'l}} }  & {\displaystyle 0}  \botstrut\\
	\end{block}
\end{blockarray}$} = -\sum_{\substack{i,j \in \beta_{1}\\ i\neq j }} \frac{\sigma_{i}\, s_{ij}}{\sigma_{ij}} = -\sum_{i,j \in \beta_1} s_{ij} + \sum_{\substack{i,j \in \beta_1\\ i \neq j}} \frac{\sigma_j\, s_{ij}}{\sigma_{ij}}.
\ee
Since the final factor is equal to minus the Pfaffian, we have
\be
\text{Pf}^\prime \Pi_{\substack{\beta_1, \beta_2\\ \bar{\beta}=\varnothing}} = - \frac{1}{2} \sum_{i,j \in \beta_1} s_{ij} = - \frac{1}{2} s_{\beta_1},
\ee
in agreement with the results for double-trace contributions in YM$\oplus$BA theory \cite{Cachazo:2014nsa}.

We confirmed that the above CHY formulae reproduce the all the amplitudes given in Section~\ref{sec:Extended-NLSM-examples} and Appendix~\ref{app:example-amplitudes}. As another consistency check, we considered multi-trace Einstein--Yang--Mills relations derived in \cite{Nandan:2016pya,Du:2017gnh}, which also apply to YM$\oplus$BA and NLSM$\oplus$BA amplitudes after using replacement rules \eqref{eq:replacement-rules}, as is clear from their CHY representations. They relate NLSM$\oplus$BA amplitudes to those of only bi-adjoint scalars on external states. For instance, for amplitudes with two adjacent pions and $n{-}2$ bi-adjoint scalars we have:
\begin{align}\label{EYM1}
\mathcal{A}_{n}^{\text{NLSM}\oplus\text{BA}}(\alpha \| \mathbb{I}_{n-2}) =&\, \frac{1}{2}\sum_{1 = i\leq j}^{n-3} k_{n-1} \cdot k_{12\cdots i}\; k_{n} \cdot k_{12\cdots j} \;m(\alpha\,|\,1,\ldots,i,n{-}1,i{+}1,\ldots,j,n,j{+}1,\ldots,n{-}2)\nn \\
&-\frac{s_{n-1,n}}{2}\sum_{i=1}^{n-3} k_{n-1}\cdot k_{12\cdots i} \!\!\!\!\!\!\!\!\!\! \sum_{\beta\in\left\{n\right\}\shuffle\left\{1,\ldots,i\right\}} \!\!\!\!\!\!\!\!\!\! m(\alpha \,|\, \beta,n{-}1,i{+}1,\ldots,n{-}2) \;+\; (n{-}1 \leftrightarrow n),\nn
\end{align}
which is valid for any left ordering $\alpha$ and the symmetrization $(n{-}1 \leftrightarrow n)$ does not act on this ordering. For example, let us apply this relation to the case $n=5$ and $\alpha = \mathbb{I}_5$. The right-hand side becomes:
\begin{align}
\frac{1}{2} \bigg[ &s_{41} s_{51}\, m(\mathbb{I}_5|14523)+ s_{41}(s_{51}{+}s_{52})\,m(\mathbb{I}_n | 14253) +(s_{41}{+}s_{42})(s_{51}{+}s_{52})\,m(\mathbb{I}_5|12453) \\
&- s_{45}s_{41}\Big(m(\mathbb{I}_5|15423)+m(\mathbb{I}_5|51423)\Big)
 -s_{45}(s_{41}{+}s_{42})\Big(m(\mathbb{I}_5|51243)+m(\mathbb{I}_5|15243)+m(\mathbb{I}_5|12543)\Big) \nn\\
& + s_{51} s_{41}\, m(\mathbb{I}_5|15423)+ s_{51}(s_{41}{+}s_{42})\,m(\mathbb{I}_n | 15243) +(s_{51}{+}s_{52})(s_{41}{+}s_{42})\,m(\mathbb{I}_5|12543)\nn\\
&- s_{54}s_{51}\Big(m(\mathbb{I}_5|14523)+m(\mathbb{I}_5|41523)\Big)
 -s_{54}(s_{51}{+}s_{52})\Big(m(\mathbb{I}_5|41253)+m(\mathbb{I}_5|14253)+m(\mathbb{I}_5|12453)\Big)\nn\bigg].\label{eq:EYM-example}
\end{align}
One can obtain expressions for the pure bi-adjoint amplitudes by, for example, using the recursions \eqref{eq:bi-adjoint-amplitude},
\begin{gather}
    m(\mathbb{I}_5|14523)  = -\frac{1}{s_{23}s_{45}},\quad m(\mathbb{I}_5 | 14253) = 0, \quad m(\mathbb{I}_5|12453) = -\frac{1}{s_{12}s_{45}}, \nn\\
    m(\mathbb{I}_5 | 15423) = \frac{1}{s_{23}}\left( \frac{1}{s_{45}} + \frac{1}{s_{51}} \right), \quad m(\mathbb{I}_5 | 51423) = -\frac{1}{s_{23}s_{51}}, \quad m(\mathbb{I}_5 | 51243) = - \frac{1}{s_{34}} \left( \frac{1}{s_{51}} + \frac{1}{s_{12}}\right)\nn\\
    m(\mathbb{I}_5 | 15243) = \frac{1}{s_{34}s_{51}}, \quad m(\mathbb{I}_5 | 12543) =  \frac{1}{s_{12}}\left( \frac{1}{s_{34}} + \frac{1}{s_{45}}\right), \quad m(\mathbb{I}_5 | 41523) = -\frac{1}{s_{51}}\left( \frac{1}{s_{23}} + \frac{1}{s_{34}}\right),\nn\\
    \quad m(\mathbb{I}_5 | 41253) = - \frac{1}{s_{12}s_{34}}.
\end{gather}
Plugging them into \eqref{eq:EYM-example} we obtain:
\be
\mathcal{A}_5^{\text{NLSM}\oplus\text{BA}}(\mathbb{I}_5 \| 123) = \frac{1}{2}\Big(\frac{s_{34}+s_{45}}{s_{12}}+\frac{s_{45}+s_{51}}{s_{23}}-1\Big).
\ee
This result agrees with the computation in \eqref{eq:A-extended-345} after a relabeling.

\subsection{\label{sec:dimensional-reduction}Dimensional Reduction of the $F^3$ Theory}

Given the discussion of \cite{Cheung:2017yef} and Section~\ref{sec:extended-NLSM}, where amplitudes of NLSM and NLSM coupled to bi-adjoint scalars were understood as coming from those of gluons in a higher dimensions, we can ask if this phenomenon holds more generally. A natural playground for these considerations is the so-called Z-theory \cite{Carrasco:2016ldy,Mafra:2016mcc,Carrasco:2016ygv}, which contains the above theories in the low-energy limit. Let us focus on the abelian sector with respect to the group $\text{U}(\widetilde{N})$. We can write down an effective action:
\be
\mathcal{L}^{\text{ab.\! Z-theory}} = \mathcal{L}^{\text{NLSM}} + \alpha'^2 \zeta_2\, \mathcal{L}^{\text{subl.}} + \ldots,
\ee
where $\mathcal{L}^{\text{NLSM}}$ is known to be the NLSM Lagrangian given in \eqref{eq:NLSM-Lagrangian}. Our goal is to find the subleading correction $\mathcal{L}^{\text{subl.}}$. It was conjectured in \cite{Mizera:2017sen} that amplitudes at this order can be computed within the CHY formalism using the replacement rules analogous to \eqref{eq:replacement-rules} applied to the $F^3$ gauge theory. Here, we extend this proposal to an off-shell statement on the level of the Lagrangian. Starting with the $F^3$ Lagrangian:
\be
\mathcal{L}^{F^3} = \Tr \Big( {\F}_{\mu\nu} {\F}_{\nu\rho} {\F}_{\rho\mu} \Big),
\ee
we perform the substitutions \eqref{eq:A-into-XY} and \eqref{eq:metric-sub} as before, to obtain:
\begin{align}
\mathcal{L}^{F^3} \bigg|_{\substack{\scriptscriptstyle\eqref{eq:A-into-XY}\\ \scriptscriptstyle\eqref{eq:metric-sub}}} = \Tr \bigg( \sum_{\alpha,\beta,\gamma = \uparrow, \downarrow} \F^{\alpha\beta}_{\mu\nu}\, \F^{\beta\gamma}_{\nu\rho} \,\F^{\gamma\alpha}_{\rho\mu} \bigg) =: \mathcal{L}^{\text{subl.}},
\end{align}
which can be expressed in term of the $\X$ and $\Z$ fields as follows:
\begin{align}
\F^{\uparrow\uparrow}_{\mu\nu} &= \frac{1}{\sqrt{2}} \partial_\mu (\Z_\nu {+} \X_\nu) - \frac{1}{\sqrt{2}} \partial_\nu (\Z_\mu {+} \X_\mu) - \frac{1}{2} [\Z_\mu {+} \X_\mu, \Z_\nu {+} \X_\nu],\nn\\
\F^{\uparrow\downarrow}_{\mu\nu} &= \frac{i}{\sqrt{2}} \partial_\mu (\Z_\nu {-} \X_\nu) - \frac{i}{2} [\Z_\mu {+} \X_\mu, \Z_\nu {-} \X_\nu],\\
\F^{\downarrow\uparrow}_{\mu\nu} &= -\frac{i}{\sqrt{2}} \partial_\nu (\Z_\mu {-} \X_\mu) - \frac{i}{2} [\Z_\mu {-} \X_\mu, \Z_\nu {+} \X_\nu],\nn\\
\F^{\downarrow\downarrow}_{\mu\nu} &=  \frac{1}{2} [\Z_\mu {-} \X_\mu, \Z_\nu {-} \X_\nu].\nn
\end{align}
This defines the Lagrangian for abelian Z-theory at the subleading order. Notice that the field $\Y$ does not appear at this order. With the above Lagrangian one can compute equations of motion and substitute the perturbiner ansatz \eqref{eq:XYZ-perturbiners}. Given the length of the resulting perturbiners, we will not display them explicitly.

Amplitudes in the abelian Z-theory at the subleading order are suppressed by two powers of $\alpha'$ compared to the leading one. In particular, we find that the boundary conditions for the $\X\Y\Z$-system need to be modified by considering only external $\Z$-states in order to account for this difference in mass dimension. Using perturbiners we compute the contribution to abelian Z-theory amplitudes $Z_\times(\mathbb{I}_n)$ at order $\alpha'^2$:\footnote{Normalization conventions are tailored to this work and are not necessarily natural from the point of view of Z-theory amplitudes \cite{Carrasco:2016ldy}.}
\be
Z_\times(\mathbb{I}_4) \bigg|_{\alpha'^2} = -  \frac{3\sqrt{2} \alpha'^2 \zeta_2}{4} s_{13} (s_{12}^2 + s_{23}^2 + s_{13}^2),
\ee
as well as
\begin{align}
Z_\times(\mathbb{I}_6)\bigg|_{\alpha'^2} = -\frac{3 \alpha'^2 \zeta_2}{2\sqrt{2}}\Bigg(& -\frac{(s_{12} + s_{23})(s_{12}^2 + s_{12} s_{23} + s_{23}^2)(s_{45}+s_{56})}{s_{123}} + 4 s_{12} s_{23} s_{234} + 4 s_{12} s_{23} s_{345}\nn\\
&- 4 s_{12} s_{23} s_{34}+ 2 s_{12} s_{23} s_{56} + 2 s_{12} s_{23} s_{45} + 2 s_{12} s_{34} s_{123} + 2 s_{12} s_{34} s_{234} \nn\\
&+ s_{12} s_{34} s_{345} + s_{12}^3 + 2 s_{12}^2 s_{45} + 2 s_{12}^2 s_{234} - 2 s_{12} s_{234}^2 - 4 s_{12} s_{123} s_{234}\nn\\
&- 2 s_{23} s_{123} s_{234} - 4 s_{34} s_{123} s_{234} - \frac{1}{2} s_{12} s_{45} s_{123} - \frac{1}{2} s_{12} s_{45} s_{345} + s_{123}^2 s_{234}\nn\\
&+ s_{123}s_{234}^2 +  \frac{1}{3}s_{12} s_{34} s_{56} + \frac{4}{3} s_{123} s_{234} s_{345} +\; {\rm cyc}(1,2,\ldots,6)\;\Bigg),\qquad
\end{align}
which matches the results of \cite{Carrasco:2016ldy} up to an overall constant. We also checked numerically up to $9$-pt that the results agree with the ones obtained using $\alpha'$-expansion method introduced in \cite{Carrasco:2016ygv}.

Let us emphasize that results at different orders in $\alpha'$ are computed using different selections of external states in the $\X\Y\Z$-system. This leaves an open question, whether more general amplitudes, such as those at higher orders in $\alpha'$ or perhaps involving couplings to bi-adjoint scalars, can be computed within the same framework.

\section{\label{sec:EFTs-without-colors}Effective Field Theories Without Colors}

In this section we focus on theories without color degrees of freedom, such as special Galileon and Born--Infeld theories. In both cases, we formulate perturbiner expansions and used them compute example amplitudes.

\subsection{\label{sec:special-Galileon}Special Galileon Theory}

Galileon theories are scalar field theories arising in certain infrared modifications of gravity \cite{Dvali:2000hr,deRham:2010ik}. Among these, there is a specific theory called the special Galileon \cite{Cachazo:2014xea,Cheung:2014dqa} with an enhanced shift symmetry and soft behaviour \cite{Hinterbichler:2015pqa}. This theory is particularly interesting from the S-matrix point of view, since it can be written as a double-copy of the NLSM \cite{Cachazo:2014xea}. Off-shell recursion relations for Galileon theories were previously considered in \cite{Kampf:2014rka}. Recently, it was proposed in \cite{Cheung:2016prv} that this fact can be made manifest on the level of the Lagrangian analogous to the one discussed in Section~\ref{sec:NLSM}. It features a triplet of fields $(\XX^{\mu\bar{\mu}}, \YY, \ZZ^{\mu\bar{\mu}})$ with the Lagrangian \cite{Cheung:2016prv}:
\begin{align}\label{eq:sGal-Lagrangian}
\mathcal{L}^{\text{sGal}} = \XX_{\mu\bar{\mu}}\Box \ZZ^{\mu\bar{\mu}} + \frac{1}{2} \YY \Box \YY &+ 2 \left( \partial_\mu \partial_{\bar{\mu}}\XX_{\nu\bar\nu} + \partial_\nu \partial_{\bar{\nu}}\XX_{\mu\bar\mu} - \partial_\mu \partial_{\bar{\nu}}\XX_{\nu\bar\mu} - \partial_\nu \partial_{\bar{\mu}}\XX_{\mu\bar\nu} \right) \ZZ^{\mu\bar{\mu}}\ZZ^{\nu\bar{\nu}} \nn\\
&+ \left(\YY \partial_\mu \partial_{\bar{\nu}} \YY - \partial_\mu\YY  \partial_{\bar{\nu}} \YY\right) \ZZ^{\mu\bar{\nu}}.
\end{align}
It is straightfoward to derive the corresponding equations of motion:
\begin{align}
\Box \XX^{\mu\bar{\mu}} &= 4 \ZZ^{\nu\bar{\nu}} \big( \partial^\mu \partial_{\bar{\nu}} {\XX_{\nu}}^{\bar{\mu}} + \partial_\nu \partial^{\bar{\mu}} {\XX^{\mu}}_{\bar{\nu}} - \partial^\mu \partial^{\bar{\mu}} \XX_{\nu\bar{\nu}} - \partial_\nu \partial_{\bar{\nu}} \XX^{\mu\bar{\mu}} \big) + \partial^\mu \YY\, \partial^{\bar{\mu}} \YY - \YY \partial^\mu \partial^{\bar{\mu}} \YY,\nn\\
\Box \YY &= -4 \ZZ^{\nu\bar{\nu}} \partial_{\nu} \partial_{\bar{\nu}} \YY - 2\partial_\nu \YY\, \partial_{\bar{\nu}} \ZZ^{\nu\bar{\nu}} - 2 \partial_{\bar{\nu}} \YY\, \partial_{\nu} \ZZ^{\nu\bar{\nu}}, \label{eq:sGal-recursions}\\
\Box \ZZ^{\mu\bar{\mu}} &= 4 \partial_{\nu} \partial_{\bar{\nu}} \big( \ZZ^{\mu\bar{\nu}}\ZZ^{\nu\bar{\mu}} - \ZZ^{\mu\bar{\mu}} \ZZ^{\nu\bar{\nu}}\big).\nn
\end{align}
We see from the Lagrangian \eqref{eq:sGal-Lagrangian} that the $\XX$-field has a gauge redundancy,
$\XX^{\mu\bar\mu} \to \XX^{\mu\bar\mu} + k^\mu \mathbf{\lambda}^{\bar\mu} + \mathbf{\lambda}^{\mu}\, k^{\bar\mu}$. Since sGal is a colorless theory, we can write the perturbiner expansion of the type \eqref{eq:colorless-perturbiner} for each field separately:
\be\label{eq:sGal-perturbiners}
\big(\XX^{\mu\bar{\mu}},\, \YY,\, \ZZ^{\mu\bar{\mu}} \big) = 
\sum_{ \cal P } \big( \XX^{\mu\bar{\mu}}_{\cal P},\, \YY_{\cal P},\, \ZZ^{\mu\bar{\mu}}_{\cal P}\big) e^{k_{\cal P}\cdot x}.
\ee
Plugging these into the equations of motion, we obtain:
\begin{gather}
\XX_{\cal P}^{\mu\bar{\mu}} = \frac{1}{s_{\cal P}}\! \sum_{{\cal P} = {\cal Q} \cup {\cal R}} \left[ 2 \ZZ_{\cal Q}^{\nu\bar{\nu}} \Big( k_{\cal R}^\mu k_{\cal R}^{\bar{\nu}} \XX_{\cal R}^{\nu \bar{\mu}} + k_{\cal R}^\nu k_{\cal R}^{\bar{\mu}} \XX_{\cal R}^{\mu \bar{\nu}} - k_{\cal R}^\mu k_{\cal R}^{\bar{\mu}} \XX_{\cal R}^{\nu \bar{\nu}} - k_{\cal R}^\nu k_{\cal R}^{\bar{\nu}} \XX_{\cal R}^{\mu \bar{\mu}} \Big) + \frac{1}{2}\! \left( k_{\cal Q}^\mu {-} k_{\cal R}^\mu \right) k_{\cal R}^{\bar{\mu}} \YY_{\cal Q} \YY_{\cal R}\right]\!,\nn\\
\YY_{\cal P} = - \frac{2}{s_{\cal P}} \sum_{{\cal P} = {\cal Q} \cup {\cal R}} \bigg( \ZZ_{\cal Q}^{\nu\bar{\nu}} k_{\cal R}^{\nu} k_{\cal R}^{\bar{\nu}} \YY_{\cal R} \bigg),\\
\ZZ_{\cal P}^{\mu\bar{\mu}} = \frac{2}{s_{\cal P}} \sum_{{\cal P} = {\cal Q} \cup {\cal R}} \bigg( \ZZ_{\cal Q}^{\mu\bar{\nu}} k_{\cal Q}^\nu k_{\cal R}^{\bar\nu} \ZZ_{\cal R}^{\nu\bar{\mu}} - \ZZ_{\cal Q}^{\mu\bar{\mu}} k_{\cal Q}^\nu k_{\cal Q}^{\bar\nu} \ZZ_{\cal R}^{\nu\bar{\nu}}\bigg). \nn
\end{gather}
Here we used the transversality conditions in both indices of the $\ZZ^{\mu\bar{\mu}}$ field, i.e., $k^\mu_{\cal P} \ZZ^{\mu\bar{\mu}}_{\cal P} = k^{\bar\mu}_{\cal P} \ZZ^{\mu\bar{\mu}}_{\cal P} = 0$.
The $n$-point amplitudes are given by the Berends--Giele formula:
\be\label{eq:sGal-amplitude}
\mathcal{A}^{\text{sGal}}_n = \lim\limits_{k_n^2 \to 0} s_{12\cdots n{-}1} \Big( \XX_{12\cdots n{-}1}^{\mu\bar{\mu}} \ZZ_{n}^{\mu\bar{\mu}} \,+\, \ZZ_{12\cdots n{-}1}^{\mu\bar{\mu}} \XX_{n}^{\mu\bar{\mu}} \,+\, \YY_{12\ldots n-1}\YY_{n} \Big).
\ee
Similarly to the NLSM case, one can choose two types of boundary conditions for the above recursions: 
\begin{align}
(a):& \qquad \big( \XX^{\mu\bar{\mu}}_i,\; \YY_i,\; \ZZ^{\mu\bar{\mu}}_i \big) = \begin{cases}
\big( 0,\; 1,\;\, 0\,\; \big)  \qquad &\text{if } i \text{ is a } \YY\text{-state} \\
\big( 0,\; 0,\; k^\mu_i k^{\bar\mu}_i \big)  \qquad &\text{if } i \text{ is a } \ZZ\text{-state},
\end{cases}\label{eq:sGal-case-a}\\
(b):& \qquad \big( \XX^{\mu\bar{\mu}}_i,\; \YY_i,\; \ZZ^{\mu\bar{\mu}}_i \big) = \begin{cases}
\big( \overline{k}_i^\mu \overline{k}_i^{\bar\mu},\; 0,\; \,0\; \big)  \qquad &\text{if } i \text{ is an } \XX\text{-state},\\
\big( \;0\;,\; 0,\; k^\mu_i k^{\bar\mu}_i \big)  \qquad &\text{if } i \text{ is a } \ZZ\text{-state}.
\end{cases}
\end{align}
In the case $(a)$ there are exactly $n{-}2$ external $\ZZ$'s and two $\YY$'s, while in the case $(b)$ there are $n{-}1$ external $\ZZ$'s and one $\XX$. The amplitude is independent of the choice of the special states in both cases. As before, only one term in the above current \eqref{eq:sGal-amplitude} contributes depending on the initial conditions.

We use the same notation with underlined and overlined labels as in the NLSM case. For example, $\YY_{1\underline{2}3}$ denotes the $\YY$-current where particles $1$ and $3$ are in the $\ZZ$-state, while $2$ is in the $\YY$-state.

\subsubsection{Four-Point Examples}

Let us illustrate how to use \eqref{eq:sGal-recursions} in practice by computing $4$-pt amplitudes in two different ways.

\begin{itemize}[leftmargin=*]
\item
Case ($a$) with $(\ZZ_1^{\mu\bar{\mu}}, \YY_{\underline{2}}, \ZZ_3^{\mu\bar{\mu}}, \YY_{\underline{4}}) = (k_1^{\mu} k_1^{\bar{\mu}}, 1, k_3^{\mu} k_3^{\bar{\mu}}, 1)$ and all other $(\XX_i^{\mu\bar\mu}, \YY_i, \ZZ_i^{\mu\bar\mu})$ vanishing.  
We first calculate the relevant rank-$2$ currents:
\begin{gather}
\YY_{1\underline{2}} = - \frac{1}{s_{12}} \left(2 \ZZ_1^{\nu\bar{\nu}} k_2^\nu k_2^{\bar{\nu}} \YY_{\underline{2}} + \ZZ_1^{\nu\bar{\nu}} (k_1^\nu k_2^{\bar{\nu}} + k_1^{\bar{\nu}} k_2^{\nu}) \YY_{\underline{2}} \right) = -2s_{12},\\
\YY_{\underline{2}3} = - \frac{1}{s_{23}} \left(2 \ZZ_3^{\nu\bar{\nu}} k_2^\nu k_2^{\bar{\nu}} \YY_{\underline{2}} + \ZZ_3^{\nu\bar{\nu}} (k_3^\nu k_2^{\bar{\nu}} + k_3^{\bar{\nu}} k_2^{\nu}) \YY_{\underline{2}} \right) = -2s_{23},\\
\ZZ_{13}^{\mu\bar{\mu}} = \frac{2}{s_{13}} k_{13}^{\nu} k_{13}^{\bar{\nu}} \left( \ZZ_1^{\mu\bar{\nu}} \ZZ_3^{\nu\bar{\mu}} - \ZZ_1^{\mu\bar{\mu}} \ZZ_3^{\nu\bar{\nu}} + \ZZ_3^{\mu\bar{\nu}} \ZZ_1^{\nu\bar{\mu}} - \ZZ_3^{\mu\bar{\mu}} \ZZ_1^{\nu\bar{\nu}} \right) = -2s_{13} (k_1^\mu - k_3^\mu)(k_1^{\bar{\mu}} - k_3^{\bar{\mu}}),\label{eq:Z-13}\\
\YY_{13} = \ZZ_{1\underline{2}}^{\mu\bar{\mu}} = \ZZ_{\underline{2}3}^{\mu\bar{\mu}} = 0.
\end{gather}
Using decomposition of the permutation ${\cal P} = 123$ as in \eqref{eq:123-decomp} and the above results, we find:
\begin{align}
\YY_{1\underline{2}3} &= -\frac{1}{s_{123}} \bigg( 2\ZZ_{13}^{\nu\bar{\nu}} k_2^{\nu}k_2^{\bar{\nu}}\YY_2+ 2\ZZ_{1}^{\nu\bar{\nu}} k_{23}^{\nu}k_{23}^{\bar{\nu}}\YY_{23} + 2\ZZ_{3}^{\nu\bar{\nu}} k_{12}^{\nu}k_{12}^{\bar{\nu}}\YY_{12} \nn\\
&\qquad\qquad +\ZZ_3^{\nu\bar{\nu}}(k_{12}^\nu k_3^{\bar{\nu}} + k_{12}^{\bar{\nu}} k_3^{\nu}) \YY_{12} + \ZZ_1^{\nu\bar{\nu}}(k_{23}^\nu k_1^{\bar{\nu}} + k_{23}^{\bar{\nu}} k_1^{\nu}) \YY_{23} + \ZZ_{13}^{\nu\bar{\nu}}(k_{2}^\nu k_{13}^{\bar{\nu}} + k_{2}^{\bar{\nu}} k_{13}^{\nu}) \YY_{2} \bigg)\nn\\
&=\frac{4 (s_{12}+s_{13}) (s_{12}+s_{23})(s_{13}+s_{23})}{s_{123}}.
\end{align}
Finally, we can calculate the amplitude:
\be
\mathcal{A}^{\text{sGal}}_4 = \lim\limits_{k_4^2 \to 0} s_{123} \YY_{1\underline{2}3} \YY_{\underline{4}} = -4 s_{12} s_{23} s_{13}.
\ee

\item Case $(b)$ with $(\ZZ_1^{\mu\bar\mu}, \ZZ_2^{\mu\bar\mu}, \ZZ_3^{\mu\bar\mu}, \XX_{\overline{4}}^{\mu\bar\mu}) = (k_1^{\mu}k_1^{\bar\mu}, k_2^{\mu}k_2^{\bar\mu}, k_3^{\mu}k_3^{\bar\mu}, \overline{k}_4^{\mu}\overline{k}_4^{\bar\mu})$ and all other $(\XX_i^{\mu\bar\mu}, \YY_i, \ZZ_i^{\mu\bar\mu})$ vanishing. The relevant rank-$2$ current was given in \eqref{eq:Z-13}. Relabeling the result we have:
\be
\ZZ_{12}^{\mu\bar\mu} = -s_{12}(k_1^\mu {-} k_2^\mu)(k_1^{\bar\mu} {-} k_2^{\bar\mu}), \quad \ZZ_{23}^{\mu\bar\mu} = -s_{23}(k_2^\mu {-} k_3^\mu)(k_2^{\bar\mu} {-} k_3^{\bar\mu}), \quad \ZZ_{13}^{\mu\bar\mu} = -s_{13}(k_1^\mu {-} k_3^\mu)(k_1^{\bar\mu} {-} k_3^{\bar\mu}).\nn
\ee
Hence we have the rank-$3$ current:
\begin{align}
\mathbf{Z}_{123}^{\mu\bar\mu} = \frac{4}{s_{123}}\bigg\{& 
s_{12}^2 \Big(s_{13} \left(k_1^{\mu }{-}k_2^{\mu }{-}k_3^{\mu }\right) \left(k_1^{\bar\mu}{-}k_2^{\bar\mu}{-}k_3^{\bar\mu}\right)+s_{23} \left(k_1^{\mu }{-}k_2^{\mu }{+}k_3^{\mu }\right) \left(k_1^{\bar\mu}{-}k_2^{\bar\mu}{+}k_3^{\bar\mu}\right)\Big)\nn\\
& +s_{12} \bigg[s_{13}^2 \left(k_1^{\mu }{-}k_2^{\mu }{-}k_3^{\mu}\right) \left(k_1^{\bar\mu}{-}k_2^{\bar\mu}{-}k_3^{\bar\mu}\right)+2 s_{23} s_{13} \Big(k_1^{\mu }\left(k_1^{\bar\mu}{-}k_2^{\bar\mu}{-}k_3^{\bar\mu}\right)\nn\\
   &-\left(k_2^{\mu }{+}k_3^{\mu }\right)k_1^{\bar\mu}{+}\left(k_2^{\mu }{-}k_3^{\mu }\right) \left(k_2^{\bar\mu}{-}k_3^{\bar\mu}\right)\Big)+s_{23}^2 \left(k_1^{\mu }{-}k_2^{\mu }{+}k_3^{\mu }\right) \left(k_1^{\bar\mu}{-}k_2^{\bar\mu}{+}k_3^{\bar\mu}\right)\bigg]\nn\\
   &+s_{13} s_{23} \left(s_{13}{+}s_{23}\right) \left(k_1^{\mu }{+}k_2^{\mu }{-}k_3^{\mu}\right) \left(k_1^{\bar\mu}{+}k_2^{\bar\mu}{-}k_3^{\bar\mu}\right)
\bigg\}.
\end{align}
This leads to the amplitude:
\be
\mathcal{A}^{\text{sGal}}_4 = \lim\limits_{k_4^2 \to 0} s_{123} \ZZ_{123}^{\mu\bar\mu} \XX_{\overline{4}}^{\mu\bar\mu} = -4 s_{12} s_{23} s_{13} (k_{4} \!\cdot\! \overline{k}_4)^2 =  -4 s_{12} s_{23} s_{13} .
\ee

\end{itemize}
We checked up to $7$-pt that \eqref{eq:sGal-amplitude} computes the correct special Galielon amplitudes.

\subsection{Born--Infeld Theory}

The final theory we consider is that non-linearly itneracting photons, the Born--Infeld theory. It is yet another theory that can be written in terms of the triplet of fields, which are now called $(\XXX^{\mu\bar{\mu}}, \YYY^\mu, \ZZZ^{\mu\bar{\mu}})$ with the following Lagrangian \cite{Cheung:2017yef}:
\begin{align}
\mathcal{L}^{\text{BI}} = \XXX_{\mu\bar\mu} \Box \ZZZ^{\mu\bar\mu} + \frac{1}{2} \YYY_\mu \Box \YYY^\mu &+ \frac{1}{2\sqrt{2}}\ZZZ^{\mu\bar\nu} \bigg( \partial_\mu \YYY_\rho \partial_{\bar\nu} \YYY^{\rho} +  \partial_{\bar\nu}\YYY_\mu \partial_\rho \YYY^\rho - 2 \partial_\rho \YYY_\mu \partial_{\bar\nu}\YYY^\rho + \partial_\mu \ZZZ^{\rho\bar\sigma} \XXX_{\rho\bar\nu\bar\sigma}\nn \\
&- \partial^\rho {\ZZZ_{\mu}}^{\bar\sigma} \XXX_{\rho\bar\nu\bar\sigma} - \ZZZ^{\rho\bar\sigma}\partial_\mu \XXX_{\rho\bar\nu\bar\sigma} - \frac{1}{4\sqrt{2}} {\ZZZ_{\mu}}^{\bar\rho}\partial_{\bar\nu} \YYY_{\sigma} \partial_{\bar\rho}\YYY^{\sigma} - \frac{1}{4\sqrt{2}} \ZZZ^{\rho\bar\sigma} \partial_{\bar\nu} \YYY_\mu \partial_{\bar\sigma} \YYY_\rho \nn\\
& + \frac{1}{2\sqrt{2}} \ZZZ^{\rho\bar\sigma} \partial_{\bar\nu}\YYY_{\rho} \partial_{\bar\sigma} \YYY_{\mu} + \frac{1}{2\sqrt{2}} \ZZZ_{\mu\bar\alpha} \partial_{\bar\nu} \ZZZ_{\rho\bar\sigma} \XXX^{\rho\bar\sigma\bar\alpha} + \frac{1}{2\sqrt{2}} \ZZZ^{\rho\bar\sigma} \partial_{\bar\sigma} {\ZZZ_{\mu}}^{\bar\alpha} \XXX_{\rho\bar\nu\bar\alpha} \bigg),\nn
\end{align}
where $\XXX_{\mu\bar\nu\bar\rho} = \partial_{\bar\nu} \XXX_{\mu\bar\rho} - \partial_{\bar\rho} \XXX_{\mu\bar\nu}$.
We can use a perturbiner expansion equivalent to that of \eqref{eq:sGal-perturbiners} in order to find the recursion relations:
\begin{align}
\XXX_{\cal P}^{\mu\bar{\mu}} &= -\frac{1}{2s_{\cal P}} \bigg\{ \frac{1}{2\sqrt{2}}\sum_{{\cal P}={\cal Q}\cup{\cal R}} \bigg[
k_Q^{\rho } \ZZZ_{\cal Q}^{\mu  \bar\sigma} \left(k_{\cal R}^{\bar\sigma}
   \XXX_{\cal R}^{\rho \bar\mu}-k_{\cal R}^{\bar\mu} \XXX_{\cal R}^{\rho  \bar\sigma}\right)-k_{\cal Q}^{\rho } \ZZZ_{\cal Q}^{\rho  \bar\sigma}
   \left(k_{\cal R}^{\bar\sigma} \XXX_{\cal R}^{\mu \bar\mu}-k_{\cal R}^{\bar\mu} \XXX_{\cal R}^{\mu  \bar\sigma}\right)\nn\\
   &\qquad\qquad\qquad\qquad\qquad\quad+k_{\cal R}^{\rho } \ZZZ_{\cal Q}^{\mu 
   \bar\sigma} \left(k_{\cal R}^{\bar\sigma} \XXX_{\cal R}^{\bar\mu \rho
   }-k_{\cal R}^{\bar\mu} \XXX_{\cal R}^{\rho \bar\sigma}\right)-2 k_{\cal R}^{\rho }
   \ZZZ_{\cal Q}^{\rho \bar\sigma} \left(k_{\cal R}^{\bar\sigma} \XXX_{\cal R}^{\mu 
   \bar\mu}-k_{\cal R}^{\bar\mu} \XXX_{\cal R}^{\mu \bar\sigma}\right)\nn\\
   &\qquad\qquad\qquad\qquad\qquad\quad-k_{\cal R}^{\mu } \ZZZ_{\cal Q}^{\rho \bar\sigma} \left(k_{\cal R}^{\bar\mu} \XXX_{\cal R}^{\rho \bar\sigma}-k_{\cal R}^{\bar\sigma} \XXX_{\cal R}^{\rho\bar\mu }\right)+\left(k_{\cal Q}^{\mu } \ZZZ_{\cal Q}^{\rho \bar\sigma}-k_{\cal Q}^{\rho
   } \ZZZ_{\cal Q}^{\mu \bar\sigma}\right) \left(k_{\cal R}^{\bar\mu} \XXX_{\cal R}^{\rho
    \bar\sigma}-k_{\cal R}^{\bar\sigma} \XXX_{\cal R}^{\rho\bar\mu
   }\right)\nn\\
   &\qquad\qquad\qquad\qquad\qquad\quad+k_{\cal Q}^{\mu } k_{\cal R}^{\bar\mu} \YYY_{\cal Q}^{\rho } \YYY_{\cal R}^{\rho
   }+k_{\cal Q}^{\bar\mu} k_{\cal R}^{\rho } \YYY_{\cal Q}^{\mu } \YYY_{\cal R}^{\rho }-2 k_{\cal Q}^{\rho }
   k_{\cal R}^{\bar\mu} \YYY_{\cal Q}^{\mu } \YYY_{\cal R}^{\rho }
\bigg],\nn\\
&\qquad\qquad+\frac{1}{8}\sum_{{\cal P} = {\cal Q} \cup {\cal R} \cup {\cal S}} \bigg[
k_{\cal S}^{\bar\sigma} \ZZZ_{\cal R}^{\rho \bar\sigma} \ZZZ_{\cal S}^{\mu  \bar\mu} \left(k_{\cal R}^{\bar\alpha}+k_{\cal S}^{\bar\alpha}\right)
   \ZZZ_{\cal Q}^{\rho \bar\alpha}-k_{\cal S}^{\bar\sigma} \ZZZ_{\cal R}^{\mu 
   \bar\sigma} \ZZZ_{\cal S}^{ \rho \bar\mu } \left(k_{\cal R}^{\bar\alpha}+k_{\cal S}^{\bar\alpha}\right) \ZZZ_{\cal Q}^{\rho  \bar\alpha}\nn\\
& \qquad\qquad\qquad\qquad\qquad\quad  -k_{\cal S}^{\bar\alpha} \left(k_{\cal Q}^{\bar\sigma}+k_{\cal R}^{\bar\sigma}\right) \ZZZ_{\cal R}^{ \rho \bar\mu} \ZZZ_{\cal S}^{\mu 
   \bar\sigma} \ZZZ_{\cal Q}^{\rho \bar\alpha}+k_{\cal S}^{\bar\sigma} \ZZZ_{\cal Q}^{ \rho \bar\mu} \ZZZ_{\cal R}^{\mu \bar\sigma}
   \left(k_{\cal Q}^{\bar\alpha}+k_{\cal R}^{\bar\alpha}\right)
   \ZZZ_{\cal S}^{\rho \bar\alpha}
\bigg]\bigg\},\nn\\
\YYY_{\cal P}^{\mu} &= -\frac{1}{2s_{\cal P}} \bigg\{ \frac{1}{2\sqrt{2}}\sum_{{\cal P}={\cal Q}\cup{\cal R}} \bigg[-\YYY_{\cal R}^{\mu } k_{\cal Q}^{\rho } k_{\cal R}^{\bar\mu} \ZZZ_{\cal Q}^{
   \rho\bar\mu }-2 \YYY_{\cal R}^{\mu } k_{\cal R}^{\bar\mu} k_{\cal R}^{\rho }
   \ZZZ_{\cal Q}^{\rho\bar\mu }-\YYY_{\cal R}^{\rho } k_{\cal Q}^{\mu } k_{\cal R}^{\bar\mu} \ZZZ_{\cal Q}^{ \rho\bar\mu }\nn\\
   &\qquad\qquad\qquad\qquad\qquad\quad +2 \YYY_{\cal R}^{\rho } k_{\cal Q}^{\rho }
   k_{\cal R}^{\bar\mu} \ZZZ_{\cal Q}^{\mu \bar\mu} +\YYY_{\cal R}^{\rho } k_{\cal R}^{\mu }
   k_{\cal R}^{\bar\mu} \ZZZ_{\cal Q}^{ \rho \bar\mu}+\YYY_{\cal R}^{\rho }
   k_{\cal R}^{\bar\mu} k_{\cal R}^{\rho } \ZZZ_{\cal Q}^{\mu \bar\mu}\bigg]\\
   &\qquad\qquad+\frac{1}{8}\sum_{{\cal P} = {\cal Q} \cup {\cal R} \cup {\cal S}} \bigg[ \YYY_{\cal S}^{\mu } k_{\cal S}^{\bar\rho} \left(k_{\cal R}^{\bar\mu}+k_{\cal S}^{\bar\mu}\right) \ZZZ_{\cal Q}^{ \sigma \bar\mu}
   \ZZZ_{\cal R}^{\sigma \bar\rho}-2 \YYY_{\cal S}^{\rho } k_{\cal S}^{\bar\sigma} \ZZZ_{\cal Q}^{ \rho\bar\mu } \ZZZ_{\cal R}^{\mu  \bar\sigma} \left(k_{\cal R}^{\bar\mu}+k_{\cal S}^{\bar\mu}\right)\nn\\
   &\qquad\qquad\qquad\qquad\qquad\quad +\YYY_{\cal S}^{\rho } k_{\cal S}^{\bar\sigma} \ZZZ_{\cal Q}^{\mu \bar\mu} \ZZZ_{\cal R}^{\rho  \bar\sigma} \left(k_{\cal R}^{\bar\mu}+k_{\cal S}^{\bar\mu}\right)\bigg]\bigg\},\nn\\
\ZZZ_{\cal P}^{\mu\bar{\mu}} &=  -\frac{1}{2s_{\cal P}} \bigg\{ \frac{1}{2\sqrt{2}}\sum_{{\cal P}={\cal Q}\cup{\cal R}} \bigg[
-k_{\cal Q}^{\rho } k_{\cal R}^{\bar\sigma} \ZZZ_{\cal Q}^{\rho  \bar\sigma} \ZZZ_{\cal R}^{\mu \bar\mu}+k_{\cal Q}^{\rho }
   k_{\cal Q}^{\bar\sigma} \ZZZ_{\cal Q}^{ \rho \bar\mu} \ZZZ_{\cal R}^{\mu  \bar\sigma}+k_{\cal Q}^{\bar\sigma} \ZZZ_{\cal Q}^{ \rho \bar\mu} \left(2 k_{\cal R}^{\rho } \ZZZ_{\cal R}^{\mu \bar\sigma}-k_{\cal R}^{\mu } \ZZZ_{\cal R}^{\rho \bar\sigma}\right)\nn\\
    &\qquad\qquad\qquad\qquad\qquad\quad  +k_{\cal R}^{\bar\sigma} \ZZZ_{\cal Q}^{\rho \bar\sigma} \left(k_{\cal R}^{\mu } \ZZZ_{\cal R}^{ \rho\bar\mu }-2
   k_{\cal R}^{\rho } \ZZZ_{\cal R}^{\mu \bar\mu}\right)
\bigg]\nn\\
  &\qquad\qquad+\frac{1}{8}\sum_{{\cal P} = {\cal Q} \cup {\cal R} \cup {\cal S}} \bigg[
  k_{\cal S}^{\bar\sigma} \ZZZ_{\cal Q}^{ \rho \bar\alpha} \ZZZ_{\cal R}^{\rho \bar\sigma} \ZZZ_{\cal S}^{\mu \bar\mu}
   \left(k_{\cal R}^{\bar\alpha}+k_{\cal S}^{\bar\alpha}\right)+k_{\cal S}^{\bar\sigma} \ZZZ_{\cal Q}^{ \rho \bar\mu}
   \ZZZ_{\cal R}^{\mu \bar\sigma} \ZZZ_{\cal S}^{ \rho\bar\alpha } \left(k_{\cal Q}^{\bar\alpha}+k_{\cal R}^{\bar\alpha}\right)\nn\\
   &\qquad\qquad\qquad\qquad\qquad\quad-k_{\cal S}^{\bar\sigma} \ZZZ_{\cal Q}^{\rho \bar\alpha} \ZZZ_{\cal R}^{\mu  \bar\sigma} \ZZZ_{\cal S}^{\rho \bar\mu}
   \left(k_{\cal R}^{\bar\alpha}+k_{\cal S}^{\bar\alpha}\right)-k_{\cal S}^{\bar\alpha} \ZZZ_{\cal Q}^{ \rho \bar\alpha}
   \ZZZ_{\cal R}^{\rho \bar\mu} \ZZZ_{\cal S}^{\mu \bar\sigma} \left(k_{\cal Q}^{\bar\sigma}+k_{\cal R}^{\bar\sigma}\right)
  \bigg]\bigg\}.\nn
\end{align}
One can show that the transversality condition $k_{\cal P}^{\bar{\mu}} Z^{\mu\bar{\mu}}_{\cal P} = 0$ holds for barred indices, but not the unbarred ones \cite{Cheung:2017yef}. The amplitudes are computed as before:
\be
\mathcal{A}^{\text{BI}}_n = \lim\limits_{k_n^2 \to 0} s_{12\cdots n{-}1} \Big( \XXX_{12\cdots n{-}1}^{\mu\bar{\mu}} \ZZZ_{n}^{\mu\bar{\mu}} \,+\, \ZZZ_{12\cdots n{-}1}^{\mu\bar{\mu}} \XXX_{n}^{\mu\bar{\mu}} \,+\, \YYY_{12\ldots n-1}^\mu \YYY_{n}^\mu \Big).
\ee
The boundary conditions with $(a)$ $n{-}2$ external $\ZZZ$'s and two $\YYY's$, or $(b)$ $n{-}1$ external $\ZZZ's$ and one $\XXX$ are given as follows:
\begin{align}
(a):& \qquad \big( \XXX^{\mu\bar{\mu}}_i,\; \YYY_i^\mu,\; \ZZZ^{\mu\bar{\mu}}_i \big) = \begin{cases}
\big( 0,\; \varepsilon^\mu_i,\;\, 0\,\; \big)  \qquad &\text{if } i \text{ is a } \YYY\text{-state} \\
\big( 0,\; 0,\; \varepsilon^\mu_i k^{\bar\mu}_i \big)  \qquad &\text{if } i \text{ is a } \ZZZ\text{-state},
\end{cases}\\
(b):& \qquad \big( \XXX^{\mu\bar{\mu}}_i,\; \YYY_i^\mu,\; \ZZZ^{\mu\bar{\mu}}_i \big) = \begin{cases}
\big( \varepsilon_i^\mu \overline{k}_i^{\bar\mu},\; 0,\; \,0\; \big)  \qquad &\text{if } i \text{ is an } \XXX\text{-state},\\
\big( \;0\;,\; 0,\; \varepsilon_i^\mu k^{\bar\mu}_i \big)  \qquad &\text{if } i \text{ is a } \ZZZ\text{-state}.
\end{cases}
\end{align}

\subsubsection{Four-Point Example}

\begin{itemize}[leftmargin=*]
\item
Case ($a$) with $(\ZZZ_1^{\mu\bar{\mu}}, \YYY_{\underline{2}}, \ZZZ_3^{\mu\bar{\mu}}, \YYY_{\underline{4}}) = (\varepsilon_1^{\mu} k_1^{\bar{\mu}}, 1, \varepsilon_3^{\mu} k_3^{\bar{\mu}}, 1)$ and all other $(\XXX_i^{\mu\bar\mu}, \YYY_i^{\mu}, \ZZZ_i^{\mu\bar\mu})$ vanishing. We find the following rank-$2$ currents:
\begin{gather}
\YYY_{1\underline{2}}^\mu = \frac{1}{2\sqrt{2}}\Big( \varepsilon_1 \cdot k_2\, \varepsilon_2^\mu - \varepsilon_2 \cdot k_1\, \varepsilon_1^\mu + \frac{1}{2}\varepsilon_1 \cdot \varepsilon_2 (k_1^\mu - k_2^\mu)\Big),\\
   \YYY_{\underline{2}3}^\mu = \frac{1}{2\sqrt{2}}\Big( \varepsilon_3 \cdot k_2\, \varepsilon_2^\mu - \varepsilon_2 \cdot k_3\, \varepsilon_3^\mu + \frac{1}{2}\varepsilon_3 \cdot \varepsilon_2 (k_3^\mu - k_2^\mu)\Big),\\
   \ZZZ_{13}^{\mu\bar\mu} = -\frac{1}{4\sqrt{2}}\left(k_1^{\bar\mu}-k_3^{\bar\mu}\right) \Big(\left(k_1^\mu-k_3^\mu \right) \varepsilon _1\cdot \varepsilon
   _3-2 \varepsilon _1^\mu  k_1\cdot \varepsilon _3+2 \varepsilon _3^\mu k_3\cdot \varepsilon _1\Big),\\
   \YYY_{13}^\mu = \ZZZ_{1\underline{2}}^{\mu\bar\mu} = \ZZZ_{\underline{2}3}^{\mu\bar\mu} = \XXX_{1\underline{2}}^{\mu\bar\mu} = \XXX_{\underline{2}3}^{\mu\bar\mu} = \XXX_{13}^{\mu\bar\mu} = 0.
\end{gather}
The relevant rank-$3$ current becomes:
\begin{align}
\YYY_{1\underline{2}3}^{\mu} = \frac{1}{32s_{123}}\bigg\{&-k_4\!\cdot\! \varepsilon _2 \varepsilon _1\!\cdot\! \varepsilon _3 s_{12} k_1^{\mu }+k_4\!\cdot\! \varepsilon _1 \varepsilon _2\!\cdot\! \varepsilon
   _3 s_{12} k_1^{\mu }+4 k_1\!\cdot\! \varepsilon _2 \varepsilon _1\!\cdot\! \varepsilon _3 s_{13} k_1^{\mu }+2 k_4\!\cdot\! \varepsilon _2 \varepsilon
   _1\!\cdot\! \varepsilon _3 s_{13} k_1^{\mu }\nn\\
   &+4 k_3\!\cdot\! \varepsilon _1 \varepsilon _2\!\cdot\! \varepsilon _3 s_{13} k_1^{\mu }+3 k_4\!\cdot\!
   \varepsilon _1 \varepsilon _2\!\cdot\! \varepsilon _3 s_{13} k_1^{\mu }+4 k_1\!\cdot\! \varepsilon _2 \varepsilon _1\!\cdot\! \varepsilon _3 s_{23}
   k_1^{\mu }+3 k_4\!\cdot\! \varepsilon _2 \varepsilon _1\!\cdot\! \varepsilon _3 s_{23} k_1^{\mu }\nn\\
   &+4 k_3\!\cdot\! \varepsilon _1 \varepsilon _2\!\cdot\!
   \varepsilon _3 s_{23} k_1^{\mu }+2 k_4\!\cdot\! \varepsilon _1 \varepsilon _2\!\cdot\! \varepsilon _3 s_{23} k_1^{\mu }+2 \varepsilon _2\!\cdot\!
   \varepsilon _3 s_{12}^2 \varepsilon _1^{\mu }-2 \varepsilon _2\!\cdot\! \varepsilon _3 s_{13}^2 \varepsilon _1^{\mu }\nn\\
   &+4 \varepsilon _2\!\cdot\!
   \varepsilon _3 s_{12} s_{13} \varepsilon _1^{\mu }+2 \varepsilon _2\!\cdot\! \varepsilon _3 s_{12} s_{23} \varepsilon _1^{\mu }-2 \varepsilon
   _2\!\cdot\! \varepsilon _3 s_{13} s_{23} \varepsilon _1^{\mu }-2 \varepsilon _1\!\cdot\! \varepsilon _3 s_{12}^2 \varepsilon _2^{\mu }-2 \varepsilon
   _1\!\cdot\! \varepsilon _3 s_{23}^2 \varepsilon _2^{\mu }\nn\\
   &-2 \varepsilon _1\!\cdot\! \varepsilon _3 s_{12} s_{13} \varepsilon _2^{\mu }-2 \varepsilon
   _1\!\cdot\! \varepsilon _3 s_{13} s_{23} \varepsilon _2^{\mu }-2 \Big[\varepsilon _1\!\cdot\! \varepsilon _2 s_{13}^2+4 k_4\!\cdot\! \varepsilon _1
   k_4\!\cdot\! \varepsilon _2 s_{13}+\varepsilon _1\!\cdot\! \varepsilon _2 s_{12} s_{13}\nn\\
   &-\varepsilon _1\!\cdot\! \varepsilon _2 s_{23}^2+2 k_4\!\cdot\!
   \varepsilon _1 k_4\!\cdot\! \varepsilon _2 s_{12}+2 k_3\!\cdot\! \varepsilon _1 k_4\!\cdot\! \varepsilon _2 \left(s_{12}+s_{13}\right)+\big(2
   k_4\!\cdot\! \varepsilon _1 k_4\!\cdot\! \varepsilon _2\nn\\
   &-\varepsilon _1\!\cdot\! \varepsilon _2 \left(s_{12}+2 s_{13}\right)\big) s_{23}+2 k_1\!\cdot\!
   \varepsilon _2 k_4\!\cdot\! \varepsilon _1 \left(s_{12}+2 s_{13}+s_{23}\right)\Big] \varepsilon _3^{\mu }-k_4\!\cdot\! \varepsilon _2 \varepsilon
   _1\!\cdot\! \varepsilon _3 k_2^{\mu } s_{12}\nn\\
   &+k_4\!\cdot\! \varepsilon _1 \varepsilon _2\!\cdot\! \varepsilon _3 k_2^{\mu } s_{12}-4 k_1\!\cdot\! \varepsilon
   _2 \varepsilon _1\!\cdot\! \varepsilon _3 k_3^{\mu } s_{12}-k_4\!\cdot\! \varepsilon _2 \varepsilon _1\!\cdot\! \varepsilon _3 k_3^{\mu } s_{12}-4
   k_3\!\cdot\! \varepsilon _1 \varepsilon _2\!\cdot\! \varepsilon _3 k_3^{\mu } s_{12}\nn\\
   &-3 k_4\!\cdot\! \varepsilon _1 \varepsilon _2\!\cdot\! \varepsilon _3
   k_3^{\mu } s_{12}-2 k_4\!\cdot\! \varepsilon _2 \varepsilon _1\!\cdot\! \varepsilon _3 k_2^{\mu } s_{13}+3 k_4\!\cdot\! \varepsilon _1 \varepsilon
   _2\!\cdot\! \varepsilon _3 k_2^{\mu } s_{13} -4 k_1\!\cdot\! \varepsilon _2 \varepsilon _1\!\cdot\! \varepsilon _3 k_3^{\mu } s_{13}\nn\\
   &-2 k_4\!\cdot\!
   \varepsilon _2 \varepsilon _1\!\cdot\! \varepsilon _3 k_3^{\mu } s_{13}-4 k_3\!\cdot\! \varepsilon _1 \varepsilon _2\!\cdot\! \varepsilon _3 k_3^{\mu }
   s_{13}-5 k_4\!\cdot\! \varepsilon _1 \varepsilon _2\!\cdot\! \varepsilon _3 k_3^{\mu } s_{13}-k_4\!\cdot\! \varepsilon _2 \varepsilon _1\!\cdot\! \varepsilon
   _3 k_2^{\mu } s_{23}\nn\\
   &+2 k_4\!\cdot\! \varepsilon _1 \varepsilon _2\!\cdot\! \varepsilon _3 k_2^{\mu } s_{23}-k_4\!\cdot\! \varepsilon _2 \varepsilon
   _1\!\cdot\! \varepsilon _3 k_3^{\mu } s_{23}-2 k_4\!\cdot\! \varepsilon _1 \varepsilon _2\!\cdot\! \varepsilon _3 k_3^{\mu } s_{23}\nn\\
   &+k_1\!\cdot\! \varepsilon
   _3 \Big[-4 \big(k_4\!\cdot\! \varepsilon _2 \left(s_{13}+s_{23}\right)+k_1\!\cdot\! \varepsilon _2 \left(s_{12}+2
   s_{13}+s_{23}\right)\big) \varepsilon _1^{\mu }\\
   &-4 \left(k_3\!\cdot\! \varepsilon _1+k_4\!\cdot\! \varepsilon _1\right)
   \left(s_{12}+s_{13}\right) \varepsilon _2^{\mu }+\varepsilon _1\!\cdot\! \varepsilon _2 \big(2 \left(k_1^{\mu }-k_2^{\mu }+k_3^{\mu
   }\right) s_{12}\nn\\
   &+\left(k_1^{\mu }{-}3 k_2^{\mu }{+}k_3^{\mu }\right) s_{13}-\left(k_1^{\mu }{+}k_2^{\mu }{+}k_3^{\mu }\right)
   s_{23}\big)\Big]+k_2\!\cdot\! \varepsilon _3 \Big[-4 k_1\!\cdot\! \varepsilon _2 \left(s_{12}+2 s_{13}+s_{23}\right) \varepsilon _1^{\mu
   }\nn\\
   &-4 \big(k_3\!\cdot\! \varepsilon _1 \left(s_{12}+s_{13}\right)+k_4\!\cdot\! \varepsilon _1 \left(s_{12}+2 s_{13}+s_{23}\right)\big)
   \varepsilon _2^{\mu }+\varepsilon _1\!\cdot\! \varepsilon _2 \Big(2 \left(k_1^{\mu }-k_2^{\mu }-k_3^{\mu }\right) s_{12}\nn\\
   &+\left(5 k_1^{\mu
   }-3 \left(k_2^{\mu }+k_3^{\mu }\right)\right) s_{13}+\left(3 k_1^{\mu }-k_2^{\mu }-k_3^{\mu }\right) s_{23}\Big)\Big]\bigg\}.\nn
\end{align}
This gives rise to the $4$-pt amplitude:
\begin{align}
\mathcal{A}^{\text{BI}}_4 =& \lim\limits_{k_4^2 \to 0} s_{123} \YYY_{1\underline{2}3}^{\mu} \YYY_{\underline{4}}^{\mu}\\
=&\, \frac{1}{8} \Big\{s_{23} \Big[\varepsilon _1\!\cdot\! \varepsilon _2 k_1\!\cdot\! \varepsilon _3 \left(k_1\!\cdot\! \varepsilon _4+k_2\!\cdot\! \varepsilon
   _4\right)+\varepsilon _1\!\cdot\! \varepsilon _3 k_2\!\cdot\! \varepsilon _4 k_3\!\cdot\! \varepsilon _2+\varepsilon _1\!\cdot\! \varepsilon _3 k_2\!\cdot\!
   \varepsilon _4 k_4\!\cdot\! \varepsilon _2\nn\\
   &\quad+\varepsilon _1\!\cdot\! \varepsilon _4 k_3\!\cdot\! \varepsilon _2 k_4\!\cdot\! \varepsilon _3+\varepsilon _1\!\cdot\!
   \varepsilon _4 k_4\!\cdot\! \varepsilon _2 k_4\!\cdot\! \varepsilon _3+\varepsilon _2\!\cdot\! \varepsilon _3 k_2\!\cdot\! \varepsilon _1 k_2\!\cdot\! \varepsilon
   _4+k_1\!\cdot\! \varepsilon _4 \Big(\varepsilon _1\!\cdot\! \varepsilon _2 k_4\!\cdot\! \varepsilon _3\nn\\
   &\quad+\varepsilon _1\!\cdot\! \varepsilon _3 \left(k_3\!\cdot\!
   \varepsilon _2+k_4\!\cdot\! \varepsilon _2\right)+\varepsilon _2\!\cdot\! \varepsilon _3 k_2\!\cdot\! \varepsilon _1\Big)+\varepsilon _2\!\cdot\! \varepsilon _4
   k_2\!\cdot\! \varepsilon _1 k_4\!\cdot\! \varepsilon _3+\varepsilon _3\!\cdot\! \varepsilon _4 k_2\!\cdot\! \varepsilon _1 k_3\!\cdot\! \varepsilon _2 \nn\\
   &\quad+\varepsilon
   _3\!\cdot\! \varepsilon _4 k_3\!\cdot\! \varepsilon _1 k_3\!\cdot\! \varepsilon _2+\varepsilon _3\!\cdot\! \varepsilon _4 k_3\!\cdot\! \varepsilon _1 k_4\!\cdot\!
   \varepsilon _2-s_{12} \left(\varepsilon _1\!\cdot\! \varepsilon _4 \varepsilon _2\!\cdot\! \varepsilon _3-\varepsilon _1\!\cdot\! \varepsilon _3 \varepsilon
   _2\!\cdot\! \varepsilon _4+\varepsilon _1\!\cdot\! \varepsilon _2 \varepsilon _3\!\cdot\! \varepsilon _4\right)\Big]\nn\\
   &\quad+s_{12} \Big[\varepsilon _1\!\cdot\!
   \varepsilon _4 k_3\!\cdot\! \varepsilon _2 k_4\!\cdot\! \varepsilon _3+\varepsilon _1\!\cdot\! \varepsilon _4 k_4\!\cdot\! \varepsilon _2 k_4\!\cdot\! \varepsilon
   _3+\varepsilon _2\!\cdot\! \varepsilon _3 k_2\!\cdot\! \varepsilon _4 \left(k_2\!\cdot\! \varepsilon _1+k_3\!\cdot\! \varepsilon _1\right)\nn\\
   &\quad+k_1\!\cdot\! \varepsilon
   _4 \left(\varepsilon _1\!\cdot\! \varepsilon _2 k_4\!\cdot\! \varepsilon _3+\varepsilon _1\!\cdot\! \varepsilon _3 k_3\!\cdot\! \varepsilon _2+\varepsilon _2\!\cdot\!
   \varepsilon _3 k_2\!\cdot\! \varepsilon _1\right)+\varepsilon _2\!\cdot\! \varepsilon _4 k_2\!\cdot\! \varepsilon _1 k_4\!\cdot\! \varepsilon _3+\varepsilon
   _2\!\cdot\! \varepsilon _4 k_3\!\cdot\! \varepsilon _1 k_4\!\cdot\! \varepsilon _3\nn\\
   &\quad+k_1\!\cdot\! \varepsilon _3 \left(\varepsilon _1\!\cdot\! \varepsilon _2
   k_1\!\cdot\! \varepsilon _4+\varepsilon _1\!\cdot\! \varepsilon _4 k_4\!\cdot\! \varepsilon _2+\varepsilon _2\!\cdot\! \varepsilon _4 \left(k_2\!\cdot\! \varepsilon
   _1+k_3\!\cdot\! \varepsilon _1\right)\right)\nn\\
   &\quad+\varepsilon _3\!\cdot\! \varepsilon _4 k_2\!\cdot\! \varepsilon _1 k_3\!\cdot\! \varepsilon _2+\varepsilon _3\!\cdot\!
   \varepsilon _4 k_3\!\cdot\! \varepsilon _1 k_3\!\cdot\! \varepsilon _2-s_{12} \varepsilon _2\!\cdot\! \varepsilon _3 \varepsilon _1\!\cdot\! \varepsilon
   _4\Big] - s_{23}^2 \varepsilon _1\!\cdot\! \varepsilon _2 \varepsilon _3\!\cdot\! \varepsilon _4\Big\}.\nn
\end{align}
One can check that the above amplitude is invariant under gauge transformations $\varepsilon_i^\mu  \to \varepsilon_i^\mu + \alpha k_i^\mu$.

\end{itemize}

\section{\label{sec:double-copy}Double-Copy Relations for Perturbiners}

Double-copy provides a precise statement about relations between scattering amplitudes in different quantum field theories. It comes in two incarnations: color-kinematics duality due to Bern, Carrasco, and Johansson \cite{Bern:2010ue,Bern:2008qj}, and the earlier Kawai--Lewellen--Tye (KLT) relations \cite{Kawai:1985xq}. Since at tree-level, the two descriptions are equivalent, we will focus on the KLT relations. They have a geometric origin in terms of intersection theory of certain cohomology classes defined on the moduli space of punctured Riemann spheres \cite{Mizera:2017cqs,Mizera:2017rqa}. For concreteness, let us discuss a particular form of these relations:
\be\label{eq:KLT-relations}
\mathcal{A}^{\text{theory}_1 \otimes \text{theory}_2}_n = \sum_{ \rho, \tau \in S_{n-3} } \mathcal{A}^{\text{theory}_1}_n(1,\rho,n{-}1,n)\, S[\rho | \tau]_1 \, \mathcal{A}^{\text{theory}_2}_n(1,\tau,n,n{-}1).
\ee
Here the sum goes over $(n{-}3)!$ permutations $\rho$ and $\tau$ of the labels $23\cdots n{-}2$. The KLT matrix $S[\rho | \tau]_1$ is symmetric and has the recursion relations \cite{BjerrumBohr:2010hn,Carrasco:2016ldy}:
\be\label{eq:KLT-matrix}
S[Pj | QjR]_i = k_{iQ} \!\cdot\! k_j\, S[P | QR]_i, \qquad S[\varnothing|\varnothing]_i = 1,
\ee
where $j$ is a single letter and $P,Q,R$ are ordered words. For instance, we have [include whatever we'll need later on]:
\begin{align}\label{eq:KLT-matrix-examples}
& \quad S[2|2]_1 = s_{12}, \qquad S[23|23]_1 = s_{12}(s_{13} + s_{23}),\nn\\
& S[23|32]_1 = s_{12}s_{13}, \qquad S[32|32]_1 = s_{13}(s_{12} + s_{23}).
\end{align}
The KLT matrix $S[\rho | \tau]_1$ can alternatively be computed as an inverse of a matrix whose entries are bi-adjoint scalar amplitudes \cite{Cachazo:2013iea}. In terms of perturbiners $\phi_{P|Q}$ from \cite{Mafra:2016ltu} we have:
\be\label{eq:KLT-inverse}
S^{-1}[\rho | \tau]_i = \phi_{i\rho | i\tau}.
\ee
Note, however, that KLT relations \eqref{eq:KLT-relations} hold much more generally and we could have chosen arbitrary sets of $(n{-}3)!$ orderings to sum over, not necessarily being related by permutations. What the special choice corresponding to $S[\rho|\tau]_i$ gives us are the simple recursion relations \eqref{eq:KLT-matrix} and the polynomial form of the coefficients, which is not obvious from \eqref{eq:KLT-inverse}.

KLT relations \eqref{eq:KLT-relations} relevant to our discussion will be that of $\text{sGal} = \text{NLSM} \otimes \text{NLSM}$ and $\text{BI} = \text{YM} \otimes \text{NLSM}$ \cite{Cachazo:2014xea}. These hold on the level of scattering amplitudes. In the following we investigate whether this statement holds for off-shell quantities of our interest, namely perturbiner coefficients computing Berends--Giele currents. This is by no means guaranteed, as off-shell currents are defined up to field redefinitions and gauge redundancies. However, given that the Lagrangians of \cite{Cheung:2016prv,Cheung:2017yef} manifest their color-kinematics structure, it is reasonable to expect that KLT relations for perturiners should hold, at least in the $\text{sGal} = \text{NLSM} \otimes \text{NLSM}$ case.

The relations we find have the general form:
\be\label{eq:KLT-currents}
\big(J_{12\cdots m}^{\Lambda \bar{\Lambda}}\big)^{\text{theory}_1 \otimes \text{theory}_2} = \sum_{ \rho, \tau \in S_{m-1} } \big(J_{1\rho}^{\Lambda}\big)^{\text{theory}_1} S[\rho | \tau]_1 \, \big(J_{1\tau}^{\bar{\Lambda}}\big)^{\text{theory}_2} \;+\; \big(\Delta^{\Lambda \bar{\Lambda}}_{12\cdots m}\big)^{\text{theory}_1 \otimes \text{theory}_2}.
\ee
On the left-hand side we have a rank-$m$ current for a colorless theory with (optional) external Lorentz indices $\Lambda \bar{\Lambda}$, while on the right-hand side there are currents for colored theories $J_{1\rho}^{\Lambda}$ and $J_{1\tau}^{\bar\Lambda}$ of the same rank with ordered words $\rho$ and $\tau$ of length $m{-}1$. Note that the sum runs over $(m{-}1)!$ permutations and the KLT matrix is functionally the same as that given in \eqref{eq:KLT-matrix}. We also included the discrepancy factor $\Delta^{\Lambda \bar{\Lambda}}_{12\cdots m}$. We find that for fields without gauge redundancies it is exactly zero, while for gauge fields it is pure gauge $\propto k^\mu_{12\cdots m} k^{\bar\mu}_{12\cdots m}$. Naturally, this term vanishes on-shell.

Let us comment on the form of the discrepancy term $\Delta^{\Lambda \bar{\Lambda}}_{12\cdots m}$. For instance, for a tensor current (such as $\XX^{\mu\bar\mu}_{12\cdots m}$ in the case of sGal) we can have the following general types of terms that give vanishing contribution to the amplitudes:
\be
{}^{(1)}\!\Delta\, k_{12\cdots m }^\mu k_{12\cdots m }^{\bar\mu} \;+\; {}^{(2)}\!\Delta^\mu\, k_{12\cdots m }^{\bar\mu} \;+\; {}^{(3)}\!\Delta^{\bar\mu}\, k_{12\cdots m }^\mu \;+\; {}^{(4)}\!\Delta^{\mu\bar\mu}.
\ee
The first three terms vanish on-shell because the tensor current is contracted with the wavefunction of the $(m{+}1)$-st particle, $\ZZ_{m+1}^{\mu\bar\mu} = k_{m+1}^\mu k_{m+1}^{\bar\mu}$, which squares to zero as $k^2_{m+1} = 0$ on-shell. The last term, ${}^{(4)}\!\Delta^{\mu\bar\mu}$, which is not allowed to have a pole in $s_{12\cdots m}$, does not contribute to amplitudes as it is suppressed by a power of $s_{12\cdots m} \to k_{m+1}^2 = 0$ in the numerator when computing an amplitude from a current. For gauge fields, the first three terms can be removed by a suitable gauge transformation. For any type of field, the last term can in principle be removed by a field reparametrization. However, we find that for fields without gauge redundancies, such as $\ZZ^{\mu\bar\mu}, \YY, \ZZZ^{\mu\bar\mu}, \YYY^{\mu}$ no discrepancy terms are present, while for gauge fields $\XX^{\mu\bar\mu}$, $\XXX^{\mu\bar\mu}$ only one of the first three terms is present and can be removed by gauge transformations.

Obtaining amplitude relations from \eqref{eq:KLT-currents} amounts to setting $m=n{-}1$, multiplying both sides by $s_{12\cdots n{-}1}$ times the one-particle wavefunctions $J^{\Lambda\bar{\Lambda}}_{n} = J^{\Lambda}_{n} J^{\bar\Lambda}_{n}$, and taking the on-shell limit. After these operations, we obtain:
\begin{align}\label{eq:n-2-KLT}
\mathcal{A}_{n}^{\text{theory}_1 \otimes \text{theory}_2} &= \lim_{k_n^2 \to 0} s_{12\cdots n-1} \big(J_{12\cdots n-1}^{\Lambda \bar{\Lambda}} J_n^{\Lambda \bar{\Lambda}} \big)^{\text{theory}_1 \otimes \text{theory}_2} \nn\\
&= \lim_{k_n^2 \to 0}
\sum_{ \rho, \tau \in S_{n-2} } \mathcal{A}^{\text{theory}_1}_n(1,\rho,n)\, \frac{S[\rho | \tau]_1}{s_{12\cdots n-1}} \, \mathcal{A}^{\text{theory}_2}_n(1,\tau,n).
\end{align}
These relations differ from the ones presented in \eqref{eq:KLT-relations} in two aspects. First, the summations are over $(n{-}2)!$ instead of $(n{-}3)!$ terms each. Secondly, it needs to be defined with the limit outside of the sum, as the divergent part cancels only after summing over $(n{-}2)!$ terms.\footnote{In other words, the matrix $S[\rho | \tau]_1$ is in the kernel the $(n{-}2)!$ vector of amplitudes $\mathcal{A}^{\text{theory}_2}_n(1,\tau,n)$, which is why the former is often called the \emph{KLT kernel} \cite{BjerrumBohr:2010hn}.} In fact, this gives non-trivial relations between amplitudes first discussed by Bjerrum-Bohr, Damgaard, Feng, and S{\o}ndergaard \cite{BjerrumBohr:2010ta,BjerrumBohr:2010zb}, who also found the $(n{-}2)!$ form of the KLT relations \eqref{eq:n-2-KLT} between Yang--Mills and gravity amplitudes, see also \cite{Feng:2010br,Feng:2010hd}.

In the following sections we give more details behind the KLT formula for perturbiners \eqref{eq:KLT-currents} in the two cases of our interest: $\text{sGal} = \text{NLSM} \otimes \text{NLSM}$ and $\text{BI} = \text{YM} \otimes \text{NLSM}$.

\subsection{\label{sec:sGal-double-copy}$\text{sGal} = \text{NLSM} \otimes \text{NLSM}$}

As described in Sections~\ref{sec:special-Galileon} and \ref{sec:NLSM}, perturbiners for sGal and NLSM theories comprise of triplets $(\XX^{\mu\bar{\mu}}_{\cal P}, \YY_{\cal P}, \ZZ^{\mu\bar{\mu}}_{\cal P})$ and $(X^{\mu}_P,Y_P,Z^{\mu}_P)$ respectively. Double-copy is already manifest on the level of their rank-$1$ currents,
\be
\big(\,\XX^{\mu\bar{\mu}}_i,\, \YY_i,\, \ZZ^{\mu\bar{\mu}}_i\big) = \big(X^\mu_i X^{\bar\mu}_i,\; Y_i\, Y_i,\; Z^{\mu}_i Z^{\bar{\mu}}_i \big),
\ee
as long as we consistently match the external states, i.e., if the $i$-th particle is in the $\XX/\YY/\ZZ$-state in special Galielon, it should also be in the $X/Y/Z$-state in NLSM respectively.

Notice that amplitudes can be computed in four distinct ways. In the case $(a)$ we can use the currents $\YY_{12\cdots n-1}$ or $\XX^{\mu\bar\mu}_{12\cdots n-1}$, depending on the placement of the two special $\YY$-states among $n{-}2$ $\ZZ$-states. Additionally, in the first option, $\YY_{12\cdots n-1}$, we need to specify the placement of a single $\YY$-state (the other is already entering through $\YY_n$), while in the second one, $\XX^{\mu\bar\mu}_{12\cdots n-1}$, one needs to specify positions of two $\YY$-states. In the case $(b)$ the situation is similar: we can use two currents $\ZZ^{\mu\bar\mu}_{12\cdots n-1}$ or $\XX^{\mu\bar\mu}_{12\cdots n-1}$, depending on where the extra $\XX$-state has been put among the remaining $n{-}1$ $\ZZ$-states. In the first option, $\ZZ^{\mu\bar\mu}_{12\cdots n-1}$, there is no further choice, as $\XX_n$ is already saturating the allowed $\XX$-states, while for the second one, $\XX^{\mu\bar\mu}_{12\cdots n-1}$, one needs to specify the position of the $\XX$-state. To summarize, there are four types of currents we consider:
\be\label{eq:currents-sGal}
\YY_{12\cdots \underline{i} \cdots m}, \qquad \XX^{\mu\bar\mu}_{12\cdots \underline{i}\cdots \underline{j}\cdots m}, \qquad \ZZ^{\mu\bar\mu}_{12\cdots m}, \qquad  \XX^{\mu\bar\mu}_{12\cdots \overline{i}\cdots m}.
\ee
Here we used the notation in which $\underline{i}$ denotes that the $i$-th particle is in the $\YY$-picture, $\overline{i}$ that it is the in $\XX$-picture, and in the absence of an underline or overline it is in the $\ZZ$-picture. An entirely analogous analysis holds for the NLSM perturbiners and their boundary conditions, resulting in four types of currents:
\be\label{eq:currents-NLSM}
Y_{12\cdots \underline{i} \cdots m}, \qquad X^{\mu}_{12\cdots \underline{i}\cdots \underline{j}\cdots m}, \qquad Z^{\mu}_{12\cdots m}, \qquad  X^{\mu}_{12\cdots \overline{i}\cdots m}.
\ee
Let us now explore how the above currents of the special Galileon \eqref{eq:currents-sGal} and the NLSM \eqref{eq:currents-NLSM} are related to each other.

\subsubsection{Y-Currents}

We start with the simplest case of the scalar perturbiners $\YY_{\cal P}$ and $Y_P$. Using the recursions \eqref{eq:sGal-recursions} we have the possible rank-$2$ currents for special Galileon:
\be
\YY_{\underline{1}2} = \YY_{1\underline{2}} = - 2 s_{12}.
\ee
Similarly, using \eqref{eq:NLSM-recursion} we have rank-$2$ currents for the NLSM:
\be
Y_{\underline{1}2} = -Y_{1\underline{2}} = -\frac{1}{\sqrt{2}}.
\ee
Hence, recalling that $S[2|2]_1 = s_{12}$ we can relate the two types of perturbiners, for both sets of boundary conditions, through the relations:
\be
\YY_{\underline{1}2} = -4 \left( Y_{\underline{1}2}\, S[2|2]_1\, Y_{\underline{1}2} \right),\qquad \YY_{1\underline{2}} = -4 \left( Y_{1\underline{2}}\, S[2|2]_1\, Y_{1\underline{2}} \right).
\ee

At rank-$3$ the expressions become a little more interesting. Once again, we find that sGal perturbiners are permutation invariant in the labels $123$ and hence independent of the choice of the special label:
\be
\YY_{\underline{1}23} = \YY_{1\underline{2}3} = \YY_{12\underline{3}} = \frac{4(s_{12} + s_{13})(s_{12}+s_{23})(s_{13}+s_{23})}{s_{123}}.
\ee
The corresponding NLSM currents are as follows:
\be
Y_{\underline{1}23} = Y_{12\underline{3}} = \frac{s_{12} + s_{23}}{2s_{123}}, \qquad Y_{1\underline{2}3} = -\frac{s_{12} + 2s_{13} + s_{23}}{2s_{123}}.
\ee
The remaining permutations can be obtained by relabeling $2 \leftrightarrow 3$. Explicitly, we have:
\be
Y_{\underline{1}32} = Y_{13\underline{2}} = \frac{s_{13} + s_{23}}{2s_{123}}, \qquad Y_{1\underline{3}2} = -\frac{s_{13} + 2s_{12} + s_{23}}{2s_{123}}.
\ee
This leads to the following KLT relations:
\be
\YY_{\underline{1}23} = 16 \raisebox{-0.35em}{$\begin{blockarray}{c}
	\begin{block}{[c]}
{ Y_{\underline{1}23} } \topstrut\\
{ Y_{\underline{1}32} } \botstrut\\
\end{block}
\end{blockarray}^{\!\intercal}\!
\begin{blockarray}{cc}
	\begin{block}{[cc]}
		{ S[23 | 23]_1 } & { S[23 | 32]_1 } \topstrut\\
		{ S[32 | 23]_1 } & { S[32 | 32]_1 } \botstrut\\
	\end{block}
\end{blockarray}\,
\begin{blockarray}{c}
	\begin{block}{[c]}
		{ Y_{\underline{1}23} } \topstrut\\
		{ Y_{\underline{1}32} } \botstrut\\
	\end{block}
\end{blockarray}$}.\vspace{-1em}
\ee
Here we used the entries of the KLT matrix $S[\rho|\tau]_1$ from \eqref{eq:KLT-matrix-examples}. Analogous relations hold for the two other choices of boundary conditions with $2$ and $3$ underlined.

Starting at rank-$4$, the expressions for currents become rather involved, and hence we will not display them here explicitly. We confirmed up to rank-$6$ that the following KLT relations hold:
\be
\YY_{12\cdots \underline{i} \cdots m } = (-4)^{m-1} \!\!\!\sum_{ \rho, \tau \in S_{m-1} } Y_{1\rho(23\cdots \underline{i}\cdots m)}\, S[\rho | \tau]_1 \, Y_{1\tau(23\cdots \underline{i}\cdots m)},
\ee
for all assignments of the special particle $\underline{i}$.

\subsubsection{Z-Currents}

We now consider the currents $\ZZ^{\mu\bar{\mu}}_{\cal P}$ and $Z^{\mu}_P$ in the cases of sGal and NLSM, respectively. Note that even though these fields have Lorentz indices and come from Yang--Mills theory in higher dimensions, they are not gauge fields and do not have redundancies in their definition. According to our analysis in \eqref{eq:currents-sGal} and \eqref{eq:currents-NLSM}, we need to set all boundary conditions according to $\ZZ_{i}^{\mu\bar\mu} = k_i^\mu k_i^{\bar\mu}$ and $Z_i^\mu = k_i^\mu$ without special choices. With this picture in mind, we compute rank-$2$ currents for the special Galileon:
\be
\ZZ^{\mu\bar\mu}_{12} = -2 s_{12} \big( k_1^{\mu} - k_2^{\mu} \big) \big( k_1^{\bar\mu} - k_2^{\bar\mu} \big).
\ee
Similarly, for the NLSM we find:
\be
Z^{\mu}_{12} = -\frac{1}{\sqrt{2}} \left(k_1^{\mu} - k_2^{\mu}\right).
\ee
Hence we have straightforwardly:
\be
\ZZ_{12}^{\mu\bar\mu} = -4 \big( Z_{12}^{\mu}\, S[2|2]_1\, Z_{12}^{\bar\mu} \big).
\ee
At rank-$3$ we use the recursion relations to find:
\begin{align}
\ZZ^{\mu\bar\mu}_{123} = \frac{2}{s_{123}} \bigg[ &(s_{13}+s_{23}) \Big( k_1^\mu k_1^{\bar\mu}  (s_{12}+s_{13})(s_{12}+s_{23})\nn \\
&- (k_1^\mu k_2^{\bar\mu} + k_2^\mu k_1^{\bar\mu})(s_{12}(s_{12}+s_{13}+s_{23}) - s_{13}s_{23})\Big) + \text{perm}(1,2,3) \bigg].
\end{align}
In the case of NLSM we have:
\be
Z_{123}^{\mu} = \frac{1}{2s_{123}} \bigg( (k_1^\mu + k_3^\mu) (s_{12} + s_{23}) - k_2^\mu (s_{12} + 2s_{13} + s_{23}) \bigg),
\ee
and similarly $Z_{132}^\mu$ which is related to the above expression by relabeling $2 \leftrightarrow 3$. We can relate these perturbiners using the KLT relation [check factors]:
\be
\ZZ_{123}^{\mu\bar\mu} = 16 \raisebox{-0.35em}{$\begin{blockarray}{c}
	\begin{block}{[c]}
{ Z_{123}^{\mu} } \topstrut\\
{ Z_{132}^{\mu} } \botstrut\\
\end{block}
\end{blockarray}^{\!\intercal}\!
\begin{blockarray}{cc}
	\begin{block}{[cc]}
		{ S[23 | 23]_1 } & { S[23 | 32]_1 } \topstrut\\
		{ S[32 | 23]_1 } & { S[32 | 32]_1 } \botstrut\\
	\end{block}
\end{blockarray}\,
\begin{blockarray}{c}
	\begin{block}{[c]}
		{ Z_{123}^{\bar\mu} } \topstrut\\
		{ Z_{132}^{\bar\mu} } \botstrut\\
	\end{block}
\end{blockarray}$}.\vspace{-1em}
\ee
We find up to rank-$6$ that the above KLT relations generalize to the expression:
\be
\ZZ_{12 \cdots m }^{\mu\bar\mu} = (-4)^{m-1} \!\!\!\sum_{ \rho, \tau \in S_{m-1} } Z^{\mu}_{1\rho(23\cdots m)}\, S[\rho | \tau]_1 \, Z^{\bar\mu}_{1\tau(23\cdots m)}.
\ee

\subsubsection{X-Currents}

Finally, we study double-copy for currents $\XX^{\mu\bar\mu}_{\cal P}$ and $X^{\mu}_P$. Recall from Sections~\ref{sec:special-Galileon} and \ref{sec:NLSM} that these fields posses gauge redundancies, which translate to perturbiners as follows:
\be\label{eq:X-gauge-trafo}
\XX^{\mu\bar\mu}_{\cal P} \to \XX^{\mu\bar\mu}_{\cal P} + k^\mu_{\cal P}\, \mathbf{\Lambda}^{\bar\mu}_{\cal P} + \widetilde{\mathbf{\Lambda}}^{\mu}_{\cal P}\, k^{\bar\mu}_{\cal P}, \qquad\qquad X_P^{\mu} \to X_P^{\mu} + k_P^\mu\, \Lambda_P.
\ee
for some functions $\mathbf{\Lambda}^{\bar\mu}_{\cal P}, \widetilde{\mathbf{\Lambda}}^{\mu}_{\cal P}, \Lambda_P$ of labels ${\cal P}, P$. Unlike in the case of scattering amplitudes, which are gauge-invariant, these transformations act non-trivially on perturbiners. We will be working with gauges selected by their respective recursion relations in \eqref{eq:sGal-recursions} and \eqref{eq:NLSM-recursion}. As mentioned before, there are two possible choices of boundary conditions: $(a)$ with $\underline{i}$ and $\underline{j}$ in the Y-picture, and $(b)$ with $\overline{i}$ in the X-picture. We focus on the former at first.

Using the recursion \eqref{eq:sGal-recursions} for special Galileon theory, we obtain the rank-$2$ current:
\be
\XX^{\mu\bar\mu}_{\underline{1}\underline{2}} = - \frac{1}{2s_{12}} ( k_1^\mu - k_2^\mu ) ( k_1^{\bar\mu} - k_2^{\bar\mu} ).
\ee
Similarly, for NLSM we use \eqref{eq:NLSM-recursion} to find:
\be
X_{\underline{1}\underline{2}}^{\mu} = \frac{1}{2\sqrt{2} s_{12}} (k_1^\mu - k_2^\mu).
\ee
Hence the KLT relation works as before:
\be
\XX^{\mu\bar\mu}_{\underline{1}\underline{2}} = -4 \big( X_{\underline{1}\underline{2}}^{\mu}\, S[2|2]_1\, X_{\underline{1}\underline{2}}^{\bar\mu}\big).
\ee
Rank-$3$ currents have a more interesting structure. Those for sGal are no longer permutation-invariant and depend on the choice of the two special labels,
\begin{align}
\XX_{\underline{1}\underline{2}3}^{\mu\bar\mu} = \frac{1}{s_{123}} \bigg[ &\frac{4}{s_{12}} \big(s_{23} k_1^\mu - s_{13} k_2^\mu \big)\big( s_{23} k_1^{\bar\mu} - s_{13} k_2^{\bar\mu}\big) \nn\\
&+ s_{23} \big(k_1^{\mu} - k_2^{\mu} - k_3^{\mu}\big)\big(k_1^{\bar\mu} - k_2^{\bar\mu} - k_3^{\bar\mu}\big) + s_{13} \big(k_1^{\mu} - k_2^{\mu} + k_3^{\mu}\big)\big(k_1^{\bar\mu} - k_2^{\bar\mu} + k_3^{\bar\mu}\big) \bigg].
\end{align}
The other two choices $\XX_{\underline{1}2\underline{3}}^{\mu\bar\mu}$ and $\XX_{1\underline{2}\underline{3}}^{\mu\bar\mu}$ are related by relabelling of indices. For the NLSM we find the analogous rank-$3$ currents:
\begin{align}
&X_{\underline{1}\underline{2}3}^\mu = \frac{1}{4 s_{123}} \left[ -k_1^\mu + k_2^\mu + k_3^\mu - \frac{2}{s_{12}}\big( s_{23} k_1^\mu - s_{13} k_2^\mu \big) \right], \qquad X_{\underline{1}2\underline{3}}^\mu = - \frac{1}{2s_{123}} k_2^\mu,\\
&\qquad\qquad\qquad X_{1\underline{2}\underline{3}}^\mu = \frac{1}{4 s_{123}} \left[ k_1^\mu + k_2^\mu - k_3^\mu + \frac{2}{s_{23}}\big( s_{13} k_2^\mu - s_{12} k_3^\mu \big) \right].
\end{align}
This time, KLT relations do not work cleanly and one finds a discrepancy:
\be
\XX_{\underline{1}\underline{2}3}^{\mu\bar\mu} = 16 \raisebox{-0.35em}{$\begin{blockarray}{c}
	\begin{block}{[c]}
{ X_{\underline{1}\underline{2}3}^{\mu} } \topstrut\\
{ X_{\underline{1}3\underline{2}}^{\mu} } \botstrut\\
\end{block}
\end{blockarray}^{\!\intercal}\!
\begin{blockarray}{cc}
	\begin{block}{[cc]}
		{ S[23 | 23]_1 } & { S[23 | 32]_1 } \topstrut\\
		{ S[32 | 23]_1 } & { S[32 | 32]_1 } \botstrut\\
	\end{block}
\end{blockarray}\,
\begin{blockarray}{c}
	\begin{block}{[c]}
		{ X_{\underline{1}\underline{2}3}^{\bar\mu} } \topstrut\\
		{ X_{\underline{1}3\underline{2}}^{\bar\mu} } \botstrut\\
	\end{block}
\end{blockarray}$} + \Delta_{\underline{1}\underline{2}3}^{\mu\bar\mu}.\vspace{-1em}
\ee
However, we find that the last term is of pure-gauge form:
\be
\Delta_{\underline{1}\underline{2}3}^{\mu\bar\mu} = \frac{(s_{13}-s_{23})^2}{s_{123}^2} k_{123}^{\mu}\, k_{123}^{\bar\mu}.
\ee
It does not contribute to amplitudes upon a contraction with $k_4^{\mu} k_4^{\bar\mu}$ and imposing momentum conservation. Similarly, $\Delta_{\underline{1}2\underline{3}}^{\mu\bar\mu}$ and $\Delta_{1\underline{2}\underline{3}}^{\mu\bar\mu}$ are obtained by relabelling the above result. From the above expression we see that $\Delta_{\underline{1}\underline{2}3}^{\mu\bar\mu}$ is not permutation invariant and depends on the choice of boundary conditions. Recall that this is the case already for the current $\XX_{\underline{1}\underline{2}3}^{\mu\bar\mu}$ itself.

Let us consider the other type of boundary condition with a single particle in the $X$-picture. For sGal we have:
\be
\XX_{\overline{1}2}^{\mu\bar\mu} = - \frac{2}{s_{12}} \Big( \overline{k}_1 \!\cdot\! k_2\, k_1^\mu - s_{12}\, \overline{k}_1^\mu \Big) \Big( \overline{k}_1 \!\cdot\! k_2 \,k_1^{\bar\mu} - s_{12}\, \overline{k}_1^{\bar\mu} \Big).
\ee
Recall that $\overline{k}_i$ denotes the conjugate momentum such that $\overline{k}_i \!\cdot\! k_i = -1$ and is not related to other meanings of the overline notation for Lorentz indices and particle labels. For $\XX_{1\overline{2}}^{\mu\bar\mu}$ we have a similar result obtained by taking $1 \leftrightarrow 2$. In the case of NLSM, we find:
\be
X_{\overline{1}2}^\mu = \frac{1}{\sqrt{2}}\left(  \frac{\overline{k}_1 \!\cdot\! k_2}{s_{12}} k_1^\mu - \overline{k}_1^{\mu} \right), \qquad X_{1\overline{2}}^\mu = \frac{1}{\sqrt{2}}\left( - \frac{\overline{k}_2 \!\cdot\! k_1}{s_{12}} k_2^\mu + \overline{k}_2^{\mu} \right).
\ee
With these perturbiners we find:
\be
\XX_{\overline{1}2}^{\mu\bar\mu} = -4 \big( X_{\overline{1}2}^{\mu}\, S[2|2]_1\, X_{\overline{1}2}^{\bar\mu} \big),
\ee
and a similar relation for the $\overline{2}$ boundary condition. At rank-$3$, we have:
\begin{align}
\XX_{\overline{1}23}^{\mu\bar\mu} = \frac{4}{s_{123}} \bigg[& \frac{ (\overline{k}_1 \!\cdot\! k_2)^2 }{s_{12}} \Big( s_{23}\, k_1^\mu - s_{13}\, k_2^\mu \Big) \Big( s_{23}\, k_1^{\bar\mu} - s_{13}\, k_2^{\bar\mu} \Big) + \Big( (k_3 \!\cdot\! \overline{k}_1)^2\, s_{12} + s_{23}\, k_2 \!\cdot\! \overline{k}_1\, k_{23} \!\cdot\! \overline{k}_1  \Big) k_1^\mu k_1^{\bar\mu}\nn\\
&+k_3 \!\cdot\! \overline{k}_1 \left( s_{12}\, k_3 \!\cdot\! \overline{k}_1 + (s_{23}-s_{13})\, k_2 \!\cdot\! \overline{k}_1 \right) \left( k_1^\mu k_2^{\bar\mu} + k_2^\mu k_1^{\bar\mu} \right) \nn\\
&+ k_3 \!\cdot\! \overline{k}_1 \left( s_{12}\, k_3 \!\cdot\! \overline{k}_1 - 2 s_{13}\, k_2 \!\cdot\! \overline{k}_1 \right) k_2^\mu k_2^{\bar\mu}  - (s_{12}+s_{23})(s_{13}+s_{23}) k_2 \!\cdot\! \overline{k}_1 \left( k_1^\mu \overline{k}_1^{\bar\mu} + \overline{k}_1^\mu k_1^{\bar\mu}\right) \nn\\
&- (s_{13} + s_{23}) \left( s_{12}\, k_3 \!\cdot\! \overline{k}_1 - s_{13}\, k_2 \!\cdot\! \overline{k}_1 \right) \left( k_2^\mu \overline{k}_1^{\bar\mu} + \overline{k}_1^\mu k_2^{\bar\mu}\right)\nn\\
& + \frac{1}{2} (s_{12}+s_{13})(s_{12}+s_{23})(s_{13}+s_{23}) \overline{k}_1^\mu \overline{k}_1^{\bar\mu} + (2 \leftrightarrow 3) \bigg]
\end{align}
and for the NLSM:
\begin{align}
X_{\overline{1}23}^{\mu} &= \frac{1}{2s_{123}} \bigg[ (s_{12}+s_{23}) \overline{k}_1^\mu - k_2 \!\cdot\! \overline{k}_1\, k_1^\mu - k_3 \!\cdot\! \overline{k}_1\, k_2^\mu + \frac{k_2 \!\cdot\! \overline{k}_1}{s_{12}} \Big( s_{13}\, k_2^\mu - s_{23}\, k_1^\mu \Big) \bigg],\\
X_{1\overline{2}3}^{\mu} &= \frac{1}{2s_{123}} \bigg[ k_3 \!\cdot\! \overline{k}_2 k_{12}^\mu + k_1 \!\cdot\! \overline{k}_2 k_{23}^\mu - (s_{12}+2s_{13}+s_{23})\overline{k}_2^\mu + \frac{k_1 \!\cdot\!\overline{k}_2}{s_{12}} \Big(s_{13} k_2^\mu - s_{23} k_1^\mu \Big) \\
&\qquad\qquad\qquad\qquad\qquad\qquad\qquad\qquad\qquad\qquad\qquad\quad + \frac{k_3 \!\cdot\!\overline{k}_2}{s_{23}} \Big(s_{13} k_2^\mu - s_{12} k_3^\mu \Big) \bigg],\nn\\ 
X_{12\overline{3}}^{\mu} &= \frac{1}{2s_{123}} \bigg[ (s_{12}+s_{23}) \overline{k}_3^\mu - k_1 \!\cdot\! \overline{k}_3\, k_2^\mu - k_2 \!\cdot\! \overline{k}_3\, k_3^\mu + \frac{k_2 \!\cdot\! \overline{k}_3}{s_{23}} \Big( s_{13}\, k_2^\mu - s_{12}\, k_3^\mu \Big) \bigg].
\end{align}
Using the above results we find that, once again, the KLT relations hold up to a pure-gauge term:
\be
\XX_{\overline{1}23}^{\mu\bar\mu} = 16 \raisebox{-0.35em}{$\begin{blockarray}{c}
	\begin{block}{[c]}
{ X_{\overline{1}23}^{\mu} } \topstrut\\
{ X_{\overline{1}32}^{\mu} } \botstrut\\
\end{block}
\end{blockarray}^{\!\intercal}\!
\begin{blockarray}{cc}
	\begin{block}{[cc]}
		{ S[23 | 23]_1 } & { S[23 | 32]_1 } \topstrut\\
		{ S[32 | 23]_1 } & { S[32 | 32]_1 } \botstrut\\
	\end{block}
\end{blockarray}\,
\begin{blockarray}{c}
	\begin{block}{[c]}
		{ X_{\overline{1}23}^{\bar\mu} } \topstrut\\
		{ X_{\overline{1}32}^{\bar\mu} } \botstrut\\
	\end{block}
\end{blockarray}$} + \Delta_{\overline{1}23}^{\mu\bar\mu}\vspace{-1em}
\ee
with
\be
\Delta_{\overline{1}23}^{\mu\bar\mu} = \frac{4 \big( s_{12}\, k_3 \!\cdot\! \overline{k}_1 - s_{13}\, k_2 \!\cdot\! \overline{k}_1 \big)^2 }{s_{123}^2}\, k_{123}^{\mu}\,k_{123}^{\bar\mu}.
\ee
Notice that $\Delta_{\overline{1}23}^{\mu\bar\mu}$ has the same symmetry properties as $\XX_{\overline{1}23}^{\mu\bar\mu}$, i.e., is symmetric in $2 \leftrightarrow 3$. The discrepancy terms for other sets of boundary conditions are obtained by relabeling of the above result. We find up to rank-$5$ that the pattern continues and we have the KLT relations:
\begin{align}
\XX_{12 \cdots \underline{i} \cdots \underline{j} \cdots m }^{\mu\bar\mu} &= (-4)^{m-1} \!\!\!\sum_{ \rho, \tau \in S_{m-1} } X^{\mu}_{1\rho(2 \cdots \underline{i} \cdots \underline{j} \cdots m)}\, S[\rho | \tau]_1 \, X^{\bar\mu}_{1\tau(2 \cdots \underline{i} \cdots \underline{j} \cdots m)} + \Delta^{\mu\bar\mu}_{12 \cdots \underline{i} \cdots \underline{j} \cdots m},\\
\XX_{12 \cdots \overline{i} \cdots m }^{\mu\bar\mu} &= (-4)^{m-1} \!\!\!\sum_{ \rho, \tau \in S_{m-1} } X^{\mu}_{1\rho(2 \cdots \overline{i} \cdots m)}\, S[\rho | \tau]_1 \, X^{\bar\mu}_{1\tau(2 \cdots \overline{i} \cdots m)} + \Delta^{\mu\bar\mu}_{12 \cdots \overline{i} \cdots m},
\end{align}
where the additional terms are proportional to the sum of momenta involved:
\be
\Delta^{\mu\bar\mu}_{12 \cdots \underline{i} \cdots \underline{j} \cdots m} = \frac{f^{(i,j)}_m(\{k_p\})}{s_{12\cdots m}^2}\, k_{12\cdots m}^{\mu}\, k_{12\cdots m}^{\bar\mu},\qquad \Delta^{\mu\bar\mu}_{12 \cdots \overline{i} \cdots m} = \frac{g^{(i)}_m(\{k_p, \overline{k}_p\})}{s_{12\cdots m}^2}\, k_{12\cdots m}^{\mu}\, k_{12\cdots m}^{\bar\mu}.
\ee
Here $f^{(i,j)}_m$ and $g^{(i)}_m$ are scalar rational functions that depend on the specification of the boundary conditions. For example, in the next simplest case we have:
\begin{align}
f^{(1,2)}_4 =\,& \frac{8 s_{14} s_{23} \left(s_{13} s_{24}-s_{14} s_{23}\right)}{s_{12}}-2 \Big[\left(s_{14}+s_{34}\right)
   s_{13}^2+\left(s_{34}^2+\left(s_{14}-2 s_{23}\right) s_{34}-2 s_{23} \left(s_{14}+s_{24}\right)\right) s_{13}\nn\\
   &\qquad\qquad\qquad+s_{23}
   \left(s_{14}^2+\left(s_{23}-2 s_{34}\right) s_{14}+\left(s_{23}+s_{34}\right) \left(s_{24}+s_{34}\right)\right)\Big] + (3 \leftrightarrow 4),\\
   g^{(1)}_4 =\,& 4 k_2\cdot \overbar{k}_1 \Bigg[2 \Big(s_{12} s_{34}^2+\left(s_{12} \left(s_{13}{+}s_{14}\right)+s_{14}
   \left(s_{24}{-}s_{23}\right)\right) s_{34}\nn\\
   &\qquad\qquad+\left(s_{14}{+}s_{24}\right) \left(s_{13} \left(s_{12}{+}s_{24}\right)-s_{14}
   s_{23}\right)\Big)\, k_3\cdot \overbar{k}_1\nn\\
   &\qquad\qquad+\Bigg(\left(-\left(s_{14}{+}s_{24}{+}s_{34}\right)
   s_{13}^2-\left(s_{14}{+}s_{34}\right)^2 s_{13}-s_{14} \left(s_{34}^2+s_{14}
   \left(s_{23}{+}s_{34}\right)\right)\right)\\
   &\qquad\qquad-\frac{\left(s_{14} s_{23}{-}s_{13} s_{24}\right)^2}{s_{12}}\Bigg) k_2\cdot
   \overbar{k}_1\Bigg] + \text{perm}(2,3,4).\nn
\end{align}
We notice that these functions have the same symmetry properties as the corresponding currents, i.e., $f^{(i,j)}_m$ is symmetric in $i$ and $j$, as well as permutation invariant in the remaining labels, and $g^{(i)}_m$ is permutation invariant in all labels except for $i$.

Clearly, the additional discrepancy terms can be removed using gauge transformations \eqref{eq:X-gauge-trafo}. Therefore, the above KLT relations can be simplified after a sufficient gauge fixing for the fields $\XX^{\mu\nu}$ and $\X^\mu$ in the Lagrangians \eqref{eq:sGal-Lagrangian} and \eqref{eq:NLSM-Lagrangian}. An interesting question is whether there exist such gauge-fixing terms introducing only a finite number of vertices.

\subsection{$\text{BI} = \text{YM} \otimes \text{NLSM}$}

In this section we consider Born--Infeld currents, which can be written as a double-copy between currents of Yang--Mills theory and the NLSM. The construction is entirely analogous to the one given in Section~\ref{sec:sGal-double-copy}. Notice that at the level of rank-$1$ currents we have:
\be
\big(\,\XXX^{\mu\bar{\mu}}_i,\, \YYY_i^\mu,\, \ZZZ^{\mu\bar{\mu}}_i\big) = \big(A_i^\mu X^{\bar\mu}_i,\; A_i^\mu Y_i,\; A^{\mu}_i Z^{\bar{\mu}}_i \big),
\ee
where $A_i^\mu = \varepsilon_i^\mu$ are the wavefunctions of the gluons and $(X^{\bar\mu}_i, Y_i, Z^{\bar\mu}_i)$ those of the NLSM.

Before giving the general form of the proposed KLT relations, let us consider examples of rank-$2$ currents. The gauge theory current was already computed in \eqref{eq:A12-example} and reads:
\be
A^{\mu}_{12} = \frac{1}{s_{12}} \Big( \varepsilon_1 \cdot k_2\, \varepsilon_2^\mu - \varepsilon_2 \cdot k_1\, \varepsilon_1^\mu + \frac{1}{2}\varepsilon_1 \cdot \varepsilon_2 (k_1^\mu - k_2^\mu)\Big).
\ee
Recall that we consider Yang--Mills currents in the Lorenz gauge. The relevant currents in NLSM were given in the previous section. The $\YYY$-current in BI theory can be written as:
\begin{gather}
\YYY_{\underline{1}2}^{\mu} = -\frac{1}{2\sqrt{2}}\Big( \varepsilon_1 \cdot k_2\, \varepsilon_2^\mu - \varepsilon_2 \cdot k_1\, \varepsilon_1^\mu + \frac{1}{2}\varepsilon_1 \cdot \varepsilon_2 (k_1^\mu - k_2^\mu)\Big) = \frac{1}{2} \left( A^\mu_{12}\, S[2|2]_1\, Y_{\underline{1}2}\right),
\end{gather}
and similarly for the other choice of boundary conditions, $\YYY_{1\underline{2}}^{\mu} = -\YYY_{\underline{1}2}^{\mu}$. For the $\ZZZ$-current we have:
\be
\ZZZ_{12}^{\mu\bar\mu} = -\frac{1}{2\sqrt{2}}\left(k_1^{\bar\mu}-k_2^{\bar\mu}\right) \Big( \varepsilon_1 \cdot k_2\, \varepsilon_2^\mu - \varepsilon_2 \cdot k_1\, \varepsilon_1^\mu + \frac{1}{2}\varepsilon_1 \cdot \varepsilon_2 (k_1^\mu - k_2^\mu)\Big) = \frac{1}{2} \left( A_{12}^\mu \,S[2|2]_1\, Z^{\bar\mu}_{12} \right),
\ee
which is also written as a KLT of Yang--Mills and NLSM currents. For the $\XXX$-current with two labels in the $\YYY$-states we find:
\be
\XXX^{\mu\bar\mu}_{\underline{12}} = \frac{1}{4\sqrt{2} s_{12}} \Big(  2\, k_1 \!\cdot\! \varepsilon_2 \varepsilon_1^\mu k_2^{\bar\mu} + 2\, k_2 \!\cdot\! \varepsilon_1 \, \varepsilon_2^{\mu} k_1^{\bar\mu} -\varepsilon_1\!\cdot\!\varepsilon_2\, (k_1^\mu k_2^{\bar\mu} + k_2^\mu k_1^{\bar\mu}) \Big) = \frac{1}{2} \left( A^\mu_{12} \,S[2|2]_1\, X_{\overline{1}2}^{\bar\mu} \right) + \delta_{\underline{12}}^{\mu\bar\mu}\nn
\ee
with the pure gauge term:
\be
\delta_{\underline{12}}^{\mu\bar\mu} = \frac{k_{12}^{\bar\mu}}{4\sqrt{2} s_{12}} \left( k_1\cdot\varepsilon_2\, \varepsilon_1^\mu + k_2 \cdot \varepsilon_1\, \varepsilon_2^\mu - \frac{1}{2} \varepsilon_1 \cdot \varepsilon_2 (k_1^\mu + k_2^\mu) \right).
\ee
Finally, for the $\XXX$-current with a single label in the $\XXX$-state we have:
\begin{align}
\XXX^{\mu\bar\mu}_{\overline{1}2} &= \frac{1}{2\sqrt{2} s_{12}} \left(k_2 \!\cdot\! \overline{k}_1 \, k_1^{\bar\mu} - s_{12}\, \overline{k}_1^{\bar\mu}\right) \Big( \varepsilon_1 \cdot k_2\, \varepsilon_2^\mu - \varepsilon_2 \cdot k_1\, \varepsilon_1^\mu + \frac{1}{2}\varepsilon_1 \cdot \varepsilon_2 (k_1^\mu - k_2^\mu)\Big)\nn\\
&= \frac{1}{2} \left( A^{\mu}_{12} \, S[2|2]_1 \, X^{\bar\mu}_{\overline{1}2}\right).\label{eq:BI-X-12}
\end{align}
Similar expression holds for the other choice, $\XXX_{1\overline{2}}$.

We checked that similar pattern continues to higher-multiplicity. They involve more lengthy expressions and hence we will not spell them out here. For the $\YYY$ and $\ZZZ$-current we find KLT relations without any corrections:
\begin{gather}
\YYY_{12\cdots \underline{i} \cdots m }^\mu = 2^{1-m} \!\!\!\sum_{ \rho, \tau \in S_{m-1} } A^{\mu}_{1\rho(23\cdots \underline{i}\cdots m)}\, S[\rho | \tau]_1 \, Y_{1\tau(23\cdots \underline{i}\cdots m)},\\
\ZZZ_{12 \cdots m }^{\mu\bar\mu} = 2^{1-m} \!\!\!\sum_{ \rho, \tau \in S_{m-1} } A^{\mu}_{1\rho(23\cdots m)}\, S[\rho | \tau]_1 \, Z^{\bar\mu}_{1\tau(23\cdots m)},
\end{gather}
For the two types of $\XXX$-currents, we find that they hold up to pure-gauge terms, as follows:
\begin{gather}
\XXX_{12 \cdots \underline{i} \cdots \underline{j} \cdots m }^{\mu\bar\mu} = 2^{1-m} \!\!\!\sum_{ \rho, \tau \in S_{m-1} } A^{\mu}_{1\rho(2 \cdots \underline{i} \cdots \underline{j} \cdots m)}\, S[\rho | \tau]_1 \, X^{\bar\mu}_{1\tau(2 \cdots \underline{i} \cdots \underline{j} \cdots m)} + \delta^{\mu\bar\mu}_{12 \cdots \underline{i} \cdots \underline{j} \cdots m},\\
\XXX_{12 \cdots \overline{i} \cdots m }^{\mu\bar\mu} = 2^{1-m} \!\!\!\sum_{ \rho, \tau \in S_{m-1} } A^{\mu}_{1\rho(2 \cdots \overline{i} \cdots m)}\, S[\rho | \tau]_1 \, X^{\bar\mu}_{1\tau(2 \cdots \overline{i} \cdots m)} + \delta^{\mu\bar\mu}_{12 \cdots \overline{i} \cdots m}.
\end{gather}
Note that both types in general involve pure-gauge terms, even though it did not appear for the rank-$2$ case in \eqref{eq:BI-X-12}. They take the form:
\be
\delta^{\mu\bar\mu}_{12 \cdots \underline{i} \cdots \underline{j} \cdots m} = \frac{h^{\mu,(i,j)}_m (\{ k_p\})}{s_{12\cdots m}^2} k_{12\cdots m}^{\bar\mu}, \qquad \delta^{\mu\bar\mu}_{12 \cdots \overline{i} \cdots m } = \frac{q^{\mu,(i)}_m (\{ k_p, \overline{k}_p\})}{s_{12\cdots m}^2} k_{12\cdots m}^{\bar\mu}
\ee
with $h^{\mu,(i,j)}_m$ and $q^{\mu,(i)}_m$ being rational functions in the kinematic invariants. We confirmed that the above relations hold for all choices of boundary conditions up to rank-$4$.

\section{\label{sec:discussion}Discussion}

The results of this paper can be extended in multiple directions. As explained in Section~\ref{sec:perturbiner-methods}, perturbiner methods can be applied to any quantum field theory, with or without color degrees of freedom. Perturbiner expansions give a generating function of Berends--Giele currents and systematizes the derivation of the off-shell recursion relations. We applied this tool to streamline calculations and study properties of currents.

We generalized the results of \cite{Cheung:2016prv,Cheung:2017yef}, who found a concise Lagrangian for the NLSM in terms of the $(\X,\Y,\Z)$-system, in two different ways. First, we found a similar Lagrangian for the theory of NLSM pions coupled to bi-adjoint scalars in the multi-trace sector. The specific couplings between these fields are selected by the CHY representation of these amplitudes. Given how natural this Lagrangian arises in the $(\X,\Y,\Z)$-formulation, it is reasonable to expect that similar constructions can be made for other mixed theories. A general CHY formula for amplitudes in such theories was proposed in \cite{Cachazo:2016njl}. It would be interesting to explore whether Lagrangians for other types of interactions, for instance $\text{NLSM}\oplus\text{sGal}$, can be written down. Secondly, we showed that inclusion of higher-dimensional operators in the $(\X,\Y,\Z)$-system gives the first correction to the abelian Z-theory Lagrangian. One can ask whether other operators can be constructed systematically, perhaps along the lines of \cite{Elvang:2018dco}, as a parallel to the standard chiral perturbation theory \cite{Leutwyler:2012}.

In Section~\ref{sec:double-copy} we studied the notion of double copy for off-shell currents. We found that the $(\X,\Y,\Z)$-representation of NLSM, sGal, and BI theories leads to currents that satisfy KLT relations without further field redefinitions. In the case of the $\X$-field, which possesses an additional gauge redundancy, the KLT relations hold up to pure gauge terms. One question is whether there exists a finite number of gauge-fixing terms in the Lagrangian that make KLT relations exact.

It is expected that similar KLT relations hold for currents in other theories. This questions is especially important in the view of the recent surge of interest in off-shell double-copy formulations between Yang--Mills and Einstein gravity, see, e.g., \cite{Monteiro:2014cda,Luna:2015paa,Ridgway:2015fdl,Luna:2016due,Cardoso:2016amd,Luna:2016hge,Carrillo-Gonzalez:2017iyj,Shen:2018ebu}. A natural strategy is to turn to string theory, where KLT relations are most cleanly understood. Unfortunately, off-shell currents do not have a straightforward worldsheet formulation. One can, however, define various off-shell continuations of amplitudes by breaking the $\text{SL}(2)$ invariance and relaxing momentum conservation, see, e.g., \cite{Fu:2018hpu}. It is straightforward to prove that this gives an $(n{-}2)!$-dimensional basis of integrands and integration cycles, and hence allows to write down off-shell KLT relations. In the field-theory limit they give precisely the relations \eqref{eq:intro}, as well as a CHY formula supported on $(n{-}2)!$ solutions of the off-shell scattering equations. These results can be obtained in a way entirely analogous to \cite{Mizera:2017cqs,Mizera:2017rqa}. It is however not clear what the physical meaning of such off-shell quantities is. It would be interesting to construct them in such a way that they coincide with Berends--Giele currents in a specific gauge.

\acknowledgments
We are grateful to Freddy Cachazo and especially Oliver Schlotterer for guidance throughout all stages of this work. We thank the organizers and participants of the PSI Winter School 2018, during which this project was initiated, as well as Alfredo Guevara, Chia-Hsien Shen, and Fei Teng for useful discussions and correspondence. This research was supported in part by Perimeter Institute for Theoretical Physics. Research at Perimeter Institute is supported by the Government of Canada through the Department of Innovation, Science and Economic Development Canada and by the Province of Ontario through the Ministry of Research, Innovation and Science.

\appendix
\section{\label{app:example-amplitudes}Further Examples of Amplitudes}

In this appendix we list all non-trivial amplitudes (excluding amplitudes equal to those in pure NLSM or pure BA theories) in the extended NLSM theory computed with \eqref{Extended-NLSM-perturbiner} up to $6$-pt. Recall that amplitudes with exactly two external bi-adjoint scalars are equal to those of all external pions, up to a sign, and hence we do not list them here. Also recall that an amplitude is identically zero if it involves an odd number of NLSM pions, and we skip those cases as well.

At $4$-pt we have two inequivalent double-trace amplitudes:
\begin{align}
&\mathcal{A}^{\text{NLSM}\oplus\text{BA}}_4(\mathbb{I}_4||12|34) = -\frac{s_{23}}{s_{12}}-1,\\
&\mathcal{A}^{\text{NLSM}\oplus\text{BA}}_4(\mathbb{I}_4||13|24)=1.
\end{align}
At $5$-pt there are two single-trace amplitudes:
\begin{align}
&\mathcal{A}^{\text{NLSM}\oplus\text{BA}}_5(\mathbb{I}_5||345)=\frac{1}{2} \left(\frac{s_{12}+s_{23}}{s_{45}}+\frac{s_{51}+s_{12}}{s_{34}}-1\right),\\
&\mathcal{A}^{\text{NLSM}\oplus\text{BA}}_5(\mathbb{I}_5||245)=\frac{1}{2} \left(\frac{s_{12}+s_{23}}{s_{45}}-1\right),
\end{align}
as well as two double-trace ones:
\begin{align}
&\mathcal{A}^{\text{NLSM}\oplus\text{BA}}_5(\mathbb{I}_5||123|45)=-\frac{1}{s_{45}}\left(\frac{s_{34}+s_{45}}{s_{12}}+\frac{s_{45}+s_{51}}{s_{23}}-1\right)\\
&\mathcal{A}^{\text{NLSM}\oplus\text{BA}}_5(\mathbb{I}_5||124|35)=\frac{1}{s_{12}}.
\end{align}
Inequivalent single-trace amplitudes at $6$-pt are given by:
\begin{align}
&\mathcal{A}^{\text{NLSM}\oplus\text{BA}}_6(\mathbb{I}_6||3456)=\frac{1}{2s_{45}}\left(\frac{s_{12}{+}s_{61}}{s_{345}}-\frac{\left(s_{45}{+}s_{56}\right) s_{13}}{s_{56}s_{123}}\right)+\frac{1}{2s_{34}}\left(\frac{s_{12}{+}s_{61}}{s_{345}}+\frac{s_{12}{+}s_{234}}{s_{56}}-1\right),\\
&\mathcal{A}^{\text{NLSM}\oplus\text{BA}}_6(\mathbb{I}_6||2356)=\frac{1}{2}\left(\frac{s_{123}+s_{234}}{ s_{23} s_{56}}-\frac{1}{ s_{23}}-\frac{1}{s_{56}}\right),\\
&\mathcal{A}^{\text{NLSM}\oplus\text{BA}}_6(\mathbb{I}_6||2456)=-\frac{\left(s_{45}+s_{56}\right) s_{13}}{2 s_{45} s_{56} s_{123}},\\
&\mathcal{A}^{\text{NLSM}\oplus\text{BA}}_6(\mathbb{I}_6||4356)=-\frac{1}{2s_{34}}\left(\frac{s_{12}{+}s_{61}}{s_{345}}+ \frac{s_{234}{-}s_{34}}{s_{56}} + \frac{s_{23} s_{34}+s_{12} \left(s_{34}{+}s_{123}\right)}{s_{56} s_{123}}-1\right),\\
&\mathcal{A}^{\text{NLSM}\oplus\text{BA}}_6(\mathbb{I}_6||3256)=-\frac{1}{2}\left(\frac{s_{123}+s_{234}}{s_{23} s_{56}}-\frac{1}{s_{23}}-\frac{1}{s_{56}}\right),\\
&\mathcal{A}^{\text{NLSM}\oplus\text{BA}}_6(\mathbb{I}_6||4156)=-\frac{1}{2 s_{56}}\left(\frac{s_{12}+s_{23}}{s_{123}}+\frac{s_{23}+s_{34}}{s_{234}}-1\right),\\
&\mathcal{A}^{\text{NLSM}\oplus\text{BA}}_6(\mathbb{I}_6||3146)=0.
\end{align}
We also have the following double-trace amplitudes with four external bi-adjoint scalars:
\begin{align}
&\mathcal{A}^{\text{NLSM}\oplus\text{BA}}_6(\mathbb{I}_6||14|56)=\frac{1}{2s_{56}}\left( {-}\frac{\left(s_{12}{+}s_{23}\right) \left(s_{45}{+}s_{56}\right)}{s_{123}} {-}\frac{\left(s_{23}{+}s_{34}\right) \left(s_{56}{+}s_{61}\right)}{ s_{234}} {+} s_{23}{+}s_{345}{+}s_{56}\right)\!,\\
&\mathcal{A}^{\text{NLSM}\oplus\text{BA}}_6(\mathbb{I}_6||24|56)=\frac{s_{13}}{{2 s_{56}}} \left(\frac{s_{56}+s_{61}}{s_{234}}-\frac{s_{46}}{s_{123}}\right),\\
&\mathcal{A}^{\text{NLSM}\oplus\text{BA}}_6(\mathbb{I}_6||34|56)=\frac{1}{2 s_{34}}\bigg(
-\frac{s_{12}\left(s_{23}{+}s_{34}\right) \left(s_{56}{+}s_{61}\right)}{s_{56} s_{234}}-\frac{\left(s_{34}{+}s_{45}\right)\left(s_{12}{+}s_{61}\right)}{s_{345}}\\
&\qquad\qquad\qquad\qquad\qquad\qquad\qquad+\frac{s_{34} \left(s_{23}{+}s_{45}{-}s_{61}\right)+s_{12}   \left(s_{23}{+}s_{34}{-}s_{45}{+}s_{56}{+}s_{61}\right)}{s_{56}}\nn\\
&\qquad\qquad\qquad\qquad\qquad\qquad\qquad+\frac{\left(-s_{23}{-}s_{56}{+}s_{123}{+}s_{234}\right)   s_{345}}{s_{56}}-\frac{\left(s_{12}{+}s_{23}\right) s_{34} \left(s_{45}{+}s_{56}\right)}{s_{56} s_{123}}\nn\\
&\qquad\qquad\qquad\qquad\qquad\qquad\qquad -\frac{\left(s_{34}{+}s_{61}\right)   s_{123}}{s_{56}} -\frac{\left(s_{12}{+}s_{45}{-}s_{123}\right) s_{234}}{s_{56}} -s_{23}{+}s_{34}{+}s_{45}{+}s_{61}
\bigg),\nn\\
&\mathcal{A}^{\text{NLSM}\oplus\text{BA}}_6(\mathbb{I}_6||23|56)= \frac{1}{2s_{123}}\bigg( \frac{s_{12}}{s_{23}}\left(  \frac{s_{45} s_{234} }{s_{56}} +s_{234}{-}s_{45} {-}s_{56} \right) + \frac{s_{45} (s_{234}{-}s_{12}{-}s_{23})}{s_{56}} \\
&\qquad\qquad\qquad\qquad\qquad + s_{234} {-} s_{12} {-}s_{23}{-}s_{45}{-}s_{56} \bigg) + \frac{1}{2s_{234}} \bigg( \frac{s_{34}}{s_{23}}\left( \frac{s_{61} s_{123}}{
   s_{56}} + s_{123} {-} s_{56} {-} s_{61} \right) \nn\\
   &\qquad\qquad\qquad\qquad\qquad +\frac{s_{61} (s_{123}{-}s_{23}{-}s_{34})}{s_{56}}+ s_{123} {-} s_{23}{-}s_{34}{-}s_{56}{-} s_{61}\bigg)\nn\\
&\qquad\qquad\qquad\qquad\qquad
+\frac{1}{2s_{23}} \left( \frac{s_{45}
   s_{12} {+} s_{34} s_{61} {-} s_{123} s_{345} {-} s_{234}
   s_{345}}{s_{56}} -s_{123} {-} s_{234} {+} s_{345} {+}  s_{12} {+} s_{34} {+} s_{56} \right)\nn\\
&\qquad\qquad\qquad\qquad\qquad+\frac{-s_{123}{-}s_{234}{+}s_{345}{+}s_{23}{+}s_{45} {+} s_{61}}{2 s_{56}} + 2,\nn\\
&\mathcal{A}^{\text{NLSM}\oplus\text{BA}}_6(\mathbb{I}_6||13|46)=\frac{-s_{34}-s_{61}+s_{234}+s_{345}}{2 s_{123}},\\
&\mathcal{A}^{\text{NLSM}\oplus\text{BA}}_6(\mathbb{I}_6||35|46)=\frac{1}{2} \left(\frac{s_{12}+s_{23}}{s_{123}}+\frac{s_{12}+s_{61}}{s_{345}}-1\right),\\
&\mathcal{A}^{\text{NLSM}\oplus\text{BA}}_6(\mathbb{I}_6||25|46)=\frac{1}{2} \left(\frac{s_{12}+s_{23}}{s_{123}}-1\right),\\
&\mathcal{A}^{\text{NLSM}\oplus\text{BA}}_6(\mathbb{I}_6||25|36)=0,
\end{align}
as well as those with six external bi-adjoint scalars:
\begin{align}
&\mathcal{A}^{\text{NLSM}\oplus\text{BA}}_6(\mathbb{I}_6||1234|56)=
-\frac{1}{s_{123}}\left( 1 + \frac{s_{45}}{s_{56}} \right)\left( \frac{1}{s_{12}} + \frac{1}{s_{23}} \right)
-\frac{1}{s_{234}}\left( 1 + \frac{s_{61}}{s_{56}}\right)\left(  \frac{1}{s_{23}} + \frac{1}{s_{34}} \right)\\
&\qquad\qquad\qquad\qquad\qquad\qquad +\frac{1}{s_{56}} \left( \frac{s_{12}{+}s_{34}{-}s_{345}}{s_{12} s_{34}}+\frac{1}{s_{23}}\right)-\frac{1}{s_{12}s_{34}},\nn\\
&\mathcal{A}^{\text{NLSM}\oplus\text{BA}}_6(\mathbb{I}_6||1342|56)=\frac{1}{s_{34} s_{56}}\bigg(\frac{s_{56}+s_{61}}{s_{234}}+\frac{s_{123} \left(s_{56}+s_{345}\right)-s_{34} s_{46}}{s_{12}
   s_{123}}-1\bigg),\\
&\mathcal{A}^{\text{NLSM}\oplus\text{BA}}_6(\mathbb{I}_6||1324|56)=\frac{1}{s_{23}s_{56}}\left(\frac{s_{56}+s_{61}}{ s_{234}}-\frac{s_{46}}{s_{123}}\right),\\
&\mathcal{A}^{\text{NLSM}\oplus\text{BA}}_6(\mathbb{I}_6||1235|46)=\frac{1}{s_{123}}\left(\frac{1}{s_{23}}+\frac{1}{s_{12}}\right),\\
&\mathcal{A}^{\text{NLSM}\oplus\text{BA}}_6(\mathbb{I}_6||1352|46)=-\frac{1}{s_{12} s_{123}},\\
&\mathcal{A}^{\text{NLSM}\oplus\text{BA}}_6(\mathbb{I}_6||1325|46)=-\frac{1}{s_{23} s_{123}},\\
&\mathcal{A}^{\text{NLSM}\oplus\text{BA}}_6(\mathbb{I}_6||1245|36)=\frac{1}{s_{12} s_{45}},\\
&\mathcal{A}^{\text{NLSM}\oplus\text{BA}}_6(\mathbb{I}_6||1452|36)=-\frac{1}{s_{12} s_{45}},\\
&\mathcal{A}^{\text{NLSM}\oplus\text{BA}}_6(\mathbb{I}_6||1425|36)=0,\\
&\mathcal{A}^{\text{NLSM}\oplus\text{BA}}_6(\mathbb{I}_6||123|456)=\frac{1}{s_{123}}\bigg( \frac{1}{s_{45}}\left( -\frac{s_{123}{+}s_{345}}{s_{12}} -\frac{s_{61}{+}s_{123}}{s_{23}} +1 \right)\\
&\qquad\qquad\qquad\qquad\qquad\qquad\qquad + \frac{1}{s_{56}}\left( \frac{s_{56}{-}s_{34}{-}s_{123}}{s_{12}} + \frac{s_{56}{-}s_{123}{-}s_{234}}{s_{23}} + 1\right)\bigg),\nn\\
&\mathcal{A}^{\text{NLSM}\oplus\text{BA}}_6(\mathbb{I}_6||124|356)=\frac{1}{s_{12} s_{56}},\\
&\mathcal{A}^{\text{NLSM}\oplus\text{BA}}_6(\mathbb{I}_6||135|246)=0.
\end{align}
Finally, we list $6$-pt triple-trace amplitudes:
\begin{align}
&\mathcal{A}^{\text{NLSM}\oplus\text{BA}}_6(\mathbb{I}_6||14|23|56)= \frac{1}{s_{123}}\left( 1 + \frac{s_{12}}{s_{23}}\right)\left(1 + \frac{s_{45}}{s_{56}} \right) +\frac{1}{s_{234}}\left(1 + \frac{s_{34}}{s_{23}
   }\right)\left(1+\frac{s_{61}}{s_{56}} \right)\\
   &\qquad\qquad\qquad\qquad\qquad\qquad -\frac{s_{345} {+} s_{23} {+} s_{56}}{s_{23}
   s_{56}},\nn\\
&\mathcal{A}^{\text{NLSM}\oplus\text{BA}}_6(\mathbb{I}_6||13|24|56)=-\frac{1}{s_{56}}\left(\frac{s_{45}}{s_{123}}+\frac{s_{61}}{s_{234}}-1\right)-\frac{1}{s_{123}}-\frac{1}{s_{234}},\\
&\mathcal{A}^{\text{NLSM}\oplus\text{BA}}_6(\mathbb{I}_6||12|34|56)=
\frac{1}{s_{123}}{+}\frac{1}{s_{234}}{+}\frac{1}{s_{345}}{+}\frac{1}{s_{12}}\left(\frac{s_{23}}{s_{123}}+\frac{s_{61}}{s_{345}}-1\right)\\
& \qquad\qquad\qquad\qquad\qquad\qquad {+}\frac{1}{s_{34}}\left(\frac{s_{23}}{s_{234}}+\frac{s_{45}}{s_{345}}-1\right){+}\frac{1}{s_{56}}\left(\frac{s_{45}}{s_{123}}+\frac{s_{61}}{s_{234}}-1\right)\nn\\
& \qquad\qquad\qquad\qquad\qquad\qquad {+}\frac{1}{s_{12}s_{34}}\left(\frac{s_{45}s_{61}}{s_{123}}+ s_{23}{-}s_{45}{-}s_{61}{+}s_{345}\right)\nn\\
& \qquad\qquad\qquad\qquad\qquad\qquad {+}\frac{1}{s_{34}s_{56}}\left(\frac{s_{23}s_{61}}{s_{234}}+s_{23}{+}s_{45}{-}s_{61}{+}s_{234}\right)\nn\\
& \qquad\qquad\qquad\qquad\qquad\qquad {+}\frac{1}{s_{12}s_{56}}\left(\frac{s_{23}s_{45}}{s_{123}}-s_{23}{-}s_{45}{+}s_{61}{+}s_{123}\right)\nn\\
& \qquad\qquad\qquad\qquad\qquad\qquad {+}\frac{s_{123}s_{61}{+}s_{45}s_{234}{-}s_{123}s_{234}{+}s_{23}s_{345}{-}s_{123}s_{345}{-}s_{234}s_{345}}{s_{12}s_{34}s_{56}},\nn\\
&\mathcal{A}^{\text{NLSM}\oplus\text{BA}}_6(\mathbb{I}_6||13|25|46)=\frac{1}{s_{123}},\\
&\mathcal{A}^{\text{NLSM}\oplus\text{BA}}_6(\mathbb{I}_6||14|25|36)=0.
\end{align}
Other partial amplitudes can be obtained by relabeling of indices. We checked that they agree with the computation using the CHY formula \eqref{eq:CHY-multi-trace}, up to a sign.

\bibliographystyle{JHEP}
\bibliography{references}

\end{document}